\documentclass[11pt,3p]{elsarticle}
\makeatletter
\def\ps@pprintTitle{\let\@oddhead\@empty
  \let\@evenhead\@empty
  \def\@oddfoot{\reset@font\hfil\thepage\hfil}
  \let\@evenfoot\@oddfoot
}
\makeatother

\usepackage{amsmath}
\usepackage{amsfonts}
\usepackage{amssymb}
\usepackage{bbm}
\usepackage{amsfonts}
\usepackage{amsthm}
\newtheorem{theorem}{Theorem}[section]

\usepackage{amscd}
\usepackage{mathtools}
\usepackage{mathrsfs}
\usepackage{subcaption}
\usepackage{commath}
\usepackage{lmodern}
\usepackage{color}
\usepackage{a4wide}
\usepackage{bm} 
\usepackage{physics}
\usepackage{enumitem}  
\usepackage{bbm}
\usepackage{booktabs}
\usepackage{hyperref}
\usepackage{float}
\usepackage{graphicx}
\usepackage{svg} 
\hypersetup{ colorlinks=true, urlcolor  = blue, linkcolor = blue,
citecolor = blue1,}
\usepackage{comment}
\usepackage[english]{babel}
\usepackage[IL2]{fontenc}
\usepackage[utf8]{inputenc}

\usepackage[linesnumbered,ruled,vlined]{algorithm2e}
\usepackage[noend]{algpseudocode}

\usepackage{amsmath,amssymb,mathtools,bm}
\usepackage{enumitem}
\usepackage{physics}

\newcommand{\JJ}{\mathbf{J}}
\newcommand{\RR}{\mathbf{R}}

\SetKwInput{KwInput}{Input}                
\SetKwInput{KwOutput}{Output}              

\SetCommentSty{mycommfont}

\newtheorem{lemma}[theorem]{Lemma}

\theoremstyle{definition}
\newtheorem{definition}[theorem]
{Definition}
\theoremstyle{remark}
\theoremstyle{remark}
\newtheorem{remark}{\upshape\bfseries Remark}

\usepackage[dvipsnames]{xcolor}
\newcommand{\marius}[1]{\textcolor{black}{ #1} }

\begin{document}

\begin{frontmatter}

\title{\marius{A Lyapunov stability proof and a port-Hamiltonian physics-informed neural network for chaotic synchronization in memristive neurons}}
\author[1]{Behnam Babaeian}
\ead{behnam.babaeian@fau.de}
\address[1]{Department Informatik, Friedrich-Alexander-Universit\"at Erlangen-N\"urnberg, Martens Str. 3, 91058 Erlangen, Germany}
\author[2]{Marius E. Yamakou}
\ead{marius.yamakou@fau.de}
\address[2]{Department of Data Science, Friedrich-Alexander-Universit\"at Erlangen-N\"urnberg, N\"urnberger Str. 74, 91052 Erlangen, Germany}

\begin{abstract}
\marius{We study chaotic synchronization in a 5D Hindmarsh--Rose neuron model augmented with electromagnetic induction and a switchable memristive autapse. For two diffusively coupled identical neurons, we derive the transverse error dynamical system and analyze local synchronization via the linearized error system around the synchronization manifold. A quadratic Lyapunov function yields explicit sufficient conditions for (i) asymptotic stability when the memristive switching remains dissipative and (ii) practical stability with an explicit ultimate bound under non-dissipative switching. We complement this with a Hamiltonian-based viewpoint: a Helmholtz decomposition of the linearized error vector field provides a closed-form synchronization Hamiltonian and its rate identity. Numerical simulations corroborate convergence or ultimate boundedness of the synchronization errors and an overall decay of the synchronization Hamiltonian and its instantaneous rate toward zero after transients, and show consistent trends between Lyapunov- and Hamiltonian-based diagnostics across parameters. Finally, we propose the first port-Hamiltonian physics-informed neural network (pH-PINN) that learns this synchronization Hamiltonian and its rate from data while preserving conservative/dissipative structure, achieving close agreement with the analytical expressions.}
\end{abstract}

\begin{keyword}
\marius{memristive neurons}, \marius{chaos}, synchronization,  Lyapunov function, \marius{Hamiltonian function}, port-Hamiltonian neural networks, physics-informed neural networks
\end{keyword}
\end{frontmatter}

\section{Introduction}
\label{sec:introduction}

Synchronization refers to the process by which interacting dynamical systems adjust aspects of their motion—such as amplitude, phase, or frequency—through coupling. In neural systems, this adjustment reflects coordinated electrical activity among neurons or brain regions and plays a key role in both physiological and pathological function. Coherent synchronization supports essential cognitive processes, including perception, memory, and attention~\cite{neustadter2016eeg}, whereas excessive or uncontrolled synchronization can underlie epileptic seizures and related disorders~\cite{lehnertz2009synchronization}. Experimental and modeling studies show that changes in synaptic strength or network connectivity can induce transitions between asynchronous, weakly synchronized, and hyper-synchronized states~\cite{borges2023intermittency,protachevicz2019bistable}. Comprehensive reviews appear in~\cite{boccaletti2006complex,osipov2007synchronization}.

Among the various notions of synchrony—such as phase synchronization~\cite{rosenblum1996phase,pikovsky1996synchronization,parlitz1996experimental,pietras2019network,fell2011role}, cluster synchronization~\cite{dahms2012cluster}, phase-synchronized clusters~\cite{jalan2003self,amritkar2003self}, generalized synchronization~\cite{abarbanel1996generalized,zheng2000generalized}, reduced-order synchronization~\cite{femat2002synchronization,bowong2006adaptive,bowong2004stability,kobiolka2025reduced}, and increased-order synchronization~\cite{qing2009increasing,al2011adaptive}—we focus here on complete synchronization of same-order chaotic systems~\cite{pecora2015synchronization,yamakou2023synchronization}. In this regime, the trajectories of two coupled systems converge so that $\lvert x_1(t)-x_2(t)\rvert \to 0$ as $t \to \infty$~\cite{osipov2007synchronization,pecora2015synchronization,fujisaka1983stability,yamakou2016ratcheting}. Complete synchronization provides the most direct measure of coherence and a tractable setting for Lyapunov stability analysis. Although other types—such as phase synchronization, characterized by $2\pi$-locked phases with uncorrelated amplitudes—play essential roles in distributed brain dynamics, we emphasize complete synchronization as a foundation for understanding the energetic and structural basis of synchronization in nonlinear neurons.

Synchronization has been widely analyzed in neural models of varying complexity~\cite{tang2014synchronization,boccaletti2002synchronization,arenas2008synchronization}. Interactions between excitatory and inhibitory populations regulate both phase and anti-phase locking~\cite{protachevicz2021emergence}, while coupling strength and connection probability influence the onset of spike or bursting synchronization~\cite{borges2017synchronised}. Historically, these phenomena have been studied using classical dynamical systems theory, providing analytical and numerical tools to characterize coherent activity. Within this framework, the present work investigates complete synchronization between two electromagnetically and memristively coupled Hindmarsh--Rose (HR) neurons. We provide a Lyapunov-based proof of asymptotic and \marius{practical} stability, corroborated by numerical simulations, and introduce a physics-informed machine-learning framework that learns the associated \marius{Hamiltonian} landscape directly from data, \marius{while respecting the conservative/dissipative structure of the synchronization Hamiltonian.}

In recent years, the physics-informed neural network (PINN) framework has become a powerful approach for solving forward and inverse problems governed by differential equations~\cite{Raissi2019,Wang2022CausalityPINN,Cuomo2022PINNReview}. The key idea is to embed governing physical laws—typically expressed as ordinary or partial differential equations—directly into the learning objective. By penalizing equation residuals during training, the model enforces physical consistency and improves generalization beyond observed data~\cite{Cuomo2022PINNReview}. For example,~\cite{Wang2022CausalityPINN} demonstrated that causality-aware loss functions enable stable learning of chaotic and turbulent systems such as the Lorenz attractor and the Navier–Stokes equations. Along related lines, Savaliya and Yamakou~\cite{savaliya2026self} used a noise-augmented PINN surrogate to model and predict self-induced stochastic resonance in the stochastic FitzHugh--Nagumo neuron by embedding the governing SDE residuals together with Kramers escape-time (barrier-matching) constraints directly into the training loss.

A complementary development is the Hamiltonian neural network (HNN)~\cite{Greydanus2019HNN}, which learns the underlying Hamiltonian function of a dynamical system. An HNN represents the dynamics using a learned Hamiltonian $H_\theta(x)$ together with a fixed skew-symmetric interconnection matrix enforcing \marius{Hamiltonian} conservation. This enables unsupervised identification of conservative dynamics from trajectory data. Extensions such as~\cite{finzi2020simplifying} and~\cite{Chen2021} incorporated structural priors, coordinate transformations, and learned symplectic forms to improve identifiability and data efficiency. Nonetheless, classical HNNs apply only to conservative systems and cannot represent dissipation or environmental interactions, which are fundamental in neuronal dynamics.

To address these limitations, HNNs were extended to port--Hamiltonian neural networks (pHNNs)~\cite{vanderSchaft2014}, which generalize Hamiltonian mechanics to include energy exchange, dissipation, and external actuation. A pHNN decomposes the vector field into a skew-symmetric interconnection matrix and a symmetric positive-semidefinite dissipation matrix coupled through energy ports. The work in~\cite{Desai2021PortHNN} demonstrated that such factorizations can be learned directly from data, recovering both conservative and dissipative components. More recent studies~\cite{neary2023phnn,Moradi2025OePHNN} introduced compositional and rollout-based variants for model learning over finite prediction windows.

A parallel line of research has extended the pHNN framework to stochastic dynamical systems. The study in~\cite{di2025port} formulated a mean--square port--Hamiltonian neural network (MS--pHNN) that decomposes stochastic dynamics into Hamiltonian drift, dissipative drift, and diffusion-induced correction terms. The authors proposed a mean--square passivity condition for stability under noise and introduced a drift--diffusion loss enabling the learning of stochastic energy landscapes. While providing a principled formulation for stochastic pH systems, this line of work focuses on noisy mechanical oscillators and does not address synchronization or physics-informed residual training.

Despite these advances, existing pHNN methods remain computationally demanding for stiff or chaotic systems and face identifiability challenges: multiple combinations of Hamiltonian, interconnection, and dissipation terms may reproduce the same trajectories~\cite{ortega2024learnability,finzi2020simplifying}. This can lead to ambiguous or physically inconsistent energy representations when training is unconstrained.

Beyond general dynamical systems modeling, these neural frameworks have growing relevance in computational neuroscience, where nonlinear oscillators such as FitzHugh–Nagumo or HR neurons display multistability, bursting, and chaotic synchronization. Physics-informed and Hamiltonian-based learning methods offer an interpretable bridge between data-driven modeling and biophysical realism. By incorporating principles such as energy conservation and dissipation, they help illuminate the energetic mechanisms underlying neuronal synchrony—a key determinant of coherent neural activity and information processing. Unifying Hamiltonian structure with physics-informed regularization thus provides a principled and biologically meaningful way to study the energetics of synchronization and decoherence in complex neural systems.

Motivated by these challenges, we retain the structural advantages of the port--Hamiltonian formulation while regularizing learning through PINN-style residual constraints. To our knowledge, no prior work has unified port--Hamiltonian factorization with physics-informed residual training in a single framework. We therefore propose the port--Hamiltonian physics-informed neural network (pH--PINN), which combines the interpretability of \marius{Hamiltonian}-based modeling with the consistency of residual-driven training. Specifically, the pH--PINN enforces the governing differential equations and uses a Helmholtz decomposition of the vector field to associate the divergence-free (\marius{Hamiltonian}-conserving) component with the Hamiltonian interconnection flow and the curl-free (dissipative) component with the non-conservative flow. This dual constraint improves identifiability, prevents degenerate \marius{Hamiltonian} landscapes, and alleviates the need for long rollout windows—making it well-suited for studying the energetic and stability aspects of synchronization in chaotic systems. As shown here, the pH--PINN accurately recovers both the Hamiltonian function and its time derivative, bridging analytical dynamical systems theory with data-driven discovery.

The paper is structured as follows: Section~\ref{sec:math_model_dynamics} introduces the 5D HR neuron model and its dynamics, illustrated through bifurcation diagrams. Section~\ref{sec:sync} establishes \marius{asymptotic and practical stability} of the synchronization manifold via Lyapunov analysis. Section~\ref{sec:numerics} presents numerical simulations that corroborate the theoretical results. Section~\ref{sec:pHNN_PINN} develops the pH--PINN framework for data-driven learning of the \marius{synchronization Hamiltonian and its rate identity}. Finally, Section~\ref{sec:summary_conclusions} summarizes the findings and discusses potential extensions.

\section{Mathematical model and dynamics}
\label{sec:math_model_dynamics}
We investigate a paradigmatic neuron model of established biological relevance—the HR model~\cite{hindmarsh1984model,selverston2000reliable}—extended to incorporate additional biophysical mechanisms. Specifically, we consider the effects of electromagnetic induction~\cite{lv2016model,yamakou2020chaotic} and an autaptic feedback current governed by a switchable memristive synapse, \marius{resulting in} a new five-dimensional formulation of the HR neuron model  \cite{zhang2024switchable} is given by
\begin{equation}\label{eq:1}
\left\{
\begin{aligned}
\dot{x} &= y - a x^3 + b x^2 + k(h + f u^2) x + \rho \phi x + I,\\
\dot{y} &= c - d x^2 - y,\\
\dot{z} &= r \left[ s(x + \marius{x_r}) - z \right],\\
\dot{u} &= -u + m \left( |u + 1| - |u - 1| \right) + x,\\
\dot{\phi} &= x - q \phi.
\end{aligned}
\right.
\end{equation}
where \(x\) and \(y\) represent the membrane potential and the fast recovery variable, respectively, while \(z\) accounts for the slow adaptation current. The variable \(u\) denotes the internal state of the switchable memristor governing the autaptic feedback current, and \(\phi\) represents the magnetic flux \marius{associated with} electromagnetic induction. The parameters \(a, b, c, d, r, s, k, h, f, m, q, \rho, \marius{x_r},\) and \(I\) are real constants that control the intrinsic dynamics, coupling, and feedback strength of the system.

This formulation extends the classical three-dimensional HR model \cite{hindmarsh1984model} by introducing two additional dynamical variables—\(u\) and \(\phi\)—thereby capturing both electromagnetic and memristive effects. These extensions enable the model to reproduce a broad spectrum of dynamical behaviors, including bursting, chaotic oscillations, and self-modulated activity induced by autaptic feedback.

The nonlinear term \(m(|u + 1| - |u - 1|)\) in Eq.~\eqref{eq:1} models the current–voltage characteristic of the switchable memristor. It provides a continuous, piecewise-linear approximation of the internal switching behavior, with the parameter \(m\) determining the memory strength of the autaptic feedback~\cite{zhang2024switchable}. The state variable \(u\), driven by both the membrane potential \(x\) and its decay term \(-u\), introduces an adaptive self-coupling mechanism that enables activity-dependent modulation of neuronal excitability. Dynamically, this feedback acts as a nonlinear memory term that enriches the system’s phase-space structure and supports transitions between quiescent, periodic, and chaotic regimes.

Unless stated otherwise, parameters are fixed at their standard values~\cite{yamakou2020chaotic}: 
\(a=1.0,\; b=3.0,\; c=1.0,\; d=5.0,\; f=0.2,\; h=0.3,\; k=0.87,\; m=0.5, q=0.005,\;r=0.006,\; s=5.2,\; \marius{x_r}=-1.56,\; \rho=0.7,\; I=0.8\).
The parameters \(k\), \(\rho\), and \(m\) are treated as bifurcation parameters and systematically varied within prescribed ranges to explore qualitative changes in the dynamics via local and global bifurcations.  

\marius{These parameters play distinct roles associated with the two added biophysical mechanisms.
Electromagnetic induction is captured by the magnetic flux variable \(\phi\), whose relaxation rate is controlled by
\(q\) in \(\dot{\phi}=x-q\phi\), and whose feedback onto the membrane potential is governed by the coupling coefficient
\(\rho\) through the term \(\rho\,\phi\,x\) \cite{ma2017mode}.
Memristive/autaptic feedback is captured by the internal state \(u\), whose switching nonlinearity is controlled by
\(m\in[0,1]\) via \(m(|u+1|-|u-1|)\), and whose modulation of the membrane potential is weighted by the gain \(k\) and
parameters \(h,f\) through the term \(k(h+fu^{2})x\) \cite{zhang2024switchable}.
Accordingly, varying \(\rho\) and \(q\) primarily tunes the induction pathway, whereas varying \(m\), \(k\), \(h\), and
\(f\) tunes the strength and nonlinearity of the memristive/autaptic pathway.
} Variations in these parameters lead to transitions among equilibrium states, periodic oscillations, and chaotic attractors, revealing a rich bifurcation structure underlying the model’s nonlinear behavior.

\begin{figure}
    \centering
    \includegraphics[width=0.32\textwidth]{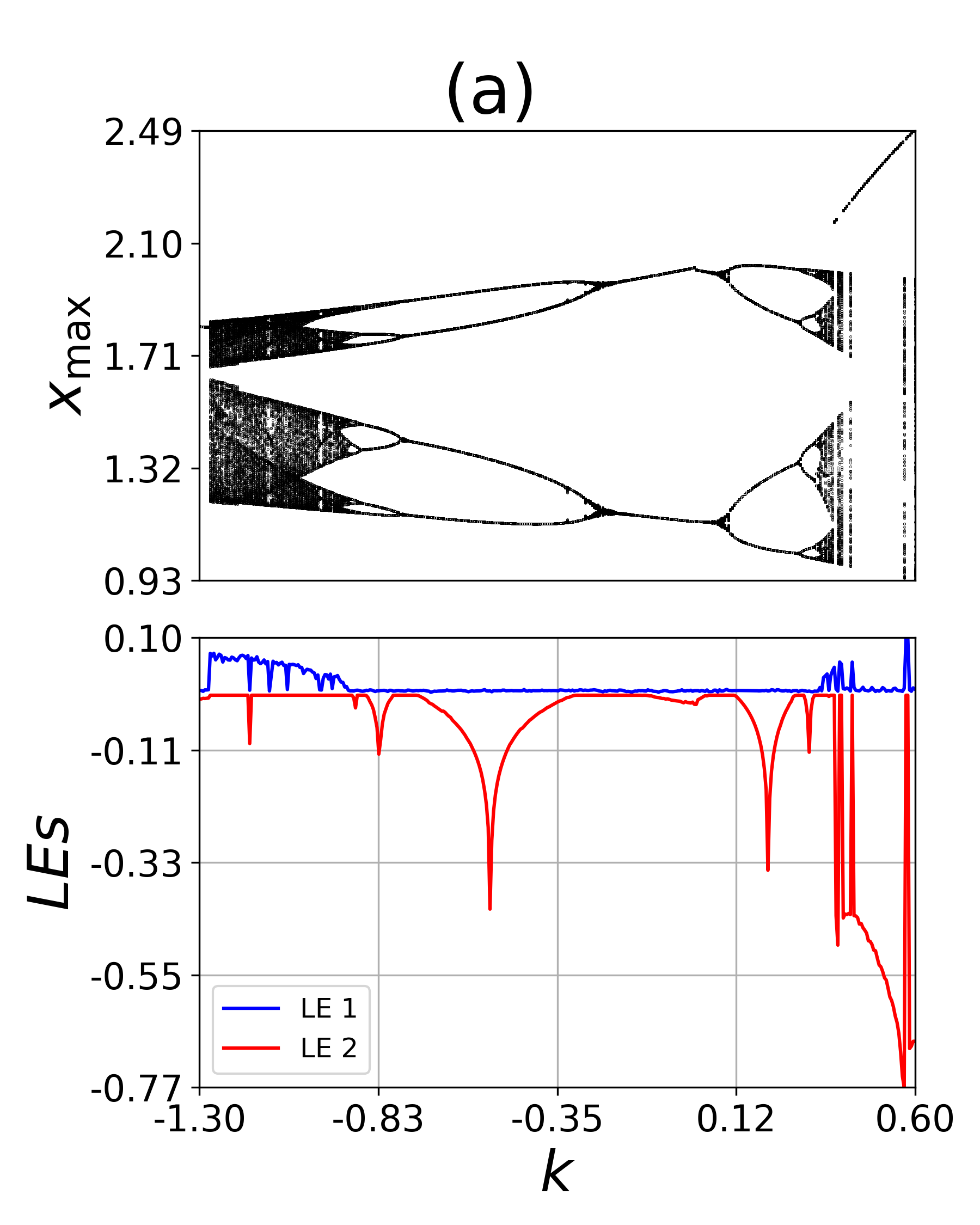}
    \includegraphics[width=0.32\textwidth]{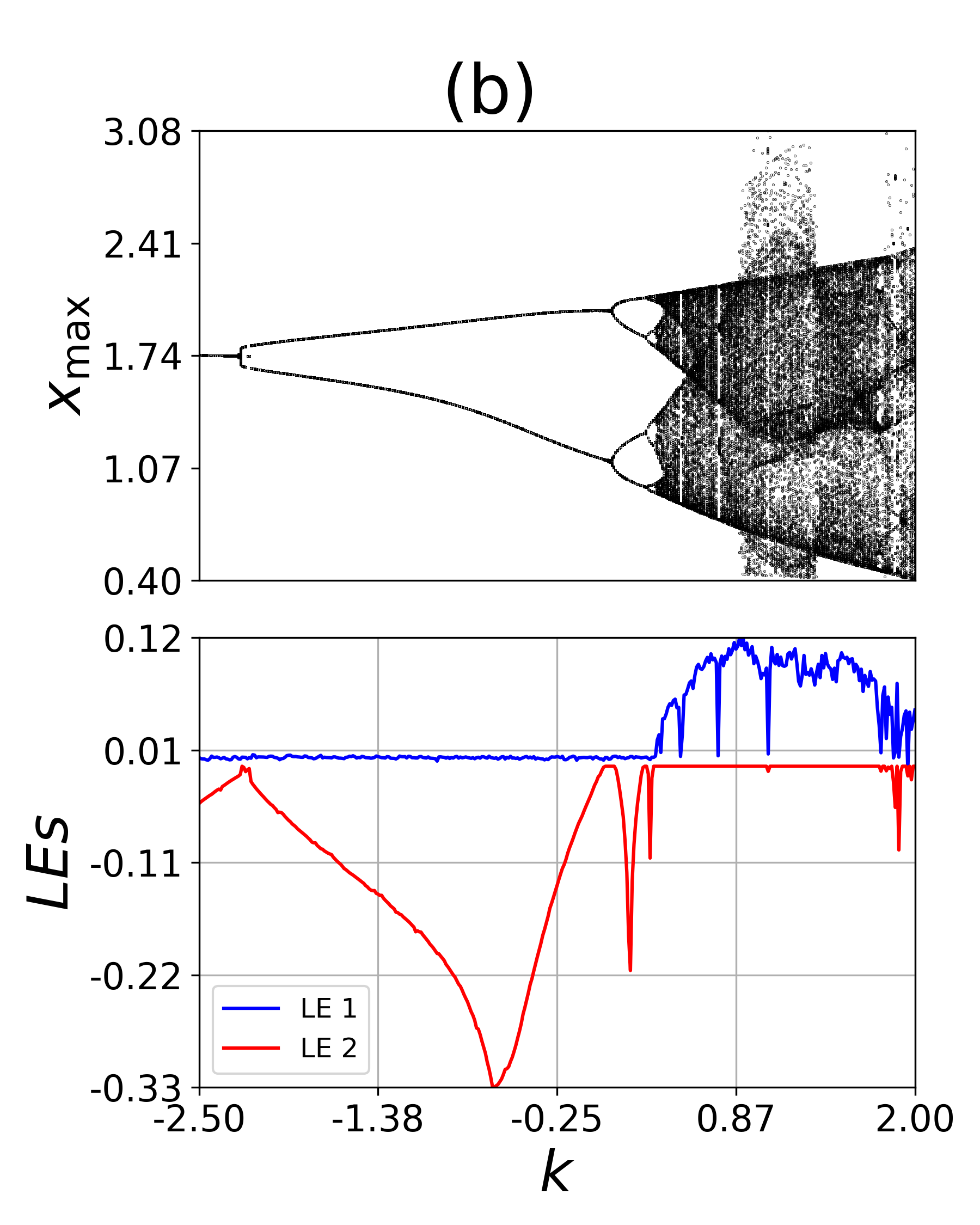}
    \includegraphics[width=0.32\textwidth]{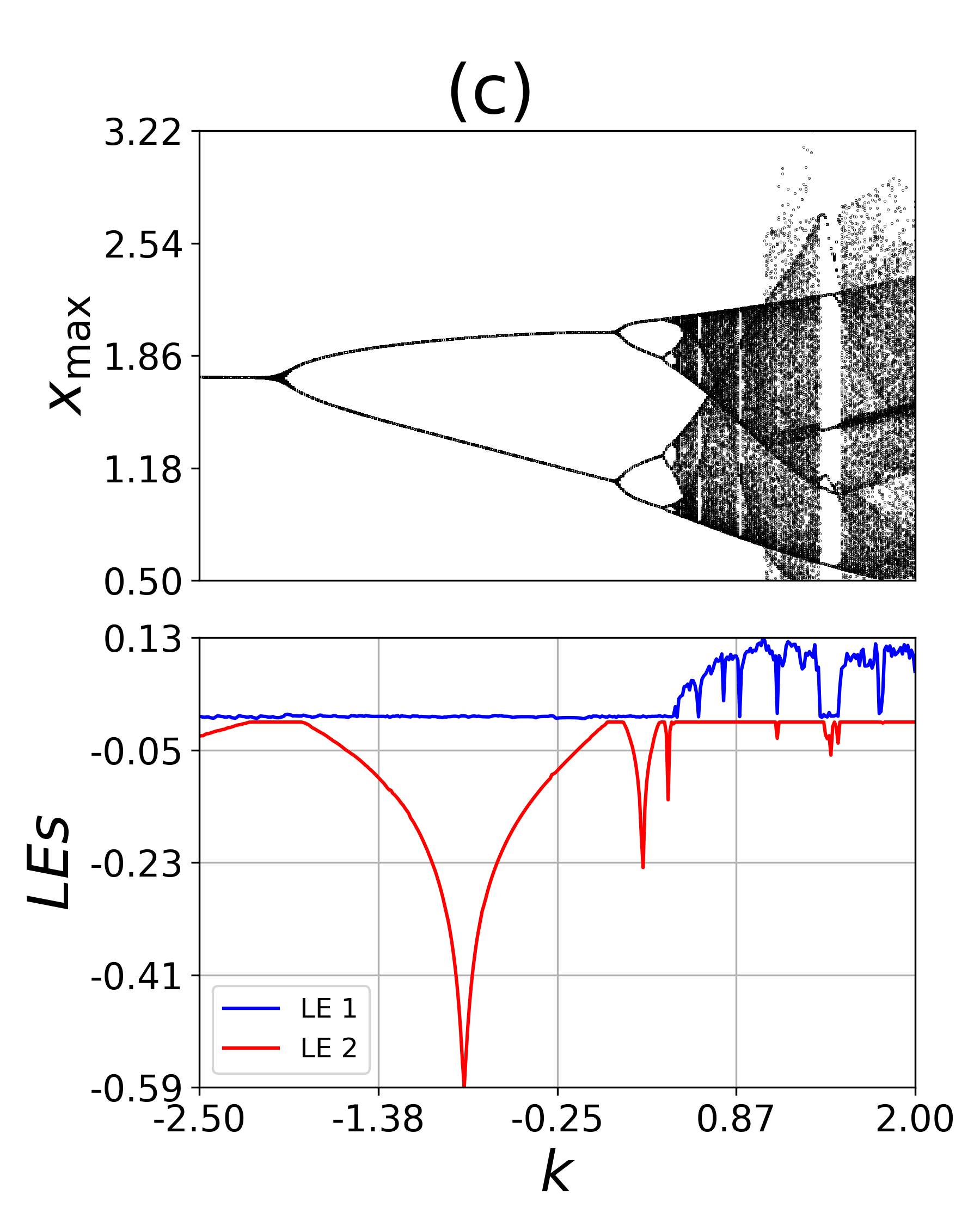}
    \caption{Bifurcation diagrams and the two largest Lyapunov exponents (LEs) versus $k$ at $\rho = 0.7$ for different values of $m$: \textbf{(a)} $m = 1$, \textbf{(b)} $m = 0.5$, \textbf{(c)} $m = 0.25$.}
    \label{fig:bifurcation_k}
\end{figure}

\begin{figure}
    \centering
    \includegraphics[width=0.32\textwidth]{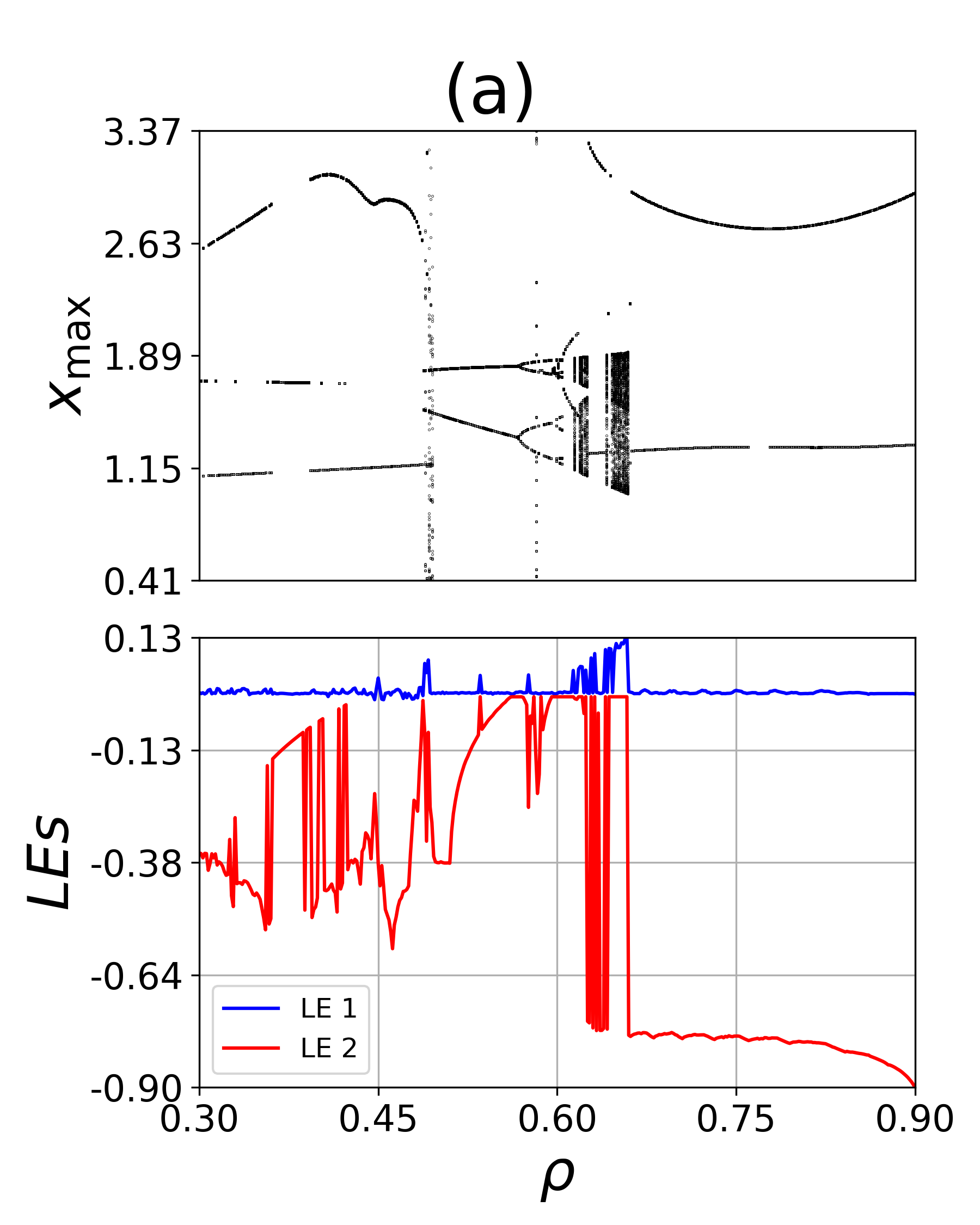}
    \includegraphics[width=0.32\textwidth]{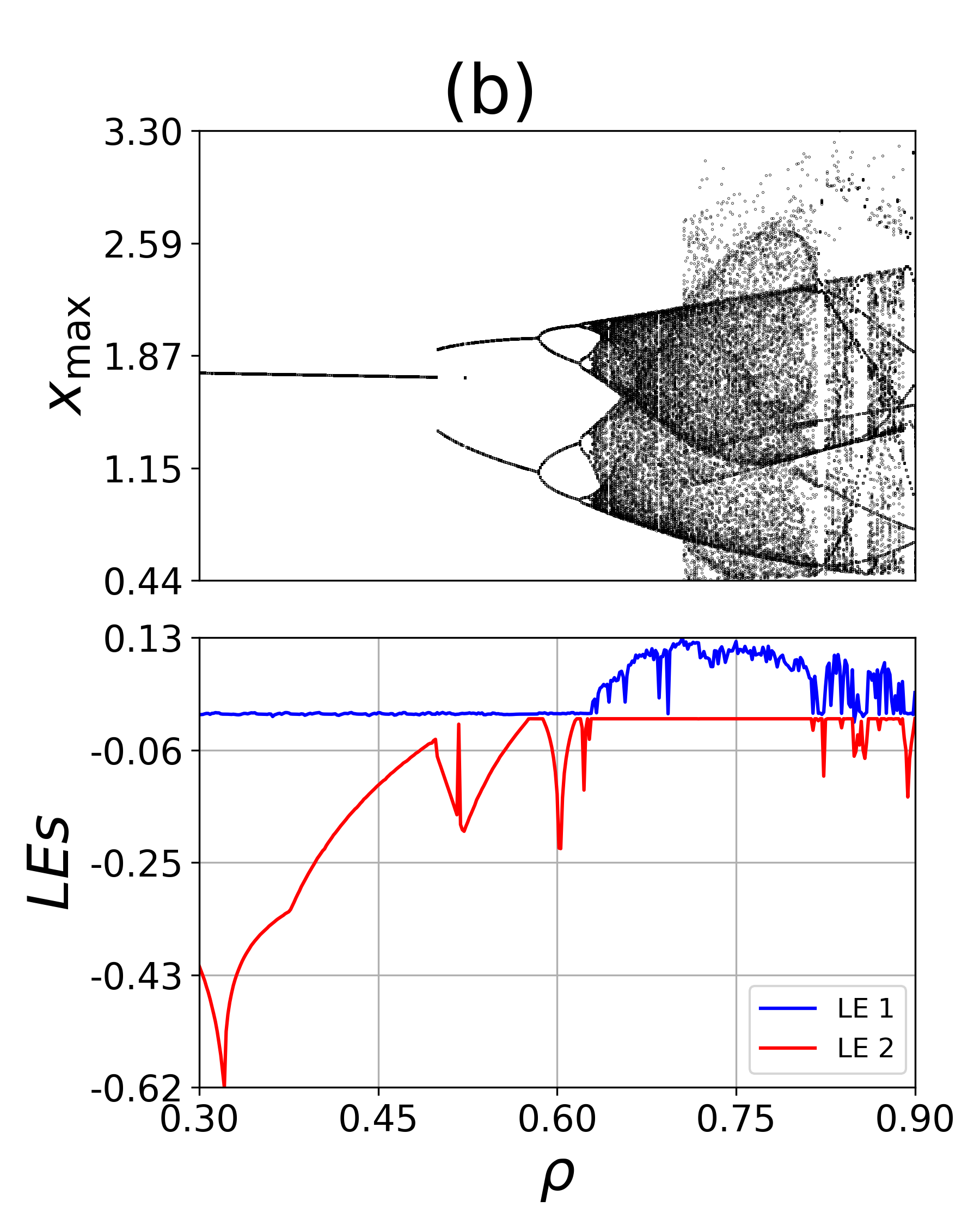}
    \includegraphics[width=0.32\textwidth]{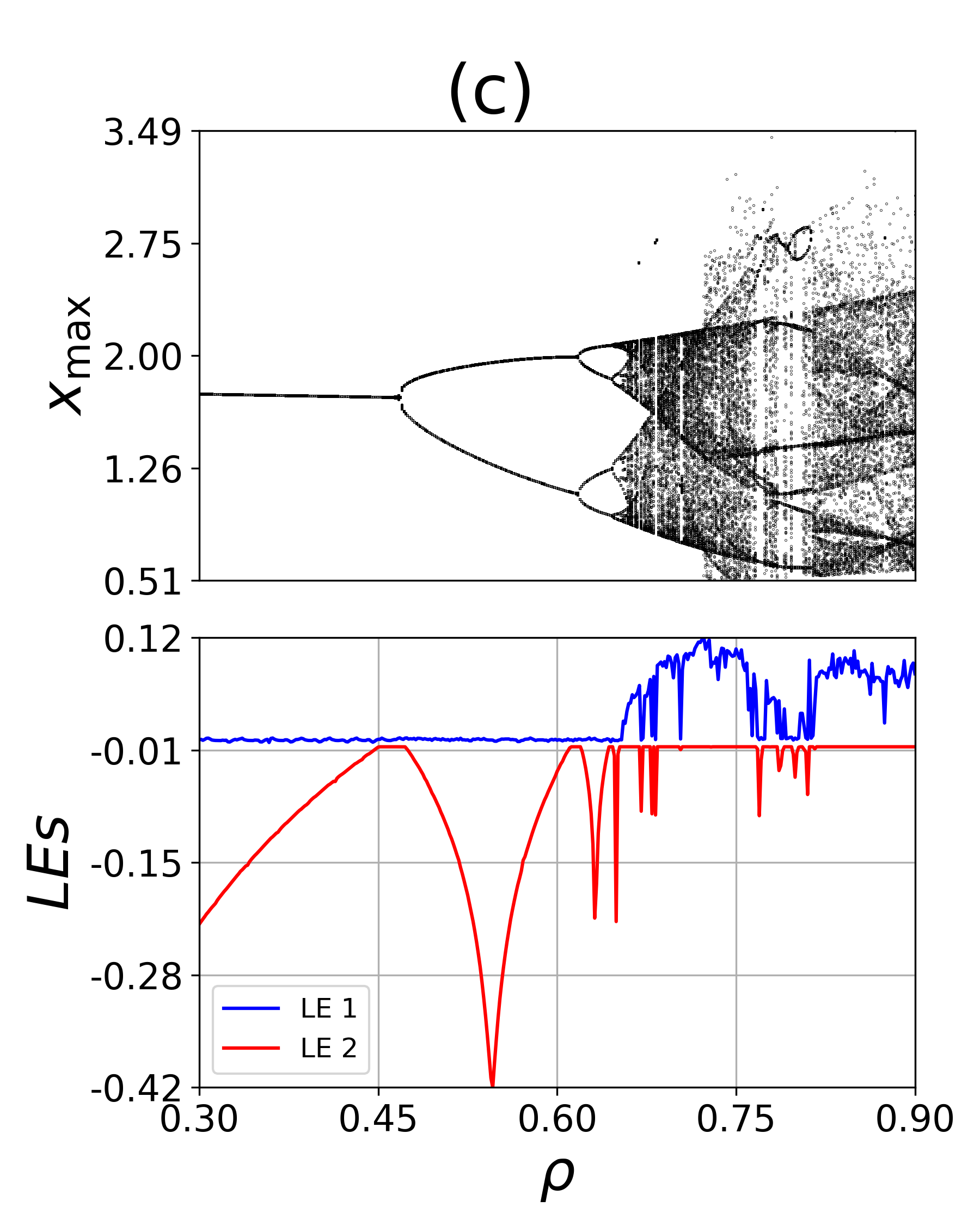}
    \caption{Bifurcation diagrams and the two largest Lyapunov exponents (LEs) versus $\rho$ at $k = 0.87$ for different values of $m$: \textbf{(a)} $m = 1$, \textbf{(b)} $m = 0.5$, \textbf{(c)} $m = 0.25$.}
    \label{fig:bifurcation_rho} 
\end{figure}

\begin{figure}[H]
    \centering
    \includegraphics[width=5.1cm,height=4.7cm]{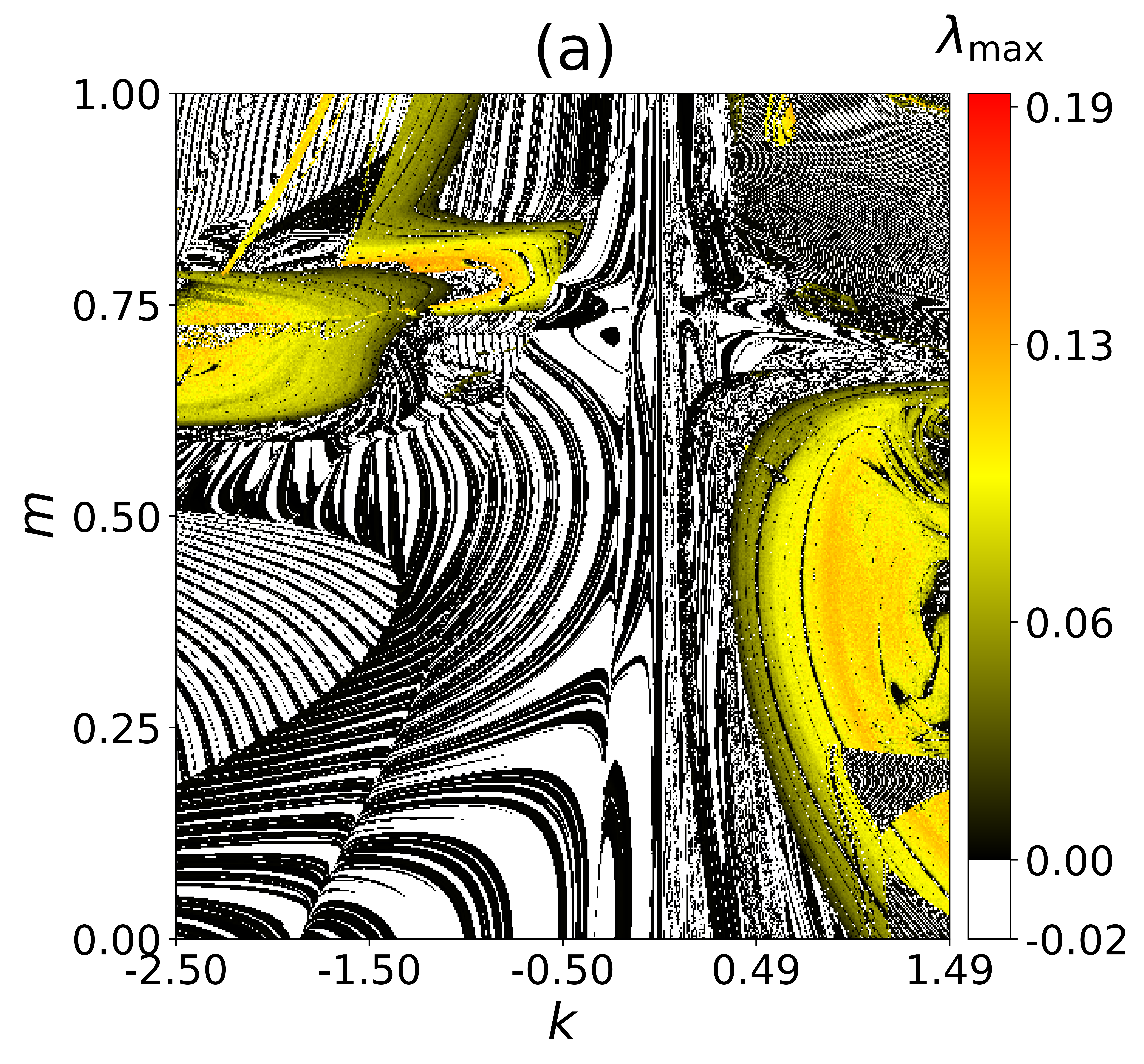}
    \includegraphics[width=5.1cm,height=4.7cm]{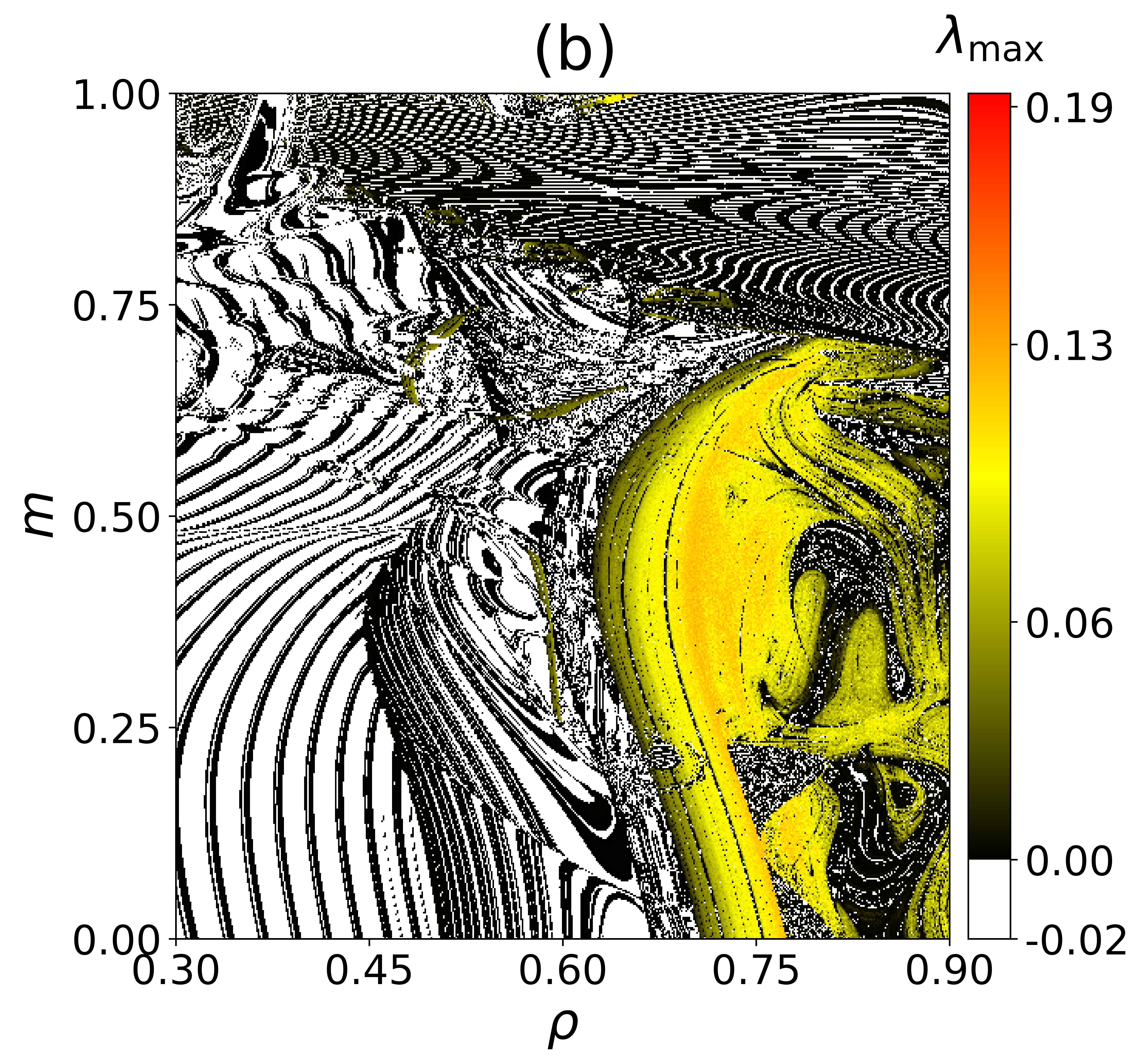}

    \includegraphics[width=5.1cm,height=4.7cm]{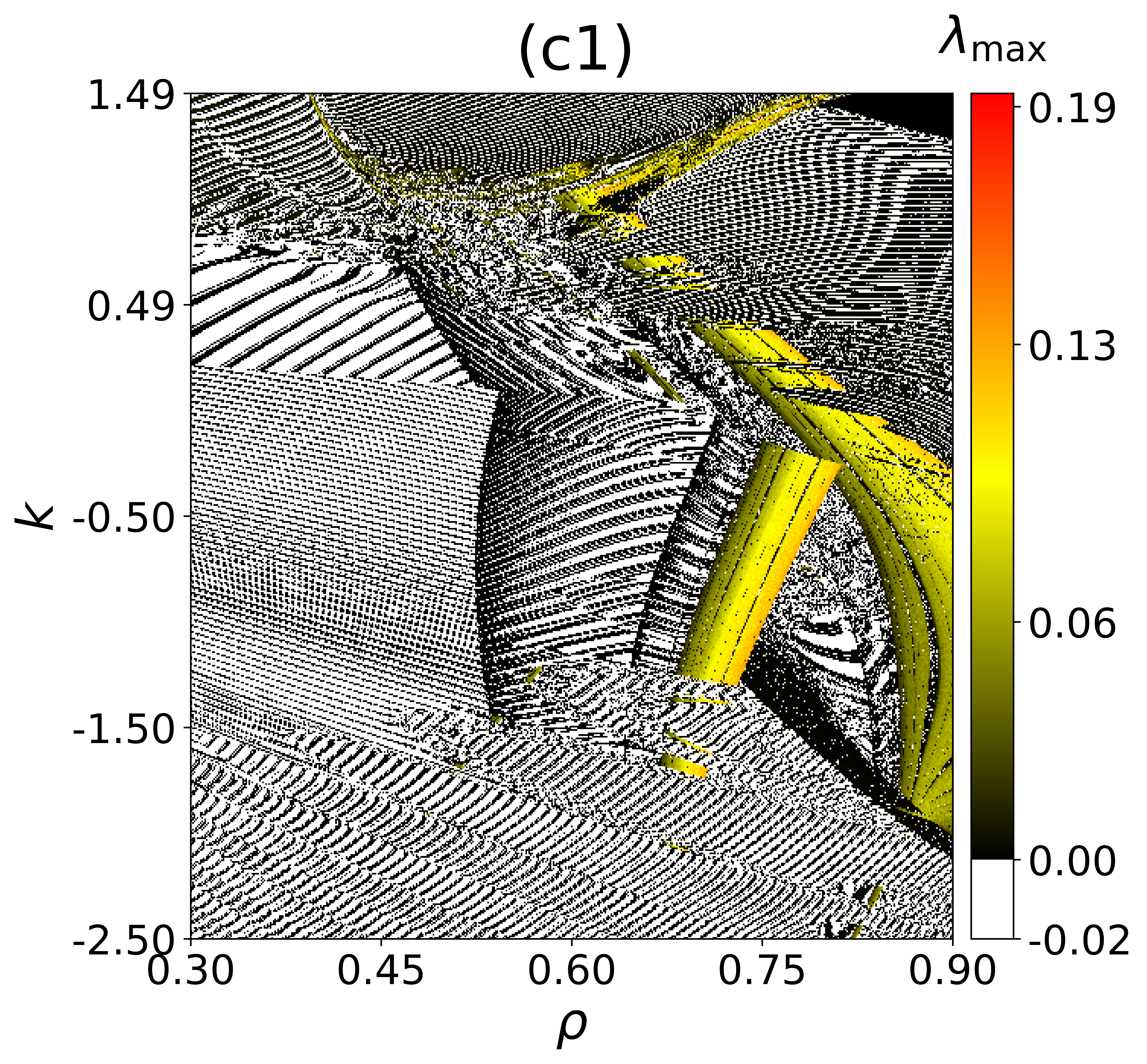}
    \includegraphics[width=5.1cm,height=4.7cm]{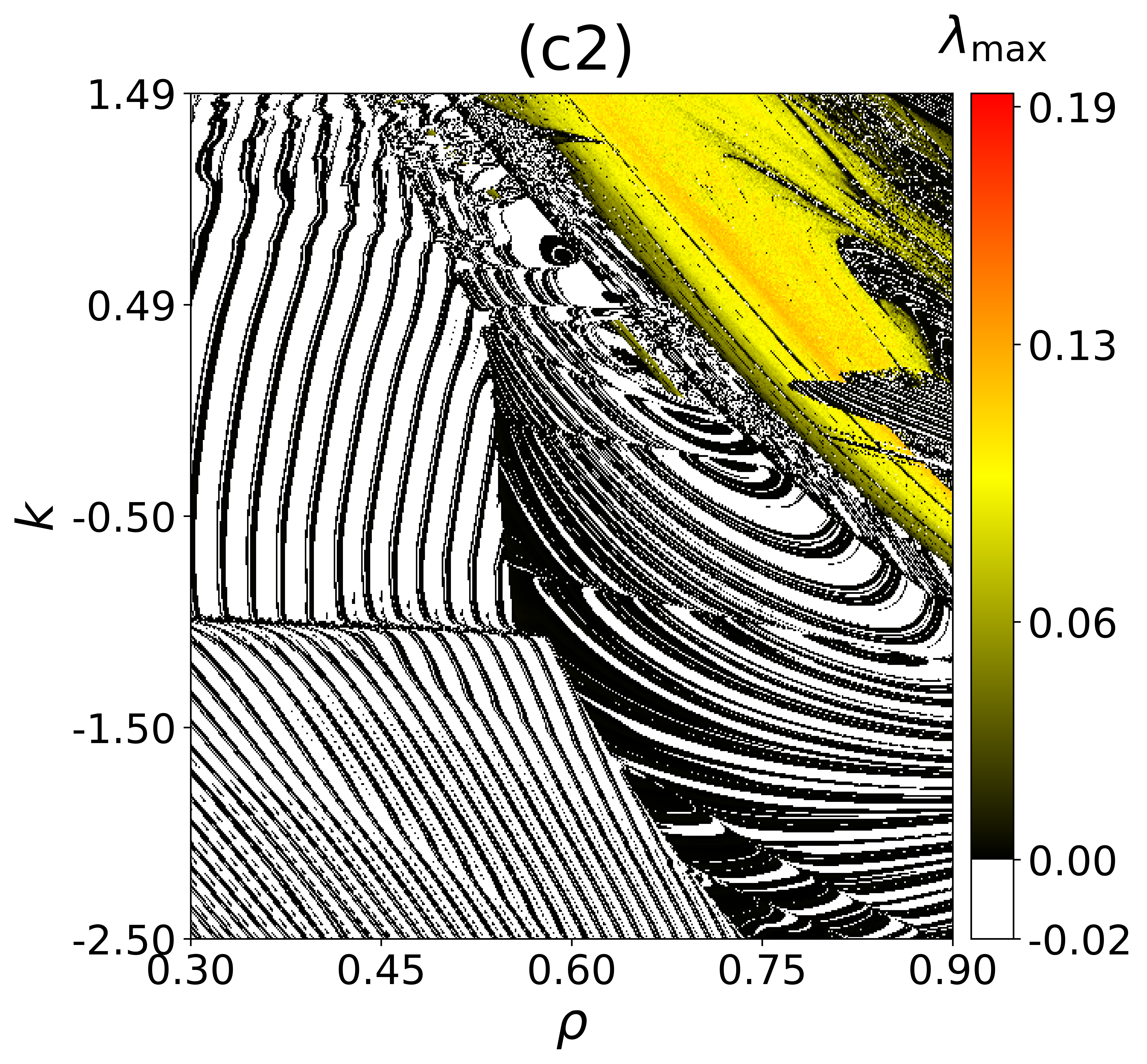}
    \includegraphics[width=5.1cm,height=4.7cm]{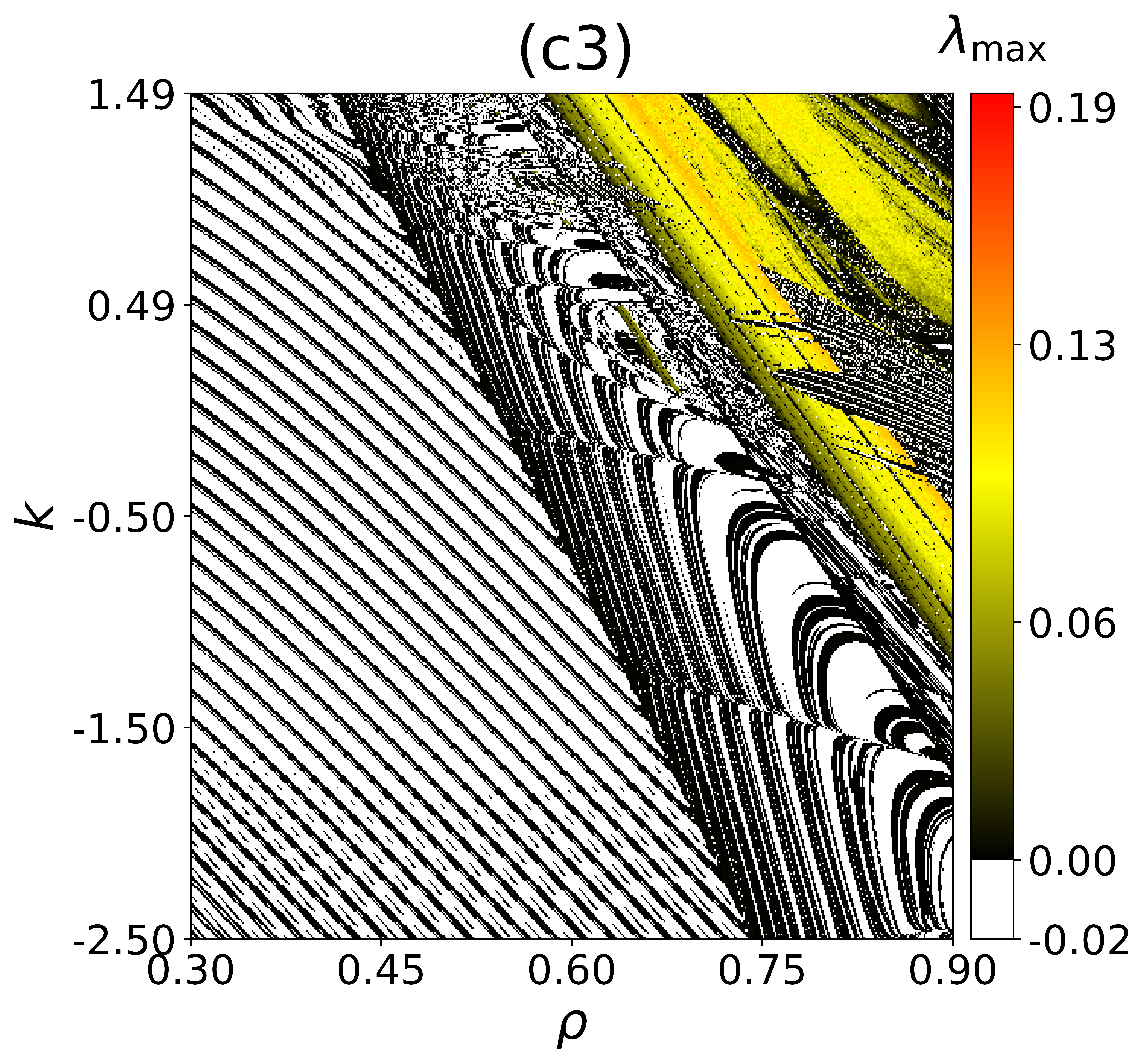}
    \caption{Two-dimensional dynamical maps for different parameter pairs: \textbf{(a)} $(k,m)$ at fixed $\rho = 0.7$; \textbf{(b)} $(\rho,m)$ at fixed $k = 0.87$; \textbf{(c1–c3)} $(\rho,k)$ at fixed $m = 1$, $0.5$, and $0.25$, respectively.}
    \label{fig:dynamic_map}
\end{figure}

Due to the strong nonlinearities in Eq.~\eqref{eq:1}, the system was analyzed via high-precision numerical integration. Simulations were carried out in Python using \texttt{JAX}~\cite{Bradbury2018JAX} and \texttt{Diffrax}~\cite{diffrax} with the adaptive Tsitouras 5/4 Runge–Kutta method (\texttt{Tsit5})~\cite{Tsitouras2011}. Computations employed 64-bit precision with tolerances of \(10^{-8}\) (relative) and \(10^{-10}\) (absolute). Each trajectory was \marius{integrated for 1000 time units from the initial state \(\mathbf{X}(0)=(0,0,0,1,0)^{\mathsf T}\)}, discarding the first 500 units to remove transients. Bifurcation diagrams and Lyapunov spectra were obtained from the post-transient local maxima of \(x(t)\), and Lyapunov exponents were computed using the Benettin algorithm with periodic orthonormalization. The resulting bifurcation structures are shown in Figs.~\ref{fig:bifurcation_k}--\ref{fig:dynamic_map}.

\begin{figure}
    \centering
    \includegraphics[width=5.5cm,height=3.5cm]{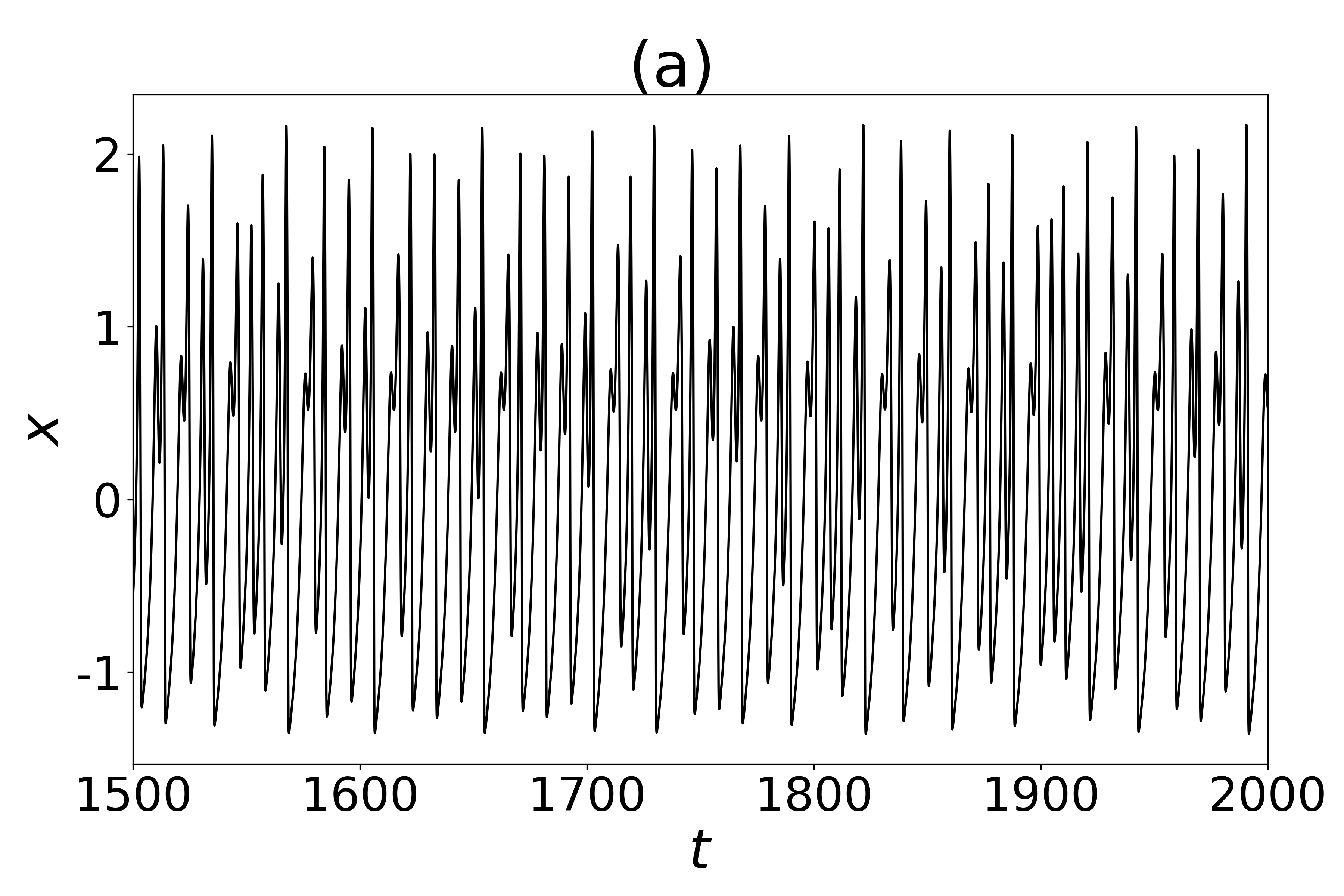}
    \includegraphics[width=5.5cm,height=3.5cm]{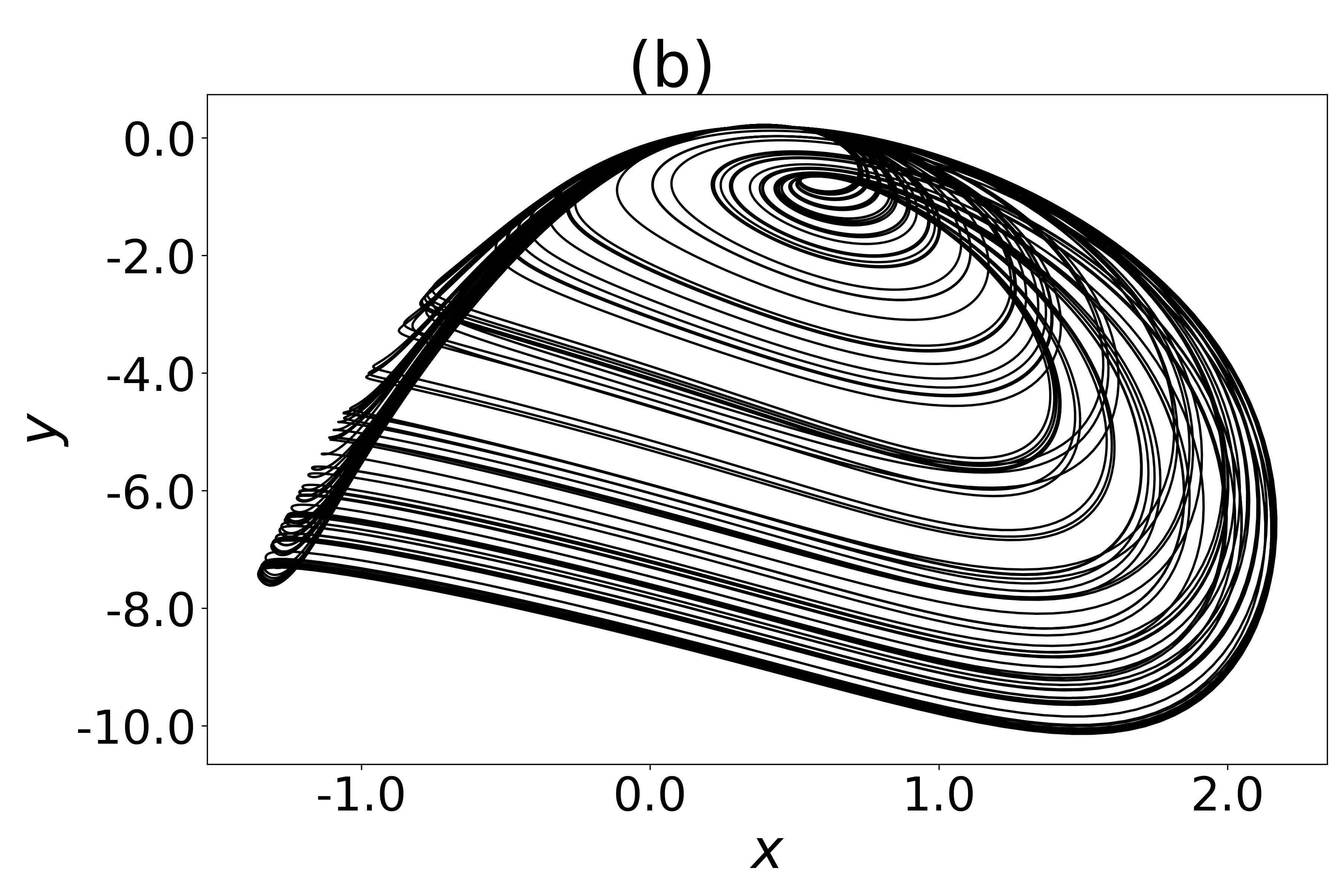}
    \caption{Chaotic time series of the $x$ variable \textbf{(a)} and corresponding phase portrait in the $x-y$ plane \textbf{(b)} with $k = 0.87$, $m = 0.5$, and $\rho = 0.7$.}  
    \label{fig:phase_k}
\end{figure}

Figure \ref{fig:phase_k} shows a chaotic time series of the $x$ variable \textbf{(a)} and corresponding phase portrait in the $x-y$ plane \textbf{(b)} for the 5D HR neuron model in Eq. \eqref{eq:1}. The model exhibits a rich spectrum of dynamical behaviors. 

Figure~\ref{fig:bifurcation_k} shows the bifurcation diagrams and the two largest Lyapunov exponents (LEs) versus \(k\) for fixed \(\rho = 0.7\) and different \(m\). As \(k\) increases, the system exhibits a period-doubling route to chaos, while smaller \(m\) values suppress chaotic oscillations. 

Figure~\ref{fig:bifurcation_rho} presents the corresponding diagrams versus \(\rho\) at fixed \(k = 0.87\). Increasing \(\rho\) induces transitions between periodic and chaotic regimes, with the range of chaos shrinking as \(m\) decreases.

Figure~\ref{fig:dynamic_map} displays two-dimensional maps of the largest Lyapunov exponent in the \((k,m)\), \((\rho,m)\), and \((\rho,k)\) planes. Reddish–yellow regions mark chaotic dynamics, delineating the global organization of periodic and chaotic attractors in parameter space.


\section{Synchronization and its Hamiltonian: A dynamical systems approach}\label{sec:sync}
In this section, we investigate the stability of complete synchronization between two diffusively coupled chaotic 5D HR  neurons governed by Eq.~\eqref{eq:1}, together with the associated \marius{synchronization Hamiltonian}. The analysis is conducted within the framework of dynamical systems theory, employing Lyapunov stability analysis \cite{krasovskii1963stability} to evaluate the onset and robustness of synchronization, and Helmholtz’s theorem \cite{kobe1986helmholtz} to characterize the energetic aspects of the synchronized state. Particular attention is devoted to the asymptotic \marius{and practical} stability of the synchronization manifold \marius{associated with the linearized error system} and to the behavior of its \marius{synchronization Hamiltonian} under variations of the key system parameters.

\subsection{Asymptotic \marius{and practical} stability in the linearized error dynamical system}\label{sec_3.1}
We consider two identical HR neurons coupled diffusively through their membrane potentials. The coupled system is written as
\begin{eqnarray}\label{coupled_eq}
\left\{
\begin{aligned}
\dot{x}_i &= y_i - a x_i^3 + b x_i^2 + k(h + f u_i^2) x_i + \rho \phi_i x_i + I_i 
+ g_e \sum_{j=1}^{2} \xi_{ij} (x_j - x_i),\\
\dot{y}_i &= c - d x_i^2 - y_i,\\
\dot{z}_i &= r \left[ s(x_i + \marius{x_r}) - z_i \right],\\
\dot{u}_i &= -u_i + m \left( |u_i + 1| - |u_i - 1| \right) + x_i,\\
\dot{\phi}_i &= x_i - q \phi_i,
\end{aligned}
\right.
\end{eqnarray}
where \(g_e\) denotes the electrical (gap-junction) coupling strength that determines the intensity of interaction between neurons; throughout this paper we fix $g_e=0.62$ unless stated otherwise.  
The coupling topology is defined by \(\xi_{ij} = 1\) for \(i \neq j\), indicating mutual coupling between neurons \(i\) and \(j\), and \(\xi_{ij} = 0\) otherwise.  
For the two-neuron system considered here, \(\xi_{12} = \xi_{21} = 1\).  
This formulation provides the mathematical basis for studying how the interplay between intrinsic neuronal dynamics and diffusive coupling gives rise to collective synchronization.

\marius{
\begin{definition}[Complete synchronization]\label{def_CS}
Two dynamical systems with states \(x(t),y(t)\in\mathbb{R}^n\) are said to achieve \emph{complete synchronization}
if there exist initial states \(x(t_0)=x_{0}\) and \(y(t_0)=y_{0}\) such that
\[
\lim_{t\to\infty}\|x(t)-y(t)\|=0.
\]
\end{definition}
\begin{definition}[Practical (\(\delta\)-)synchronization]\label{def_PS}
Two dynamical systems with states \(x(t),y(t)\in\mathbb{R}^n\) are said to achieve \emph{practical} (or \(\delta\)-)
\emph{synchronization} if there exist initial states \(x(t_0)=x_{0}\), \(y(t_0)=y_{0}\) and a tolerance
\(\delta>0\) such that
\[
\limsup_{t\to\infty}\|x(t)-y(t)\|\le \delta.
\]
\end{definition}
}

The synchronized state is represented by the invariant manifold
\begin{equation}
\mathcal{M} = \{ \mathbf{X}_1 = \mathbf{X}_2 = \mathbf{X}_s \},
\end{equation}
where \(\mathbf{X}_s = (x_s, y_s, z_s, u_s, \phi_s)^{\mathsf{T}}\) evolves according to the uncoupled dynamics
given in Eq.~\eqref{eq:1}.
To investigate the stability of \(\mathcal{M}\), we introduce the synchronization error system via
\begin{equation}
\left\{
\begin{aligned}
\dot{e}_x &= \dot{x}_2 - \dot{x}_1,\\
\dot{e}_y &= \dot{y}_2 - \dot{y}_1,\\
\dot{e}_z &= \dot{z}_2 - \dot{z}_1,\\
\dot{e}_u &= \dot{u}_2 - \dot{u}_1,\\
\dot{e}_\phi &= \dot{\phi}_2 - \dot{\phi}_1.
\end{aligned}
\right.
\end{equation}
By taking the difference between the equations of the two coupled neurons, one arrives at the corresponding full nonlinear error dynamical system given by
\marius{
\begin{eqnarray}\label{nonlinear_error}
\begin{split}
\left\{
\begin{array}{lcl}
\dot{e}_x &=& - a(e_x^3 + 3x_1e_x^2 + 3x_1^2e_x) + b(e_x^2 + 2x_1e_x)+ e_y+ k h e_x \\
&+& kf (e_x e_u^2+ 2 u_1 e_x e_u+x_1 e_u^2 + u_1^2e_x+2 u_1 x_1 e_u) +\rho (e_x e_\phi+\phi_1 e_x + x_1 e_\phi) -2 g_e e_x,\\
\dot{e}_y&=& -d(e_x^2 + 2x_1e_x) - e_y, \\
\dot{e}_z&=& r(s e_x - e_z), \\
\dot{e}_u &=& 
\begin{cases}
    e_x + (2m - 1)e_u + 2m(u_1 - 1) & \text{if } u_1 \ge 1 \text{ and } -1 < u_2 < 1, \\
    e_x - e_u - 4m & \text{if } u_1 \ge 1 \text{ and } u_2 \le -1, \\
    e_x - e_u - 2m(u_1 -1) & \text{if } -1 < u_1 < 1 \text{ and } u_2 \ge 1, \\
    e_x + (2m - 1)e_u & \text{if } -1 < u_1 < 1 \text{ and } -1 < u_2 < 1, \\
    e_x - e_u - 2m(u_1 + 1) & \text{if } -1 < u_1 < 1 \text{ and } u_2 \le -1, \\
    e_x - e_u + 4m & \text{if } u_1 \le -1 \text{ and } u_2 \ge 1, \\
    e_x + (2m - 1)e_u + 2m(u_1 + 1) & \text{if } u_1 \le -1 \text{ and } -1 < u_2 < 1,\\
    e_x - e_u & \text{otherwise},
\end{cases} \\[1em]
\dot{e}_\phi&=& e_x - q e_\phi.
\end{array}
\right.
\end{split}
\end{eqnarray}
}
Discarding second and higher-order nonlinear contributions in the errors (\(e_x^2\), \(e_u^2\), \(e_x e_\phi\), \(e_x e_u\), \(e_x^3\), \(e_xe_u^2\)), we obtain the linearized error dynamical system that describes the most significant transverse deviations from \(\mathcal{M}\) as follow:
\begin{equation}\label{linearized_error_Sys}
\begin{split}
\left\{
\begin{array}{lcl}
\dot{e}_x &=&  \big[ -3 a x_1^2 + 2 b x_1 + k(h + f u_1^2) + \rho \phi_1 - 2 g_e \big] e_x + e_y + 2 kf x_1 u_1 e_u +  \rho x_1 e_\phi,\\
\dot{e}_y &=& -2 d x_1 e_x - e_y,\\
\dot{e}_z &=& r(s e_x - e_z),\\
\dot{e}_u &=& 
\begin{cases}
e_x + (2m - 1)e_u + 2m(u_1 - 1), & \text{if } u_1 \ge 1,\; -1 < u_2 < 1,\\
e_x - e_u - 4m, & \text{if } u_1 \ge 1,\; u_2 \le -1,\\
e_x - e_u - 2m(u_1 -1), & \text{if } -1 < u_1 < 1,\; u_2 \ge 1,\\
e_x + (2m - 1)e_u, & \text{if } -1 < u_1 < 1,\; -1 < u_2 < 1,\\
e_x - e_u - 2m(u_1 + 1), & \text{if } -1 < u_1 < 1,\; u_2 \le -1,\\
e_x - e_u + 4m, & \text{if } u_1 \le -1,\; u_2 \ge 1,\\
e_x + (2m - 1)e_u + 2m(u_1 + 1), & \text{if } u_1 \le -1,\; -1 < u_2 < 1,\\
e_x - e_u, & \text{otherwise},
\end{cases}\\
\dot{e}_\phi &=& e_x - q e_\phi.
\end{array}
\right.
\end{split}
\end{equation}

\marius{
\begin{remark} Equation \eqref{linearized_error_Sys} represents the linearized transverse error dynamics around the synchronization manifold \(\mathcal{M}\). The Lyapunov/Barbalat analysis below therefore establishes asymptotic and practical convergence for this linearized error system (local transverse stability in the usual sense). A fully global nonlinear synchronization proof hinges explicit bounds on the neglected nonlinear remainder terms, and obtaining such semi-global or global estimates is beyond the scope of the current work. Nevertheless, alongside the numerics of the linearized error system (Eq. \eqref{linearized_error_Sys}) synchronization in Section \ref{sec:numerics}, we also present the synchronization of the full nonlinear error system in Eq. \eqref{nonlinear_error} as numerical evidence (Fig. \ref{fig:error_system}(a1)-(e1)).
\end{remark}
}

\marius{
\begin{theorem}[Asymptotic stability under dissipative memristive contribution]\label{Thm_lin_AS}
Consider the linearized transverse error dynamics Eq. \eqref{linearized_error_Sys} along a bounded reference trajectory
\((x_1(t),u_1(t),\phi_1(t))\) of Eq. \eqref{coupled_eq}, and assume there exists \(J>0\) such that
\(|x_1(t)|,|u_1(t)|,|\phi_1(t)|\le J\) for all \(t\ge 0\).
Let \(\mathbf{M}\) be the diagonal (hence symmetric) matrix defined in Eq. \eqref{entry}, and assume that all its diagonal entries are strictly positive so that \(\mathbf{M}\succ 0\).\\[2.0mm]
\noindent
Assume in addition that along the reference trajectory, the memristive contribution is dissipative in the sense that
\[ 
\widehat\Psi(e_u;u_1(t),u_2(t),m,J)\le -e_u^2,\qquad \forall t\ge 0,
\]
for all \(e_u\), i.e., the switching remains in the dissipative Cases \(2,3,5,8\) of Eq. \eqref{cases}.\\[2.0mm]
\noindent
Then the fixed point \(\mathbf e\equiv 0\) of Eq. \eqref{linearized_error_Sys} is asymptotically stable, and in particular
\[
\lim_{t\to\infty}\|\mathbf e(t)\|=0.
\]
\end{theorem}
}

\begin{proof}
We define the quadratic Lyapunov candidate \(V:\mathbb{R}^5\to\mathbb{R}_{\ge 0}\) by
\begin{equation}\label{lya_fucntion}
    V(e_x, e_y, e_z, e_u, e_\phi) = \tfrac{1}{2}\left(e_x^2 + e_y^2 + e_z^2 + e_u^2 + e_\phi^2\right)\ge0.
\end{equation}
Along trajectories of the error dynamics Eq. \eqref{linearized_error_Sys},
\begin{equation}
    \dot{V} = e_x \dot{e}_x + e_y \dot{e}_y + e_z \dot{e}_z + e_u \dot{e}_u + e_\phi \dot{e}_\phi.
\end{equation}
We assume solutions of the HR neuron system remain bounded (as observed numerically and consistent with the existence of a chaotic attractor). Hence there exists a constant \(J>0\) such that
\begin{equation}
    |x_1(t)| \le J, \quad |u_1(t)| \le J, \quad |\phi_1(t)| \le J, \qquad \forall t \ge 0.
\end{equation}
Inserting Eq.~\eqref{linearized_error_Sys} into the expression for \(\dot V\) and applying these bounds termwise, we obtain the upper estimate
\begin{equation}\label{dot_Lya}
\begin{split}
\dot{V} &\le \big[3|a|J^2 + 2|b|J + |kh| + |kf|J^2 + |\rho|J - 2|g_e|\big] e_x^2 - e_y^2 - |r| e_z^2 - |q| e_\phi^2 \\
&\quad + (1 + 2|d|J)|e_x||e_y| + |rs||e_x||e_z| + (2|kf|J^2 + 1)|e_x||e_u| + (|\rho|J + 1)|e_x||e_\phi| \\
&\quad + \Psi(e_u; u_1, u_2, m, J),
\end{split}
\end{equation}
where \(\Psi(e_u; u_1, u_2, m, J)\) collects the contribution coming from the memristive term and is explicitly given by
\begin{equation}\label{memristive2}
\Psi(e_u; u_1, u_2, m, J) =
\begin{cases}
|2m - 1|e_u^2 + 2|m||J - 1||e_u|, &\text{if } u_1 \ge 1,\; -1 < u_2 < 1,\\
- e_u^2 - 4|m||e_u|, &\text{if } u_1 \ge 1,\; u_2 \le -1,\\
- e_u^2 - 2|m||J - 1||e_u|, &\text{if } -1 < u_1 < 1,\; u_2 \ge 1,\\
|2m - 1|e_u^2, &\text{if } -1 < u_1 < 1,\; -1 < u_2 < 1,\\
- e_u^2 - 2|m|(J + 1)|e_u|, &\text{if } -1 < u_1 < 1,\; u_2 \le -1,\\
- e_u^2 + 4|m||e_u|, &\text{if } u_1 \le -1,\; u_2 \ge 1,\\
|2m - 1|e_u^2 + 2|m|(J + 1)|e_u|, &\text{if } u_1 \le -1,\; -1 < u_2 < 1,\\
- e_u^2, & \text{otherwise.}
\end{cases}
\end{equation}

\marius{Next, we make explicit the inequality step used to pass from the componentwise estimate
Eq.~\eqref{dot_Lya} to a compact quadratic-form bound. We invoke Young’s inequalities in the following forms:
(1) for any \(\alpha,\beta \ge 0\), $2\alpha\beta \le \alpha^2+\beta^2$ and (2) for any \(\alpha\in\mathbb{R}\), any \(c\ge 0\), and any \(\varepsilon>0\),
$c|\alpha| \le \frac{\varepsilon}{2}\alpha^2+\frac{c^2}{2\varepsilon}.$  Applying (1) to each mixed product term in Eq. \eqref{dot_Lya} yields
\begin{equation}\label{eq:young_mixed_terms}
\left\{
\begin{aligned}
(1+2|d|J)\,|e_x|\,|e_y|
&\le \frac{1}{2}\,(1+2|d|J)\,\Big(e_x^2+e_y^2\Big),\\
|rs|\,|e_x|\,|e_z|
&\le \frac{1}{2}\,|rs|\,\Big(e_x^2+e_z^2\Big),\\
(2|kf|J^2+1)\,|e_x|\,|e_u|
&\le \frac{1}{2}\,(2|kf|J^2+1)\,\Big(e_x^2+e_u^2\Big),\\
(|\rho|J+1)\,|e_x|\,|e_\phi|
&\le\frac{1}{2}\,(|\rho|J+1)\,\Big(e_x^2+e_\phi^2\Big),
\end{aligned}
\right.
\end{equation}
where $\mathbf{e} = \left( e_x, e_y, e_z, e_u, e_\phi\right)^\top$ denotes the signed error vector  components. Applying (2) to the linear \(|e_u|\)-terms arising from the memristive contribution in Eq.~\eqref{memristive2} (and, in Cases \(2,3,5\), using the sharper trivial bound \(\Psi\le -e_u^2\) since the corresponding linear term is nonpositive) yields
\begin{equation}
\Psi(e_u;u_1,u_2,m,J)\le \widehat\Psi(e_u;u_1,u_2,m,J),
\end{equation}
where \(\widehat\Psi\) is given in Eq.~\eqref{cases}.
\small{
\begin{equation}\label{cases}
\widehat\Psi(e_u;u_1,u_2,m,J):=
\begin{cases}
\Big(|2m - 1|+\dfrac{\varepsilon}{2}\Big)e_u^2 + \dfrac{2|m|^2(J-1)^2}{\varepsilon},
&\text{if } u_1 \ge 1,\; -1 < u_2 < 1,\\[4pt]
- e_u^2,
&\text{if } u_1 \ge 1,\; u_2 \le -1,\\[4pt]
- e_u^2,
&\text{if } -1 < u_1 < 1,\; u_2 \ge 1,\\[4pt]
|2m - 1|\,e_u^2,
&\text{if } -1 < u_1 < 1,\; -1 < u_2 < 1,\\[4pt]
- e_u^2,
&\text{if } -1 < u_1 < 1,\; u_2 \le -1,\\[4pt]
-\Big(1-\dfrac{\varepsilon}{2}\Big)e_u^2 + \dfrac{8|m|^2}{\varepsilon},
&\text{if } u_1 \le -1,\; u_2 \ge 1,\\[4pt]
\Big(|2m - 1|+\dfrac{\varepsilon}{2}\Big)e_u^2 + \dfrac{2|m|^2(J+1)^2}{\varepsilon},
&\text{if } u_1 \le -1,\; -1 < u_2 < 1,\\[4pt]
- e_u^2,
&\text{otherwise.}
\end{cases}
\end{equation}
}
}

\marius{
Under the strict dissipativity assumption of the theorem, we define
\begin{equation}\label{eq:psi_hathat_def}
\widehat{\widehat\Psi}(e_u;u_1,u_2,m,J):=\widehat\Psi(e_u;u_1,u_2,m,J)+e_u^2,
\end{equation}
so that \(\widehat{\widehat\Psi}(e_u;u_1(t),u_2(t),m,J)\le 0\) for all \(t\ge 0\) in the dissipative Cases \(2,3,5,8\) in Eq. \eqref{cases}. Substituting all these bounds into
Eq. \eqref{dot_Lya} and collecting terms yields the compact semi-quadratic-form estimate
\begin{eqnarray}\label{final_bound}
\dot{V} \le -\mathbf{e}^\top \mathbf{M}\mathbf{e} + \widehat{\widehat\Psi}(e_u; u_1, u_2, m, J).
\end{eqnarray}
In particular, compared to the quadratic-form bound obtained directly from Eq.~\eqref{dot_Lya}, the additional \(-e_u^2\) dissipation is absorbed into the quadratic form, which increases the \(e_u^2\)-weight by \(1\). And $\mathbf{M}$ here is the diagonal and hence symmetric matrix
\begin{equation}
\mathbf{M} =
\begin{bmatrix}
A_{xx} & 0 & 0 & 0 & 0 \\
0 & A_{yy} & 0 & 0 & 0 \\
0 & 0 & A_{zz} & 0 & 0 \\
0 & 0 & 0 & A_{uu} & 0 \\
0 & 0 & 0 & 0 & A_{\phi\phi}
\end{bmatrix},
\end{equation}
with the nonzero (diagonal) entries given by
\begin{equation}\label{entry}
\left\{
\begin{aligned}
A_{xx} &= -(3|a|+2|kf|)J^2 - \left(2|b|+|d|+\frac{3}{2}|\rho|\right)J - |kh| + 2|g_e| - \frac{1}{2}|rs| - \frac{3}{2},\\
A_{yy} &=  \frac{1}{2} - |d|J,\\
A_{zz} &= |r| - \frac{1}{2}|rs|,\\
A_{uu} &= \frac{1}{2}-|kf|J^2,\\
A_{\phi\phi} &= |q| - \frac{1}{2}|\rho|J - \frac{1}{2}.
\end{aligned}
\right.
\end{equation}
}

\marius{
Since \(\mathbf M\) is diagonal, \(\mathbf M\succ 0\) if and only if
\begin{equation}\label{eq:Mpos}
A_{xx}>0,\qquad A_{yy}>0,\qquad A_{zz}>0,\qquad A_{uu}>0,\qquad A_{\phi\phi}>0,
\end{equation}
which can be achieved with suitable parameter values of $\{a, b, d, f, h, k, m, q, r, s, \rho, g_{e},J\}$.
The fixed point \(\mathbf e\equiv 0\) of the linearized error system in Eq. \eqref{linearized_error_Sys} is Lyapunov stable once the right-hand side of Eq. \eqref{final_bound} is nonpositive, \textit{i.e.,} \(\dot V\le 0\). To ensure this, we impose the dissipativity assumption that the trajectory remains in memristive regimes for which
\begin{equation}\label{eq:dissip_repeat}
\widehat\Psi(e_u;u_1(t),u_2(t),m,J)\le -e_u^2,\qquad \forall t\ge 0,
\end{equation}
for all values of \(e_u\) (\textit{i.e.,} Cases \(2,3,5,8\) in Eq. \eqref{cases}).
Thus, under Eqs. \eqref{eq:Mpos} and \eqref{eq:dissip_repeat}, the inequality in Eq. \eqref{final_bound} reduces to
\begin{eqnarray}\label{inte}
\dot V \le -\,\mathbf e^{\mathsf T}\mathbf M \mathbf e + \widehat{\widehat\Psi}(e_u;u_1(t),u_2(t),m,J)\le -\,\mathbf e^{\mathsf T}\mathbf M \mathbf e \le 0, 
\end{eqnarray}
so that \(V(t)\) is nonincreasing; since \(V(e)=\tfrac12\|\mathbf e\|^2\) is positive definite, this implies that \(\mathbf e\equiv 0\) is Lyapunov stable.
}

To strengthen this Lyapunov stability to asymptotic stability, we must additionally show that the errors decay to zero as $t\to\infty$, \textit{i.e.,} that transverse perturbations to the synchronization manifold vanish as \(t\to\infty\). From Eq.~\eqref{lya_fucntion} and the bounds in Eqs.~\eqref{final_bound} and \eqref{cases}, the error components are uniformly bounded, \textit{i.e.,} \(e_i\in L^{\infty}\) for \(i=x,y,z,u,\phi\). \marius{Since \(x_1(t),u_1(t),\phi_1(t)\) are bounded, all time-varying coefficients appearing in Eq. \eqref{linearized_error_Sys} are bounded functions of time. Moreover, the additive terms in the piecewise definition of \(\dot e_u\) (\textit{e.g.,} \(\pm 4m\) and \(2m(u_1\pm 1)\)) are bounded because \(u_1\) is bounded. Hence the right-hand side of Eq. \eqref{linearized_error_Sys} is bounded whenever \(\mathbf e(t)\) is bounded, and therefore \(\dot{\mathbf e}(t)\in L^{\infty}\) (equivalently, each component \(\dot e_i\in L^{\infty}\)).}

\begin{lemma}\label{Rayleigh quotient} (Rayleigh bound). 
For any nonzero vector $x \in \mathbb{R}^n$ and any symmetric matrix $A \in \mathbb{R}^{n \times n}$, the quadratic form satisfies $x^\top A x \ge \lambda_{\min}(A)\,\|x\|^2,$ where $\lambda_{\min}(A)$ is the smallest eigenvalue of $A$.
\end{lemma} 
\marius{
With a positive definite symmetric \(\mathbf M\), Lemma~\ref{Rayleigh quotient} gives
\begin{equation}\label{201}
\boldsymbol{e}^\top \mathbf M \boldsymbol{e} \ge \lambda_{\min}(\mathbf M)\, \|\boldsymbol{e}\|^2,  
\end{equation}
where $\lambda_{\min}(\mathbf M)>0$ is the minimum eigenvalue of $\mathbf M$.
From Eq. \eqref{inte} and the Rayleigh bound Eq. \eqref{201}, we obtain for all \(t\ge 0\),
\begin{equation}\label{eq:Vdot_rayleigh}
\dot V(t)\le -\,\mathbf e(t)^{\mathsf T}\mathbf M\mathbf e(t)
\le -\lambda_{\min}(\mathbf M)\,\|\mathbf e(t)\|^2,
\end{equation}
where \(\lambda_{\min}(\mathbf M)>0\) denotes the smallest eigenvalue of \(\mathbf M\).
Integrating Eq. \eqref{eq:Vdot_rayleigh} from \(0\) to \(t\) yields
\begin{equation}
V(t)-V(0)\le -\lambda_{\min}(\mathbf M)\int_{0}^{t}\|\mathbf e(s)\|^{2}\,ds,    
\end{equation}
and since \(V(t)\ge 0\) for all \(t\ge 0\), it follows that
\begin{equation}
\lambda_{\min}(\mathbf M)\int_{0}^{t}\|\mathbf e(s)\|^{2}\,ds
\le V(0)-V(t)\le V(0).    
\end{equation}
Hence we obtain the uniform bound
\begin{equation}\label{bound}
\int_{0}^{t}\|\mathbf e(s)\|^{2}\,ds \le \frac{V(0)}{\lambda_{\min}(\mathbf M)} < \infty,
\qquad \forall\, t\ge 0,
\end{equation}
and therefore \(\|\mathbf e(\cdot)\|^{2}\in L^{1}([0,\infty))\).
}

\begin{lemma}\label{Barbalat's Lemma}(Barbalat's Lemma).
Let $f:[0,\infty)\rightarrow\mathbb{R}$ be a uniformly continuous function 
such that  $\lim\limits_{t\to\infty} \int_0^t f(s) ds < \infty$.
Then $\displaystyle \lim_{t \to \infty} f(t) = 0$.
\end{lemma}
\marius{To apply Barbalat's lemma, we define \(f(t) = \|\mathbf{e}(t)\|^2\). From Eq. \eqref{bound}, \(f\in L^1([0,\infty))\), \textit{i.e.,} \(\int_0^\infty f(s)\,ds < \infty\). Since \(e_i\in L^{\infty}\) and \(\dot e_i\in L^{\infty}\) for each component \((i=x,y,z,u,\phi)\), we have that
\begin{equation}
\dot f(t)=\frac{d}{dt}\|\mathbf{e}(t)\|^2 = 2 \mathbf{e}(t)^\top \dot{\mathbf{e}}(t),
\end{equation}
is also bounded. Hence $f(t)=\|\mathbf{e}(t)\|^2$ is Lipschitz and thus uniformly continuous on $[0,\infty)$. By Barbalat's lemma, $f(t)\to 0$ as $t\to\infty$, \textit{i.e.,}
\begin{equation}\label{eq:43}
\lim\limits_{t\rightarrow\infty}\|\mathbf{e}(t)\| = 0,
\end{equation}
which establishes asymptotic transverse stability of the synchronization manifold for the linearized error dynamics of Eq. \eqref{linearized_error_Sys}.}
\end{proof}

\marius{In Theorem~\ref{Thm_lin_AS} we obtain asymptotic convergence \(\|\mathbf e(t)\|\to 0\), which establishes complete synchronization in the sense of Definition~\ref{def_CS}.}

\marius{
\begin{theorem}[Practical stability under non-dissipative memristive contribution]\label{Thm_lin_UB_1467}
Consider the linearized transverse error dynamics in Eq. \eqref{linearized_error_Sys} along a bounded reference trajectory and
assume that there exists \(J>0\) such that \(|x_1(t)|,|u_1(t)|,|\phi_1(t)|\le J\) for all \(t\ge 0\).
Let \(V(\mathbf e)=\tfrac12\|\mathbf e\|^2\) and let \(\mathbf M\) be the (diagonal) matrix in Eq. \eqref{entry}, and assume that \(\mathbf M\succ 0\).
Fix any \(\varepsilon\in(0,2)\) and let \(\widehat\Psi\) be the Young upper bound in Eq. \eqref{cases}.\\
\noindent
Assume that, for all \(t\ge 0\), the memristive switching remains in the non-dissipative Cases \(1,4,6,7\) of Eq. \eqref{cases}.
Define
\begin{equation}\label{eq:gammaC_1467}
\gamma := |2m-1|+\frac{\varepsilon}{2},\qquad
C := \frac{1}{\varepsilon}\max\Big\{2|m|^2(J-1)^2,\;8|m|^2,\;2|m|^2(J+1)^2\Big\}.
\end{equation}
If \(\alpha:=\lambda_{\min}(\mathbf M)-\gamma>0\)  (with \(\lambda_{\min}(\mathbf M)=\min\{A_{xx},A_{yy},A_{zz},A_{uu},A_{\phi\phi}\}\)), then along solutions of Eq. \eqref{linearized_error_Sys} one has
\begin{equation}\label{eq:V_ineq_1467}
\dot V(t)\le -2\alpha V(t)+C,\qquad t\ge 0.
\end{equation}
Consequently,
\begin{equation}\label{eq:V_bound_1467}
V(t)\le \Big(V(0)-\frac{C}{2\alpha}\Big)e^{-2\alpha t}+\frac{C}{2\alpha},\qquad t\ge 0,
\end{equation}
and in particular
\begin{equation}\label{eq:ultimate_ball}
\limsup_{t\to\infty}\|\mathbf e(t)\|\le \sqrt{\frac{C}{\alpha}}.
\end{equation}
Thus the origin is practically stable and solutions ultimately enter and remain in the ball
\(\|\mathbf e\|\le \sqrt{C/\alpha}\).
\end{theorem}
}

\begin{proof}
\marius{
From Eq. \eqref{final_bound} we have
\begin{equation}\label{eq:Vdot_start_1467}
\dot V(t)\le -\,\mathbf e(t)^{\mathsf T}\mathbf M\,\mathbf e(t)
+\widehat\Psi\big(e_u(t);u_1(t),u_2(t),m,J\big).
\end{equation}
Since \(\mathbf M\succ 0\) is symmetric, the Rayleigh bound yields
\begin{equation}\label{eq:Rayleigh_1467}
\mathbf e^{\mathsf T}\mathbf M\,\mathbf e \ge \lambda_{\min}(\mathbf M)\,\|\mathbf e\|^2.
\end{equation}}

\marius{
Next, assume the switching remains in non-dissipative Cases \(1,4,6,7\) of Eq. \eqref{cases}. We claim that for all \(t\ge 0\),
\begin{equation}\label{eq:Psi_bound_1467}
\widehat\Psi\big(e_u(t);\cdot\big)\le \gamma\,e_u(t)^2 + C,
\end{equation}
with \(\gamma\) and \(C\) given in Eq. \eqref{eq:gammaC_1467}. Indeed:
\begin{itemize}
\item In Case \(4\), \(\widehat\Psi=|2m-1|e_u^2\le \gamma e_u^2\).
\item In Case \(1\), \(\widehat\Psi=\big(|2m-1|+\frac{\varepsilon}{2}\big)e_u^2+\frac{2|m|^2(J-1)^2}{\varepsilon}
\le \gamma e_u^2 + C\).
\item In Case \(6\), \(\widehat\Psi=-(1-\frac{\varepsilon}{2})e_u^2+\frac{8|m|^2}{\varepsilon}
\le \frac{8|m|^2}{\varepsilon}\le C \le \gamma e_u^2 + C\), \text{since\quad \(\varepsilon\in(0,2)\)}.
\item In Case \(7\), \(\widehat\Psi=\big(|2m-1|+\frac{\varepsilon}{2}\big)e_u^2+\frac{2|m|^2(J+1)^2}{\varepsilon}
\le \gamma e_u^2 + C\).
\end{itemize}
This proves Eq. \eqref{eq:Psi_bound_1467}. Since \(C\) is chosen as the maximum of the constants appearing in Cases
\(1,6,7\), the bound holds uniformly in time whenever the switching remains in Cases \(1,4,6,7\). 
}
\marius{
Using \(e_u^2\le \|\mathbf e\|^2\), we obtain
\[
\widehat\Psi\big(e_u(t);\cdot\big)\le \gamma \|\mathbf e(t)\|^2 + C.
\]}

\marius{
Substituting this and Eq. \eqref{eq:Rayleigh_1467} into Eq. \eqref{eq:Vdot_start_1467} yields
\[
\dot V(t)\le -\lambda_{\min}(\mathbf M)\|\mathbf e(t)\|^2 + \gamma \|\mathbf e(t)\|^2 + C
= -\alpha \|\mathbf e(t)\|^2 + C,
\]
where \(\alpha=\lambda_{\min}(\mathbf M)-\gamma>0\) by assumption. Since \(\|\mathbf e\|^2=2V\), we obtain Eq. \eqref{eq:V_ineq_1467}:
\[
\dot V(t)\le -2\alpha V(t) + C.
\]}

\marius{Let \(W\) solve \(\dot W=-2\alpha W + C\) with \(W(0)=V(0)\). Then
\[
W(t)=\Big(V(0)-\frac{C}{2\alpha}\Big)e^{-2\alpha t}+\frac{C}{2\alpha}.
\]
By the Comparison Principle, 
\(V(t)\le W(t)\) for all \(t\ge 0\), which proves Eq. \eqref{eq:V_bound_1467}.
Taking \(t\to\infty\) gives \(\limsup_{t\to\infty}V(t)\le \frac{C}{2\alpha}\), hence
\[
\limsup_{t\to\infty}\|\mathbf e(t)\|^2
=2\limsup_{t\to\infty}V(t)\le \frac{C}{\alpha},
\]
which yields Eq. \eqref{eq:ultimate_ball} and completes the proof.}
\end{proof}

\marius{In Theorem~\ref{Thm_lin_UB_1467} we obtain an explicit ultimate bound, which establishes practical ($\delta$-) synchronization in the sense of Definition~\ref{def_PS}, yielding \(\delta=\sqrt{C/\alpha}\), which can be arbitrary small.}

\marius{
\begin{remark}[Asymptotic versus practical stability under memristive switching]
The distinction between Theorem~\ref{Thm_lin_AS} and Theorem~\ref{Thm_lin_UB_1467} is driven by the sign structure of the
memristive contribution. In the dissipative regimes (Cases \(2,3,5,8\)), one has \(\widehat\Psi\le -e_u^2\) along the
trajectory, so the Lyapunov estimate strengthens to
\(\dot V\le -\mathbf e^{\mathsf T}\mathbf M\mathbf e - e_u^2 \le -\mathbf e^{\mathsf T}\mathbf M\mathbf e\), which implies
\(\|\mathbf e(t)\|\to 0\) as \(t\to\infty\).
In contrast, in the remaining switching regimes (Cases \(1,4,6,7\)) the Young upper bound \(\widehat\Psi\) can contain
a nonnegative quadratic term in \(e_u^2\) and/or a positive constant offset. In this case the best general estimate is of
the form \(\dot V\le -2\alpha V + C\) with \(C>0\), yielding uniform boundedness and ultimate convergence of the error to a
ball whose radius scales like \(\sqrt{C/\alpha}\), rather than asymptotic convergence to the origin unless the offset
vanishes (\textit{i.e.,} \(C=0\), which can occur only when the memristive-feedback state-switching nonlinearity parameter $m\in[0,1]$ is exactly zero).
\end{remark}
}

\subsection{\marius{\marius{Synchronization Hamiltonian} and its rate identity}}
\label{sec:synchr_dynamics_Hamilton}
According to Helmholtz’s theorem~\cite{kobe1986helmholtz}, 
any smooth vector field \(\mathbf{F}(\mathbf{r})\) of a dynamical system can be decomposed into a 
{conservative} (divergence-free) part \(\mathbf{F}_c(\mathbf{r})\) 
and a {dissipative} (curl-free) part \(\mathbf{F}_d(\mathbf{r})\) as follow:
\begin{equation}
\mathbf{F}(\mathbf{r}) = \mathbf{F}_c(\mathbf{r}) + \mathbf{F}_d(\mathbf{r}),
\label{eq:helmholtz}
\end{equation}
where the conservative component governs rotational motion, 
while the dissipative component governs the \marius{synchronization Hamiltonian} exchange with the environment. The divergence-free (conservative) and curl-free (dissipative) vector fields of the linearized error dynamical
system in Eq.~\eqref{linearized_error_Sys} are respectively given by
\begin{equation} \label{cons_diss}
\mathbf{F}_c(\mathbf{e})
=\begin{pmatrix}
e_y + 2kf\,u_1x_1\,e_u + \rho\,x_1\,e_\phi\\[3pt]
-2d\,x_1\,e_x\\[3pt]
r\,s\,e_x\\[3pt]
e_x\\[3pt]
e_x
\end{pmatrix},
\quad
\mathbf{F}_d(\mathbf{e})
=\begin{pmatrix}
N(x_1,u_1,\phi_1)\,e_x\\[3pt]
-\,e_y\\[3pt]
-\,r\,e_z\\[3pt]
\alpha(u_1,u_2)\,e_u + \beta(u_1,u_2)\\[3pt]
-\,q\,e_\phi
\end{pmatrix},  
\end{equation}
where
\begin{eqnarray}
\begin{cases}
\begin{aligned}
N(x_1,u_1,\phi_1) &= -3a\,x_1^2 + 2b\,x_1 + kh + kf\,u_1^2 + \rho\,\phi_1 - 2g_e,\\
\alpha(u_1,u_2) &= 
\begin{cases}
2m-1, &\text{if } u_1\ge1,\,-1<u_2<1,\\
-1,   &\text{if } u_1\ge1,\,u_2\le-1,\\
-1,   &\text{if } -1<u_1<1,\,u_2\ge1,\\
2m-1, &\text{if } -1<u_1<1,\,-1<u_2<1,\\
-1,   &\text{if } -1<u_1<1,\,u_2\le-1,\\
-1,   &\text{if } u_1\le-1,\,u_2\ge1,\\
2m-1, &\text{if } u_1\le-1,\,-1<u_2<1,\\
-1,   & \text{otherwise,}
\end{cases}\\
\beta(u_1,u_2) &=
\begin{cases}
2m\,(u_1-1),  &\text{if } u_1\ge1,\,-1<u_2<1,\\
-4m,          &\text{if } u_1\ge1,\,u_2\le-1,\\
-2m\,(u_1-1), &\text{if } -1<u_1<1,\,u_2\ge1,\\
0,            &\text{if } -1<u_1<1,\,-1<u_2<1,\\
-2m\,(u_1+1), &\text{if } -1<u_1<1,\,u_2\le-1,\\
4m,           &\text{if } u_1\le-1,\,u_2\ge1,\\
2m\,(u_1+1),  &\text{if } u_1\le-1,\,-1<u_2<1,\\
0,            & \text{otherwise.}
\end{cases}
\end{aligned}
\end{cases}
\end{eqnarray}

For the conservative vector field, Eq.~\eqref{cons_diss} yields a first-order linear partial differential equation for the Hamiltonian function:
\begin{equation}
\left(
e_y + 2kf\,u_1 x_1\,e_u + \rho\,x_1\,e_\phi
\right)
\frac{\partial H}{\partial e_x}
- 2d\,x_1\,e_x \frac{\partial H}{\partial e_y}
+ r s\,e_x \frac{\partial H}{\partial e_z}
+ e_x \frac{\partial H}{\partial e_u}
+ e_x \frac{\partial H}{\partial e_\phi}
= 0,
\label{eq:div_free_PDE}
\end{equation}
which, by the method of separation of variables, admits the solution
\begin{equation}
\label{eq:hamiltonian}
H(e_x, e_y, e_z, e_u, e_\phi)
= 2d\,x_1\,e_x^2 + e_y^2 - 4dkf\,u_1x_1^2\,e_u^2 - 2d\rho\,x_1^2\,e_\phi^2 + e_z - rs\,e_\phi + K,
\end{equation}
where \(K\) is an arbitrary integration constant.

\marius{
\begin{remark}
$H$ is defined up to an additive constant  $K$ (gauge freedom, which we fix  by setting $H(0)=0$), and  is not guaranteed to be positive definite or bounded below (as it includes linear terms (\textit{e.g.,} $e_z$, $-rse_{\phi}$, and an arbitrary constant $K$) as a function on $\mathbb{R}^5$. Thus, $H$ is scalar potential associated with the conservative part of the error field, and $\dot{H}=dH/dt$ gives a rate identity.
\end{remark}
}

For the dissipative vector field, Eq.~\eqref{cons_diss} gives the differential equation governing 
the time evolution of the \marius{synchronization Hamiltonian}:
\begin{align}
\label{eq:Hdot}
\frac{dH}{dt}
&= 4d\,x_1\,N(x_1,u_1,\phi_1)\,e_x^{2}
     - 2\,e_y^{2}
     - r\,e_z 
     - 8d k f\,u_1 x_1^{2}\,\alpha(u_1,u_2)\,e_u^{2}\\\nonumber
& - 8d k f\,u_1 x_1^{2}\,\beta(u_1,u_2)\,e_u 
     + 4d\rho q\,x_1^{2}\,e_\phi^{2}
     + q r s\,e_\phi.
\end{align}

\marius{\(H(e_x, e_y, e_z, e_u, e_\phi)\) represents a \marius{synchronization Hamiltonian} potential (defined up to a constant), while it temporal derivative \(dH/dt\) quantifies the instantaneous rate of change of this potential along the error dynamics and can be used as an energetic diagnostic of synchronization transients}

This formulation provides a quantitative measure of the energetic cost of synchronization: 
it defines the amount of energy per unit time required to maintain a given degree of synchrony. 
Consequently, variations in \(H\) and \(dH/dt\) under changes of system parameters 
reveal how coupling strength, dissipation, and intrinsic neuronal dynamics 
jointly determine both the stability and the energetic efficiency of the synchronized state~\cite{yamakou2020chaotic}. The numerical simulations of $H$ and $dH/dt$ are provided in Section \ref{sec:numerics}.

\section{Numerical simulations and discussion}
\label{sec:numerics}
To corroborate the stability results of Section~\ref{sec_3.1} and the Hamiltonian-based analysis of Section~\ref{sec:synchr_dynamics_Hamilton}, we performed numerical simulations of the synchronization-error dynamics in Eq.~\eqref{linearized_error_Sys} simultaneously with of the coupled memristive HR neuron system in Eq.~\eqref{coupled_eq}. The two neurons were initialized with slightly different values in all five state variables to induce synchronization transients. We then computed the time evolution of the error variables $(e_x, e_y, e_z, e_u, e_\phi)$, together with the Lyapunov and \marius{synchronization Hamiltonian} time
derivatives $\dot V$ and $\dot H$, and analyzed their dependence on the parameters $m$, $k$, $\rho$, and $g_e$.

\begin{figure}
\centering
\includegraphics[width=4cm,height=3cm]{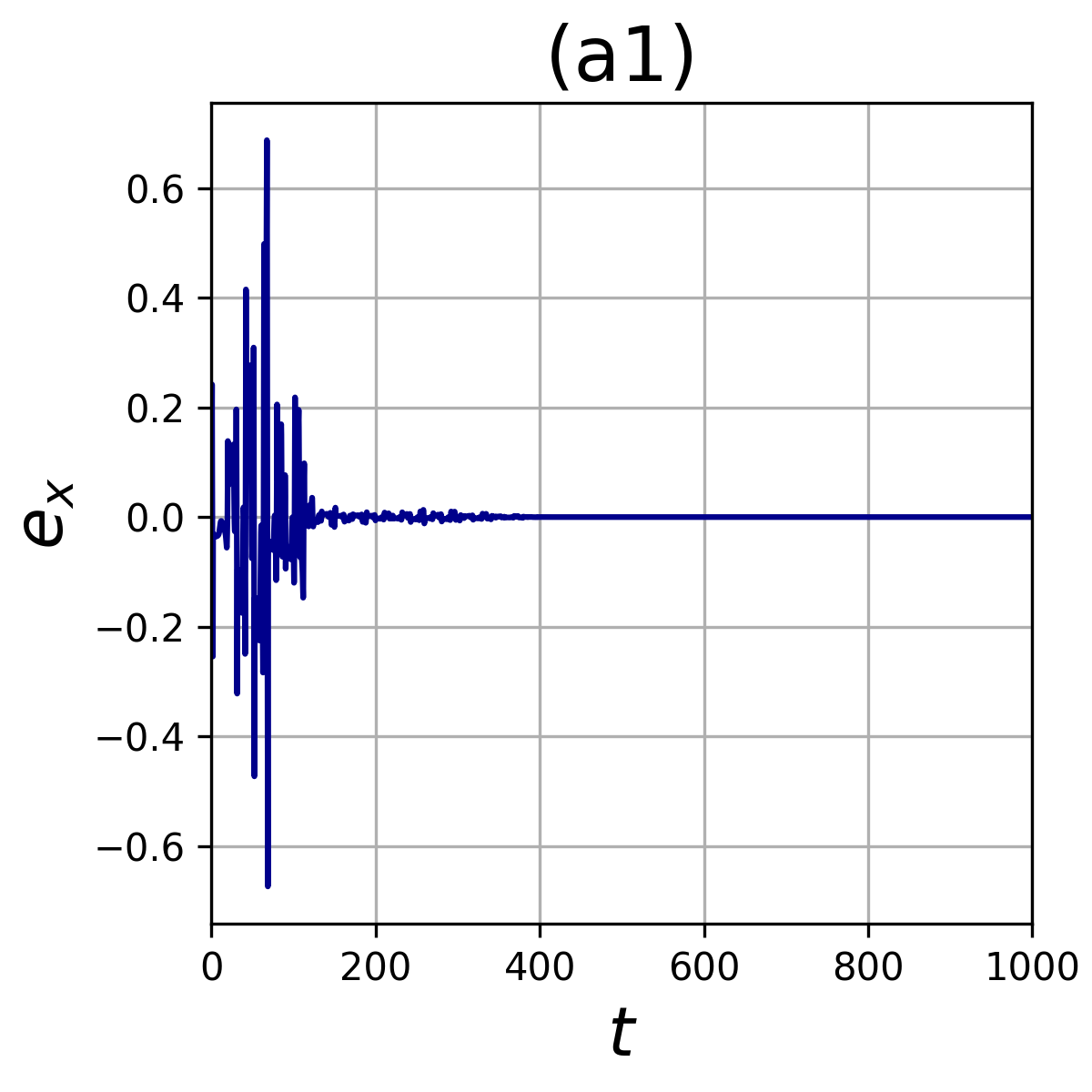}
\includegraphics[width=4cm,height=3cm]{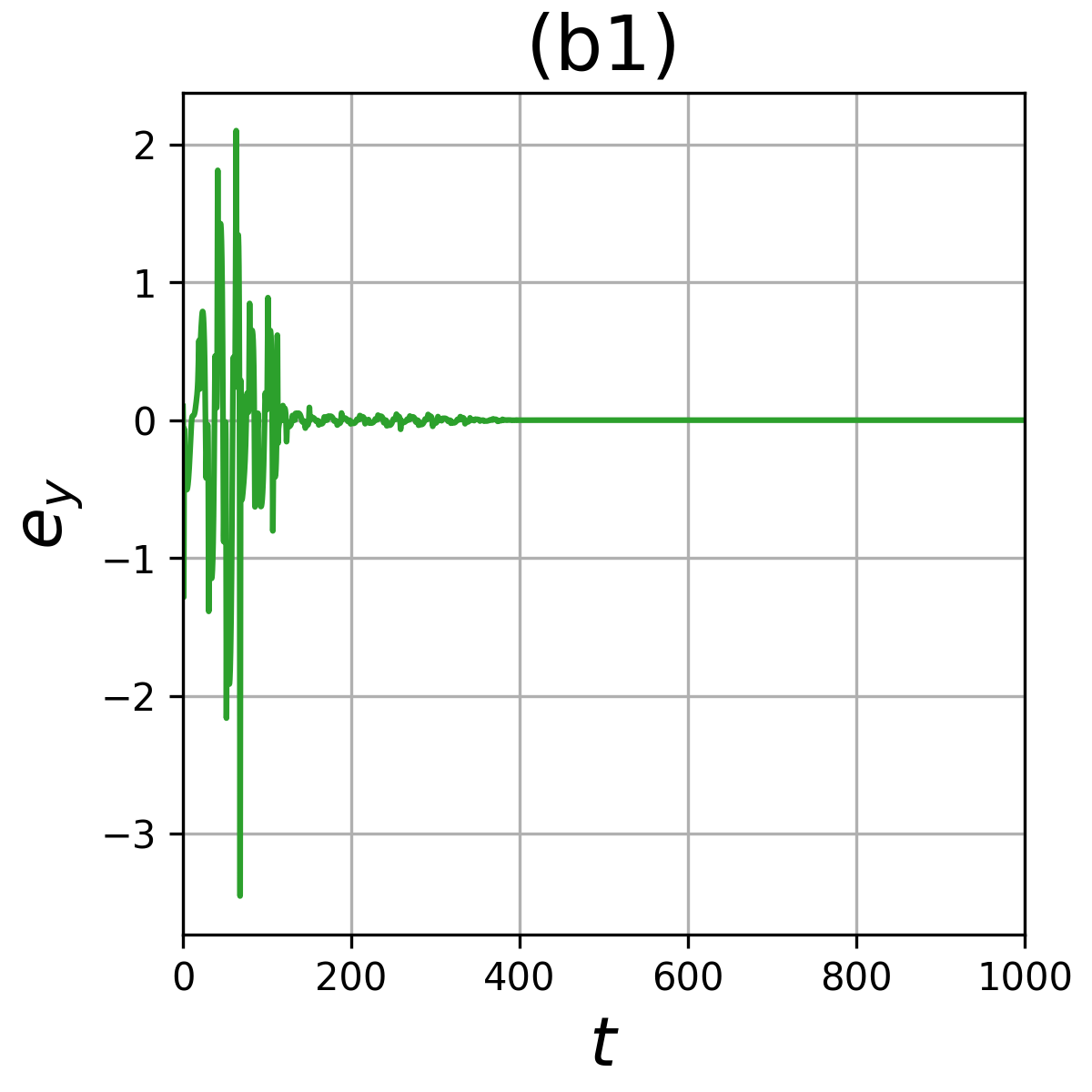}
\includegraphics[width=4cm,height=3cm]{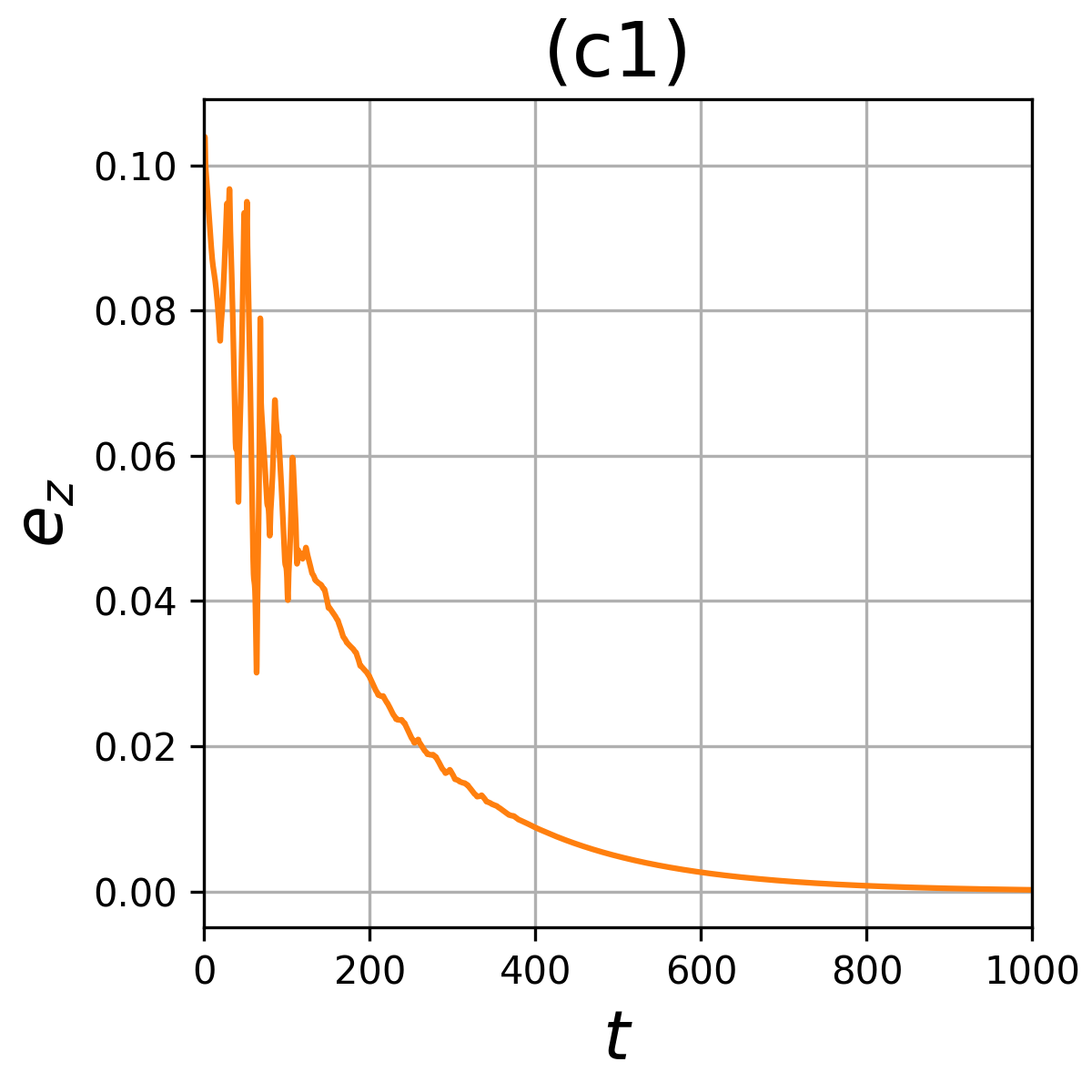}

\includegraphics[width=4cm,height=3cm]{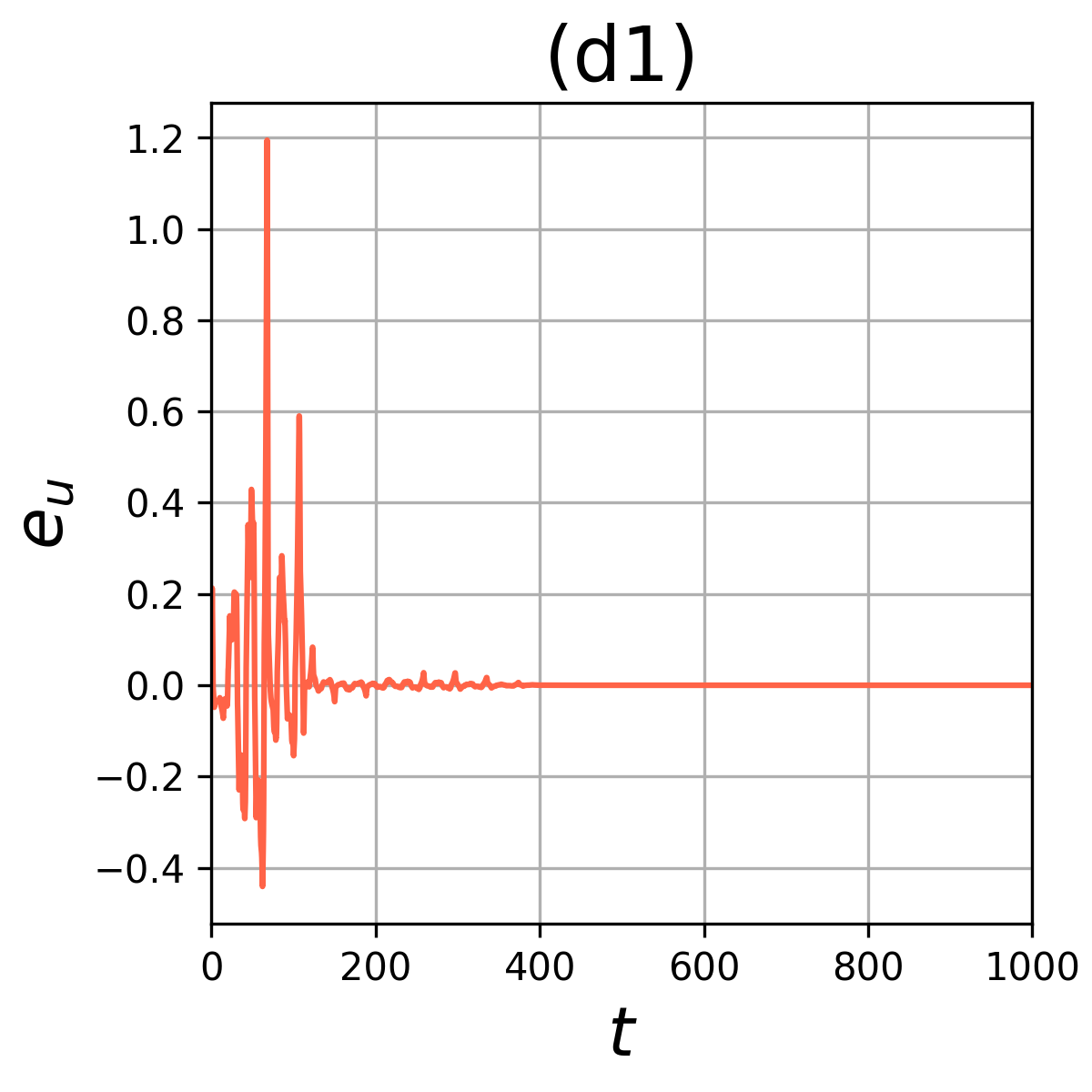}
\includegraphics[width=4cm,height=3cm]{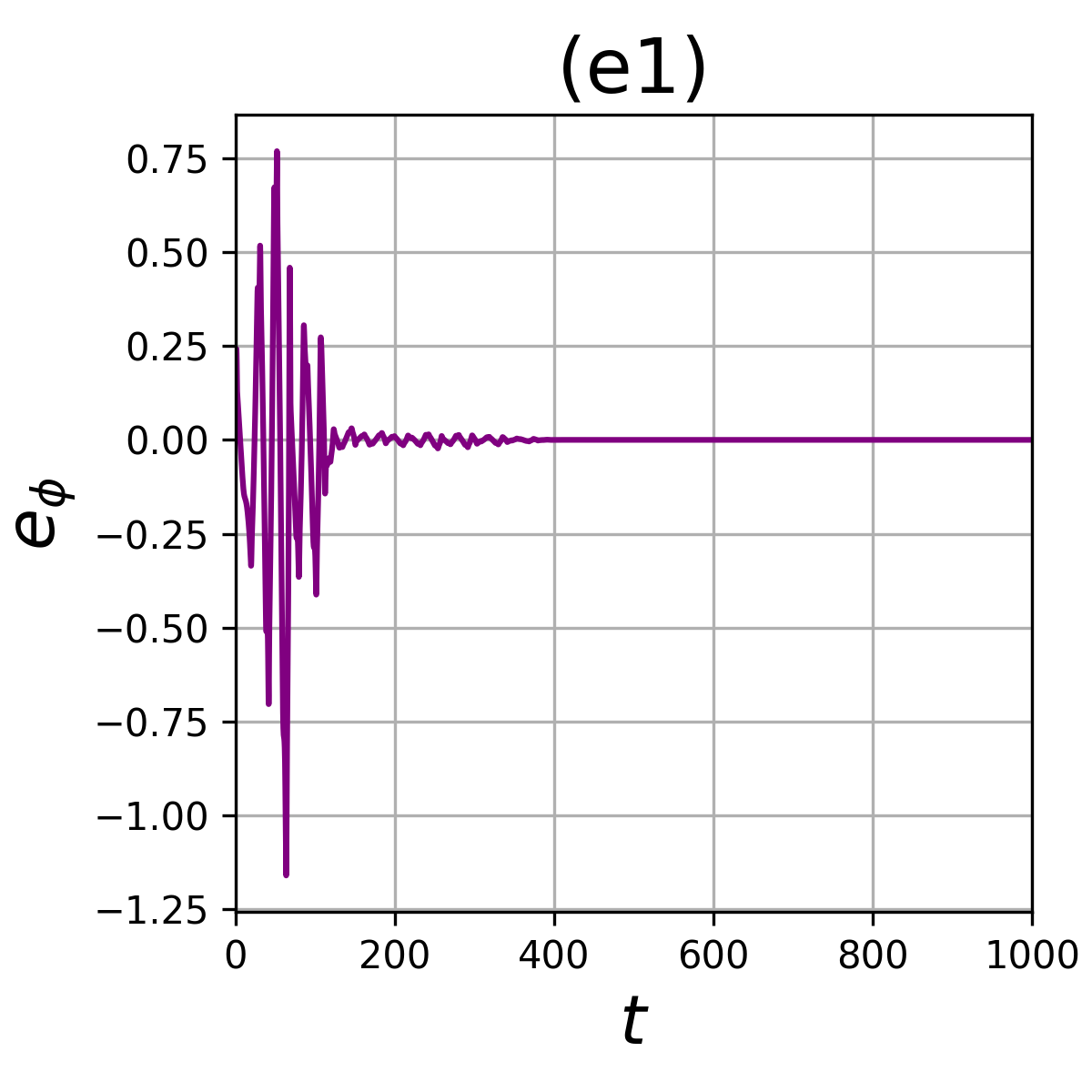}

\includegraphics[width=4cm,height=3cm]{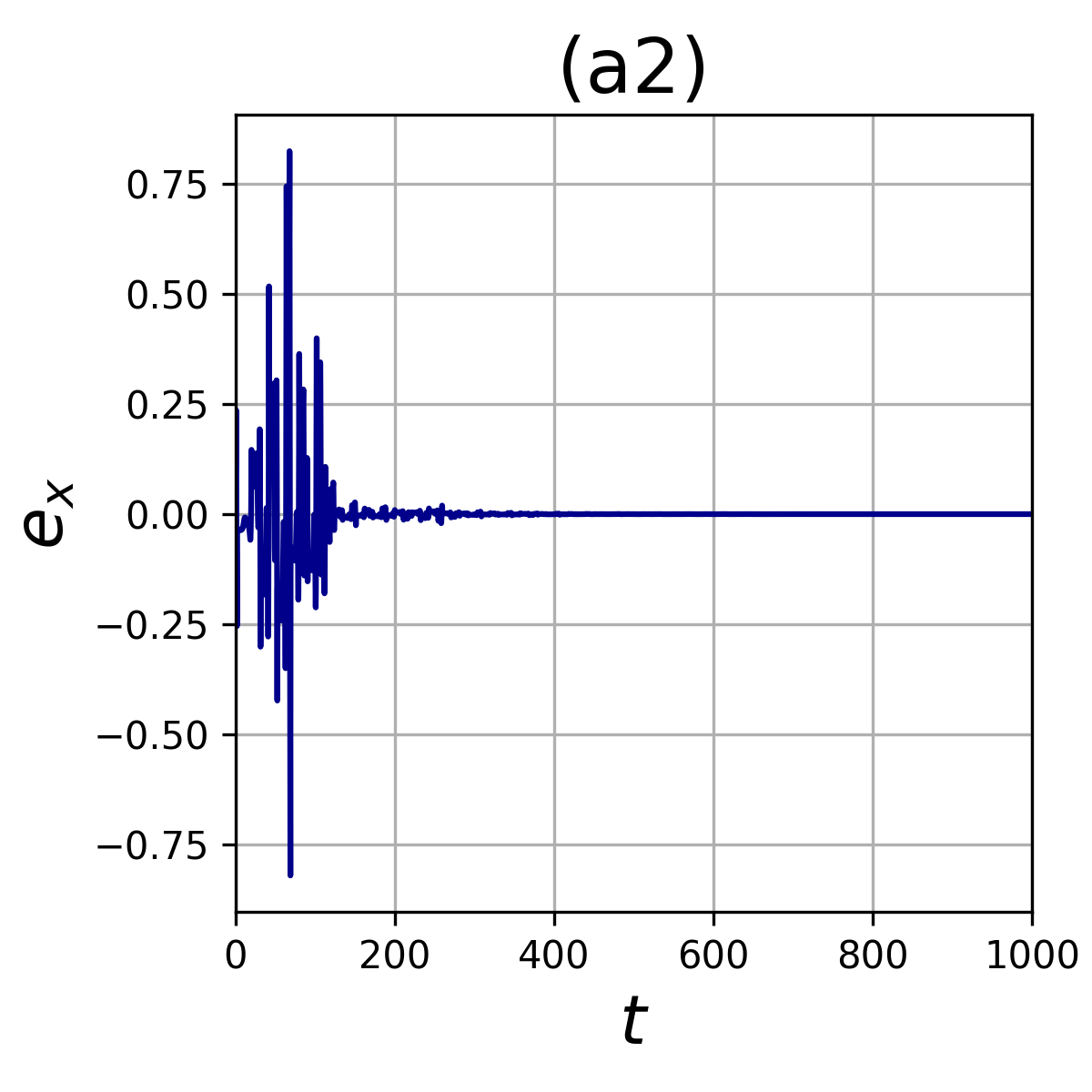}
\includegraphics[width=4cm,height=3cm]{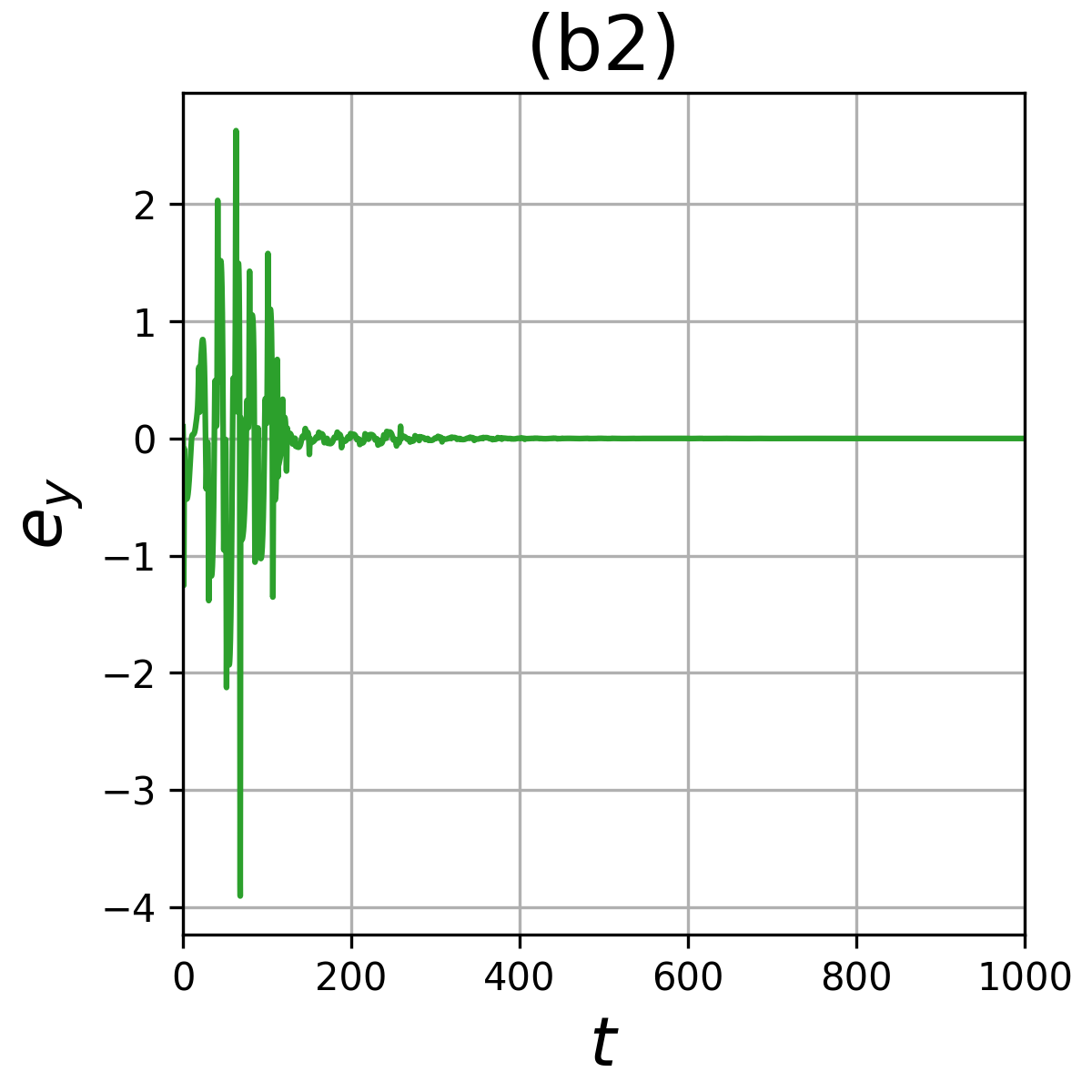}
\includegraphics[width=4cm,height=3cm]{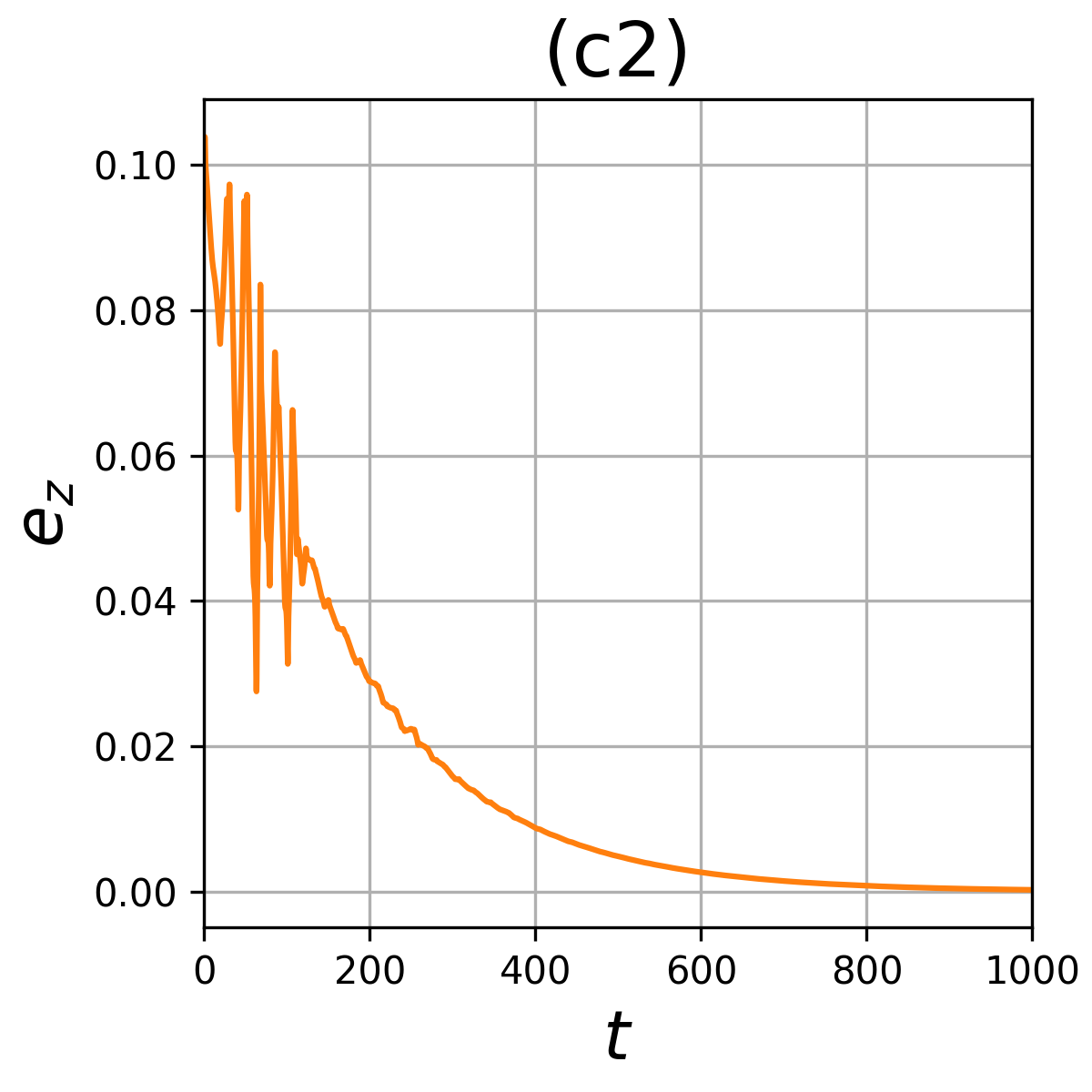}

\includegraphics[width=4cm,height=3cm]{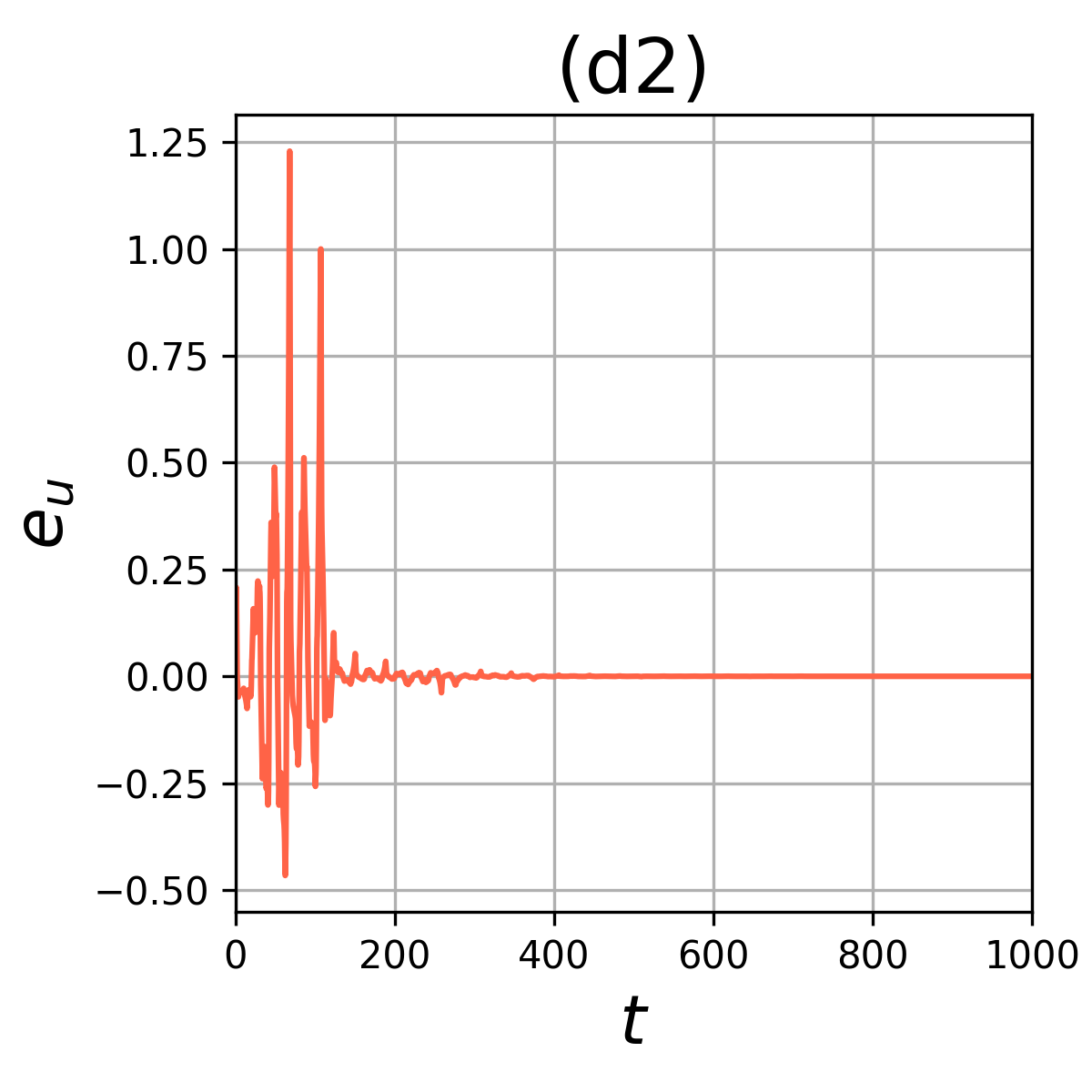}
\includegraphics[width=4cm,height=3cm]{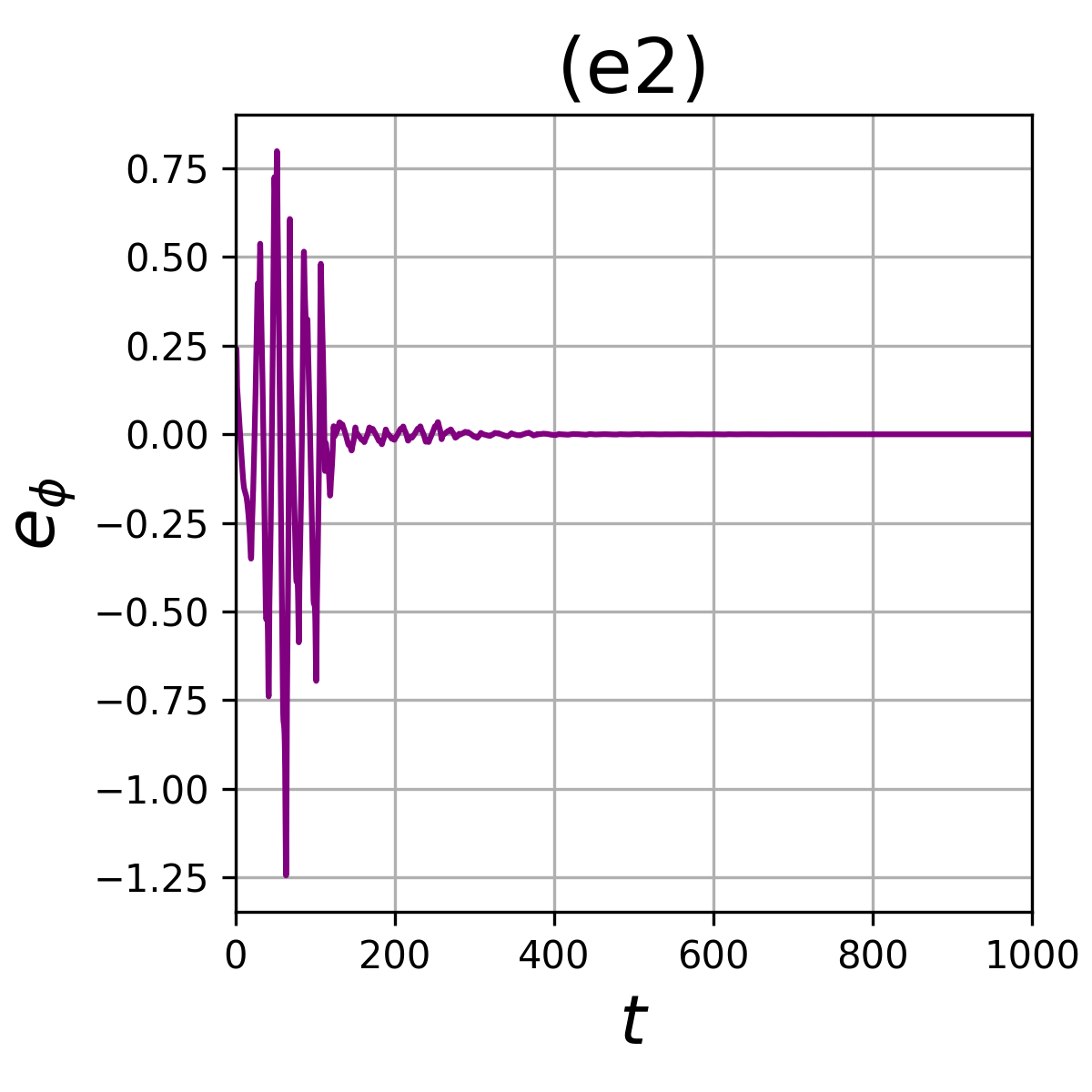}
\caption{\marius{Time evolution of the five synchronization-error components $(e_x,e_y,e_z,e_u,e_\phi)$ of the full nonlinear error system Eq. \eqref{nonlinear_error} in (a1)-(e1) and the linearized error system Eq. \eqref{linearized_error_Sys} in (a2)-(e2), all showing convergence to zero.}}
\label{fig:error_system}
\end{figure}

\marius{Figures~\ref{fig:error_system}(a2)-(e2) and (a1)-(e2) present the time series of the five synchronization--error components of the full nonlinear and linearized systems in Eqs. \eqref{fig:error_system} and \eqref{linearized_error_Sys}, respectively, each representing the difference between corresponding variables of the two coupled neurons. All error trajectories decay asymptotically to zero after a brief transient, confirming complete synchronization of the coupled system in Eq. \eqref{coupled_eq}.} One notice that the fast variables $e_x, e_y, e_u, e_\phi$ converge earlier than the slow adaptation variable $e_z$, reflecting the intrinsic separation of time scales in the neuron dynamics.
\marius{These results corroborate the linearized transverse stability established in Section \ref{sec_3.1}.}

\begin{figure}
\centering
\includegraphics[width=0.35\linewidth]{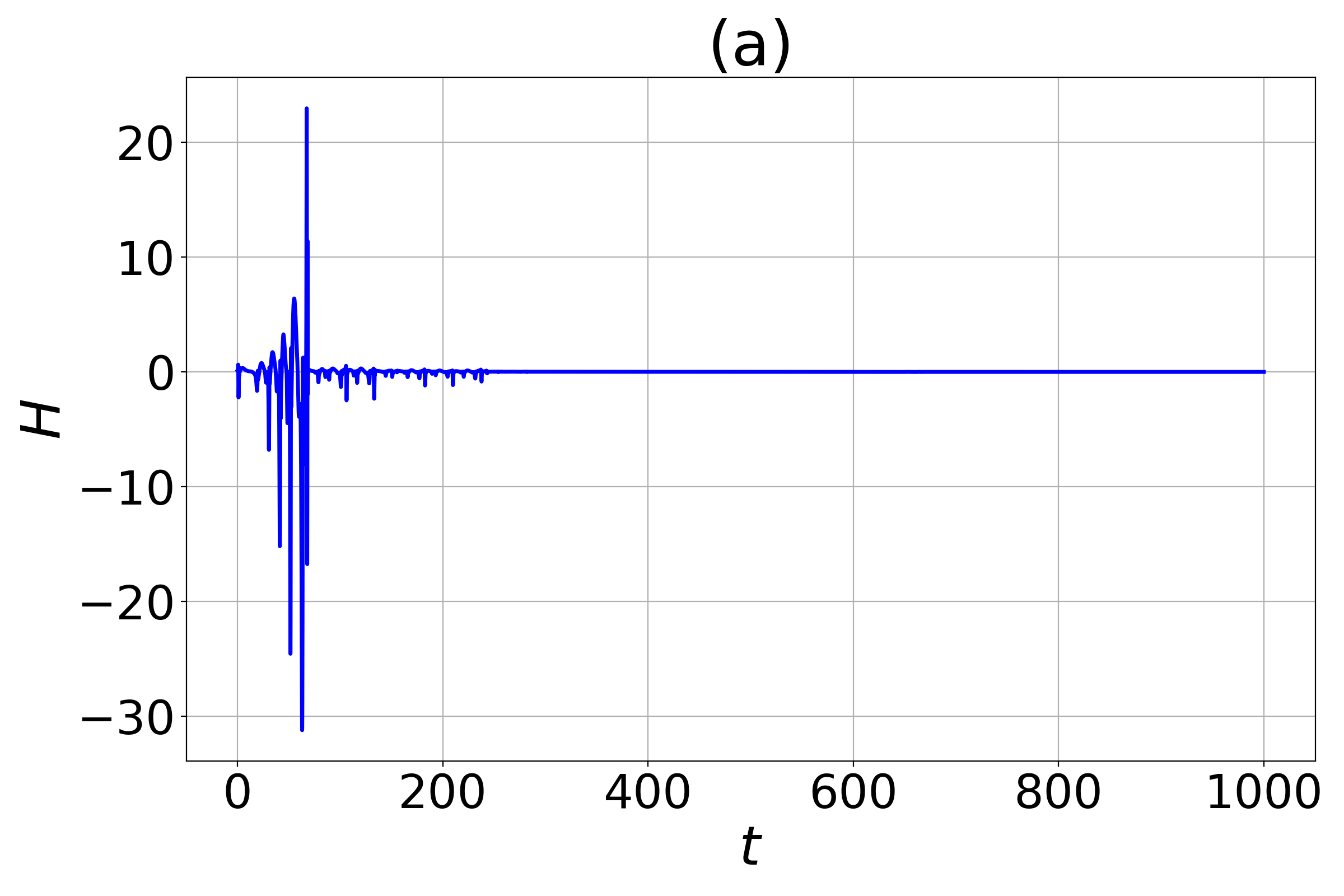}\hspace{0.01\textwidth}\includegraphics[width=0.35\linewidth]{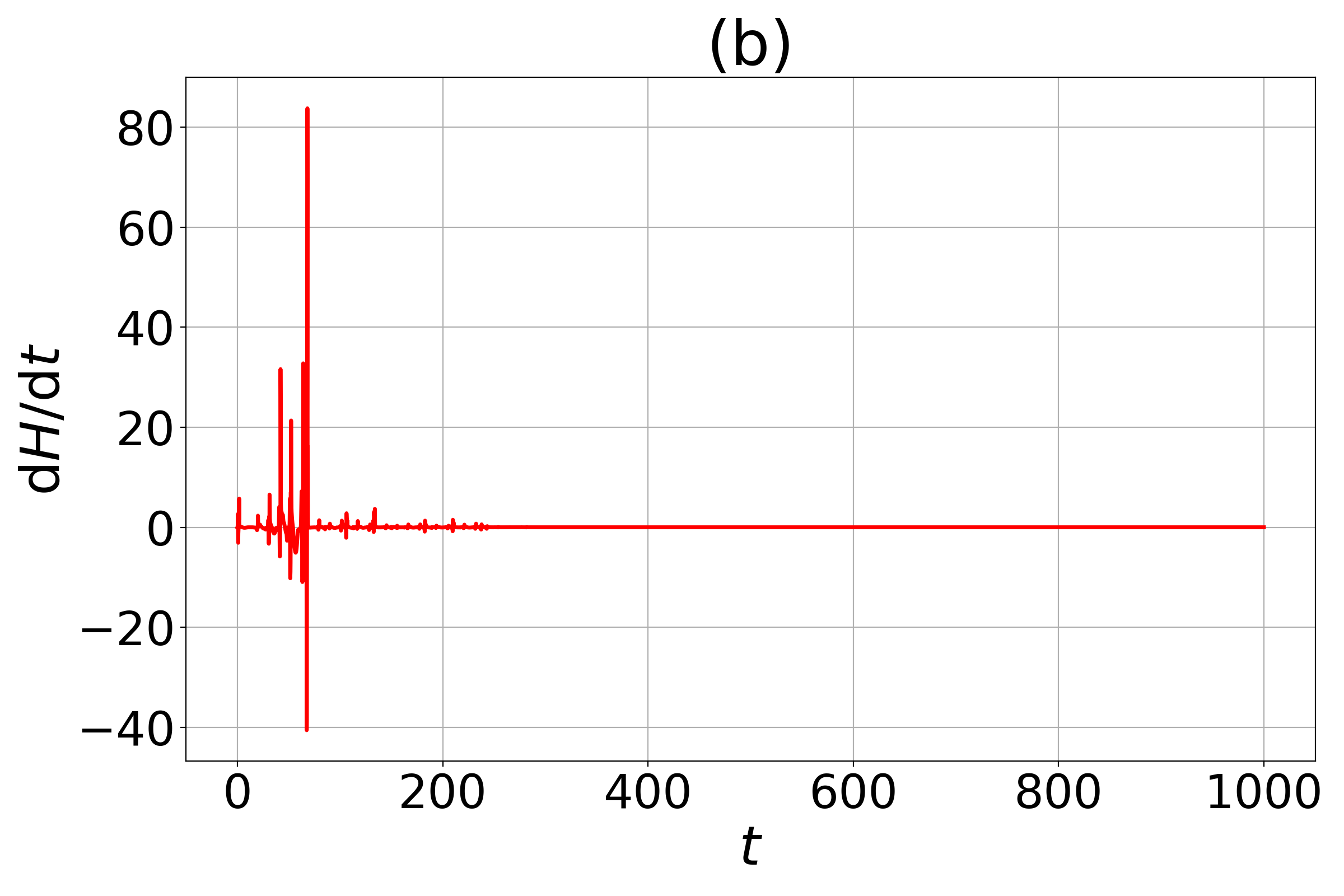}
\caption{Time evolution of the synchronization Hamiltonian function \textbf{(a)} and its time derivative \textbf{(b)}, associated with the errors shown in  \marius{Fig.~\ref{fig:error_system}(a2)-(e2)}; as the system synchronizes, both the \marius{synchronization Hamiltonian} and its derivative approach zero.}
\label{fig:hamiltonian}
\end{figure}

Figure~\ref{fig:hamiltonian} presents the temporal evolution of the \marius{synchronization Hamiltonian} $H$ and its derivative $\dot H$ \marius{associated with} the errors in \marius{Fig.~\ref{fig:error_system}(a2)-(e2)}. \marius{The Hamiltonian decays to zero as $t\to\infty$ (overall relaxation toward synchrony) while $\dot H$ can take both positive and negative values during the transient and then stabilizes near zero as the neurons approach synchrony.} The derivative $\dot H$ exhibits transient oscillations, taking both negative and positive values before stabilizing near zero; this reflects intermittent energy exchange due to memristive and flux-related terms in the dynamics. After transients, $\dot H \to 0$ and the system settles on the synchronization manifold, consistent with an overall dissipative relaxation toward synchrony.
  
\begin{figure}
\centering
\includegraphics[width=0.35\linewidth]{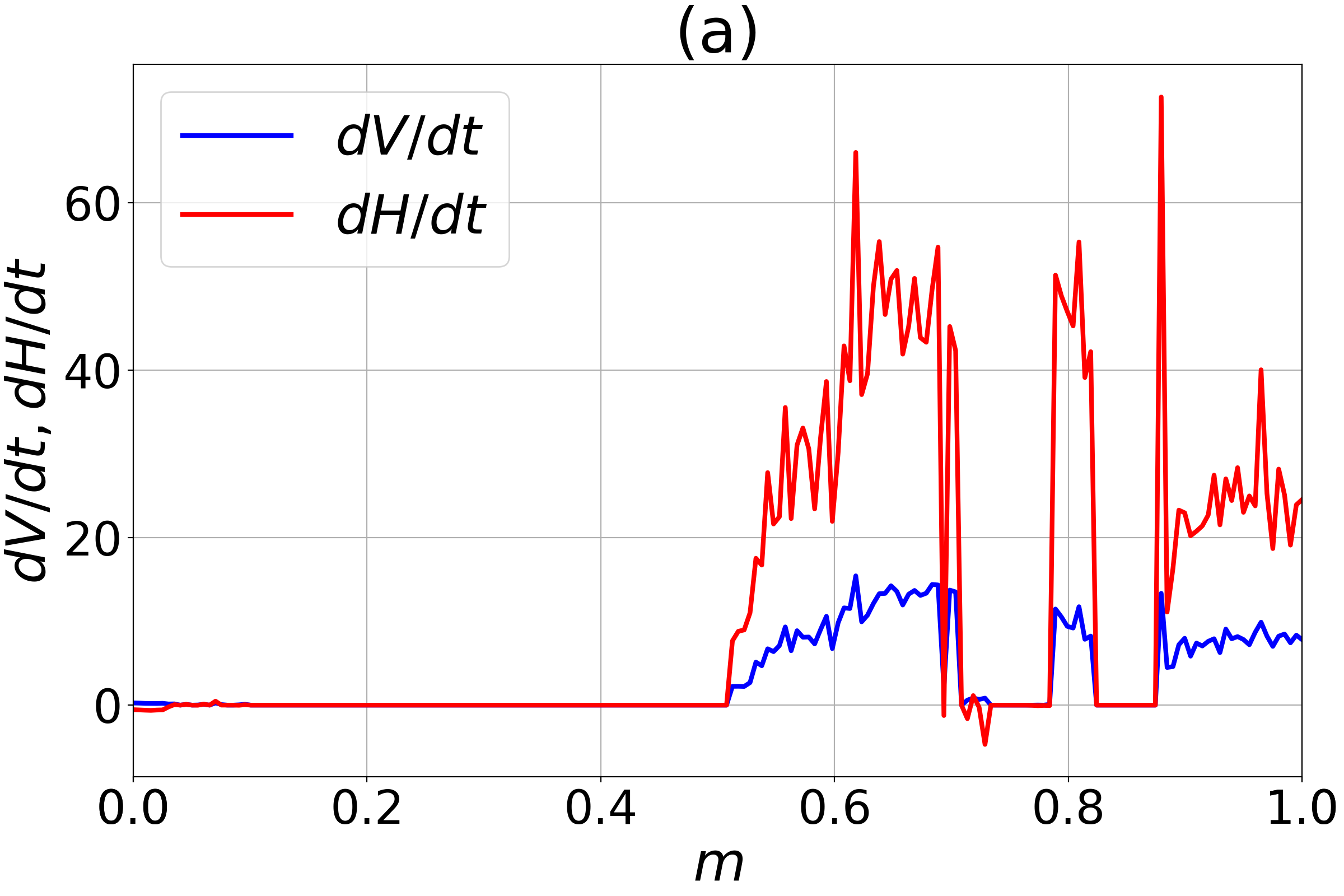}
\hspace{0.01\textwidth}\includegraphics[width=0.35\linewidth]{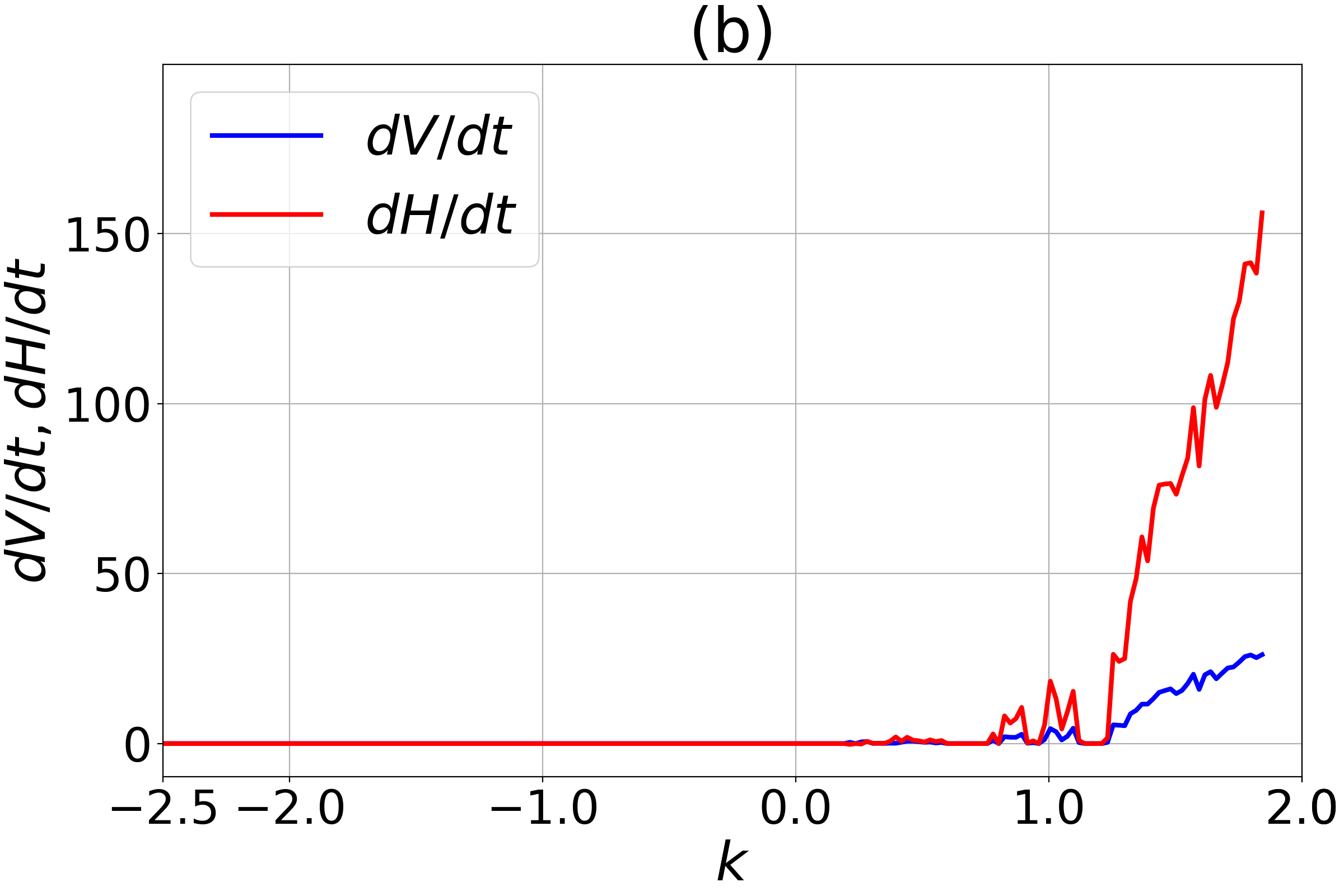}

\includegraphics[width=0.35\linewidth]{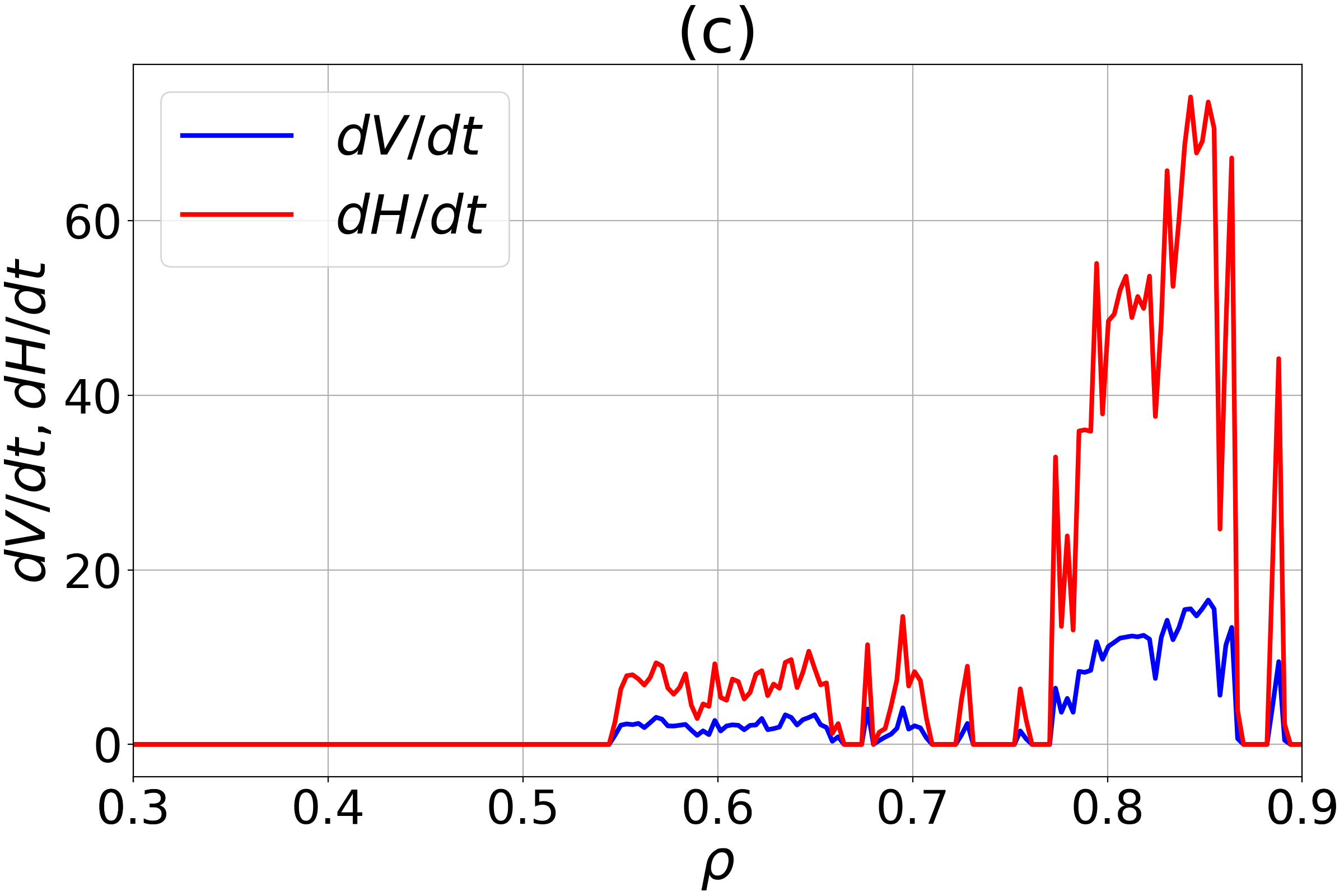}\hspace{0.01\textwidth}\includegraphics[width=0.35\linewidth]{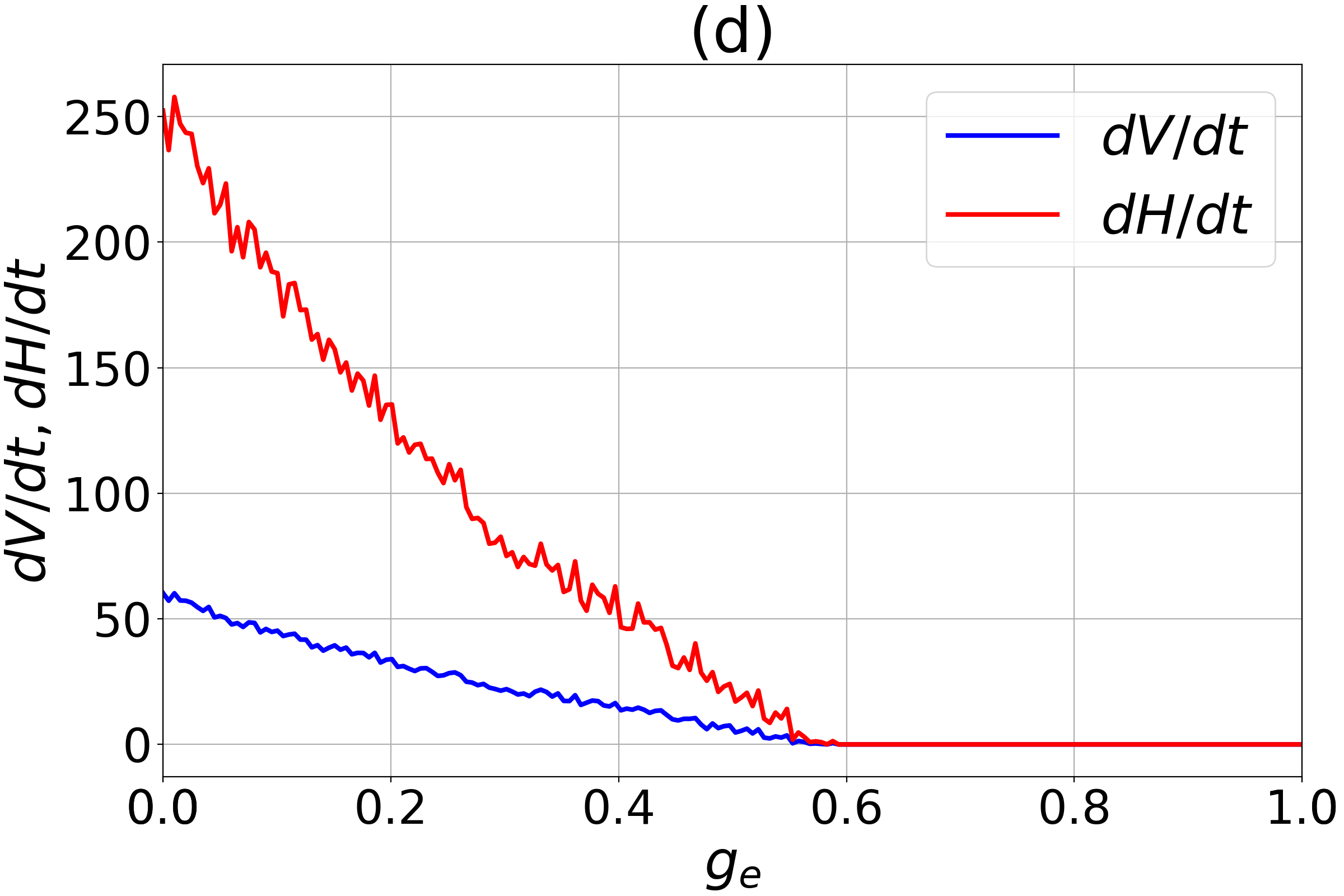}
\caption{Time derivatives of the Lyapunov and Hamiltonian functions plotted as functions of the parameters \(m\) and \(k\) in panels (a)–(b), and \(\rho\) and \(g_e\) in panels (c)–(d). In all cases, \(\dot{V}\) and \(\dot{H}\) exhibit the same qualitative trends, confirming the consistency between the Lyapunov and Hamiltonian stability assessments.
}
\label{fig:VHdot1D}
\end{figure}

Figure~\ref{fig:VHdot1D} compares the time-averaged derivatives of the Lyapunov function $\dot V$ and the Hamiltonian function $\dot H$ as functions of the main control parameters $m$, $k$, $\rho$, and $g_e$. Each data point corresponds to a post-transient average over a long trajectory. Both measures exhibit nearly identical qualitative and quantitative results across all parameters, confirming the consistency between the Lyapunov and Hamiltonian formulations. Increasing $g_e$ or reducing $m$ enhances dissipation and accelerates synchronization, whereas larger $\rho$ or $k$ increase the energy input required for synchrony. 

These findings further highlight the deep correspondence between Lyapunov-based and Hamiltonian-based characterizations of stability and energetic regulation in chaotic synchronization. In particular, the fact that $\dot{V}$ and $\dot{H}$ exhibit the same qualitative behavior implies that the Hamiltonian can serve as an effective stability certificate. This is especially valuable in situations where a suitable Lyapunov function $V$ is difficult to construct or analytically intractable, whereas the Hamiltonian structure is readily available or naturally derived from the port-Hamiltonian formulation. Thus, the Hamiltonian offers a principled and computationally accessible alternative for assessing convergence toward the synchronization manifold.

\begin{figure}
  \centering
  \includegraphics[width=5cm,height=4cm]{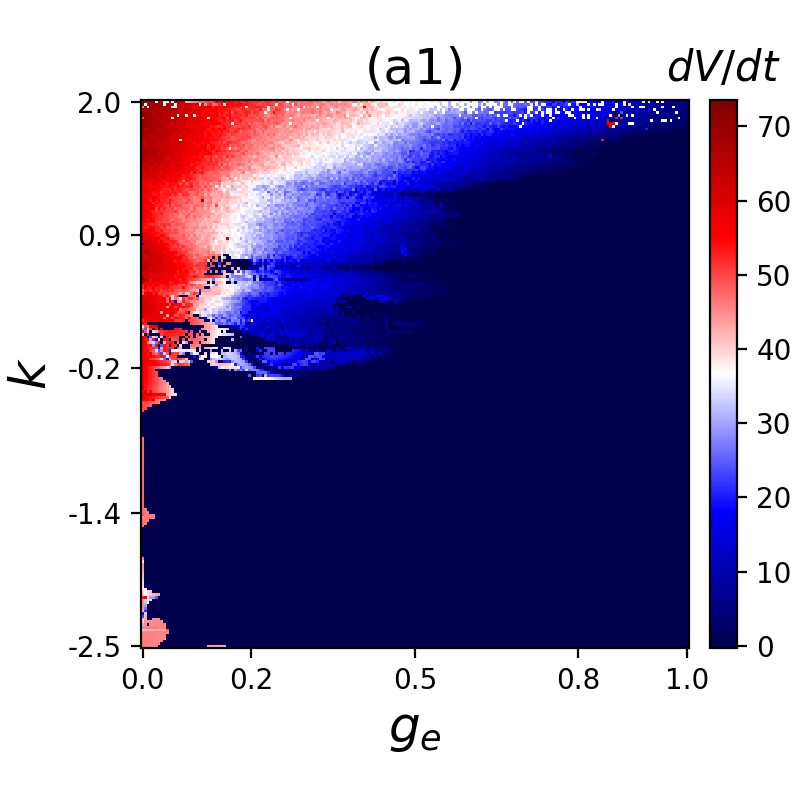}
  \hspace{0.01\textwidth}%
  \includegraphics[width=5cm,height=4cm]{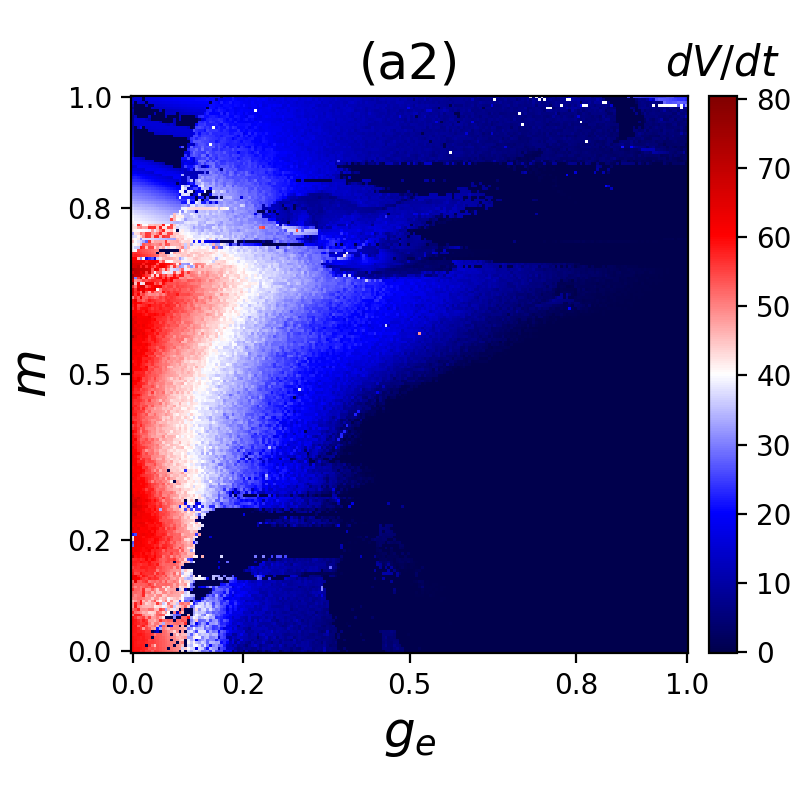}
  \hspace{0.01\textwidth}%
  \includegraphics[width=5cm,height=4cm]{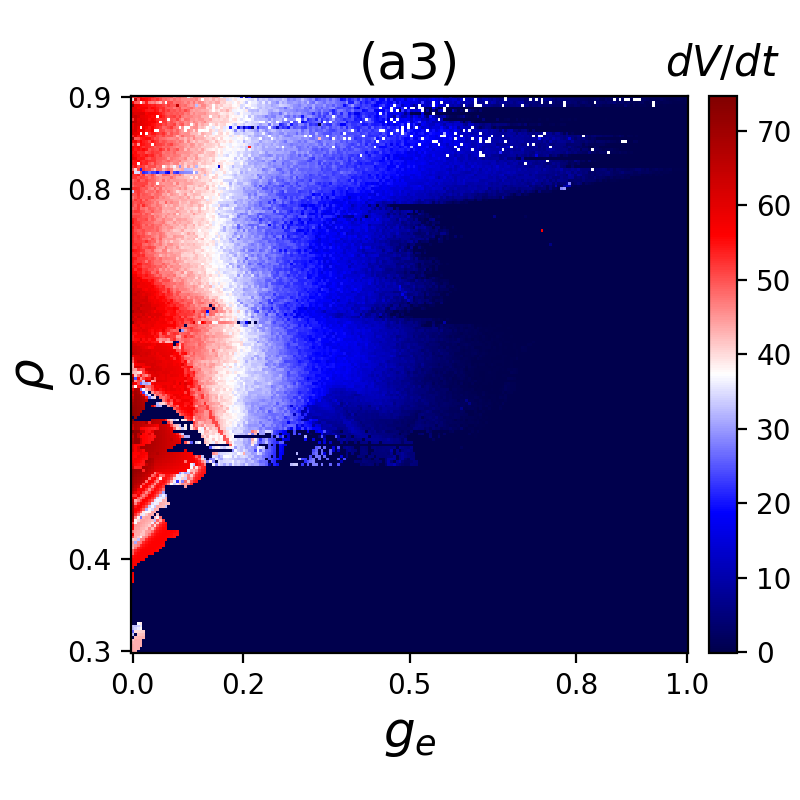}

  \includegraphics[width=5cm,height=4cm]{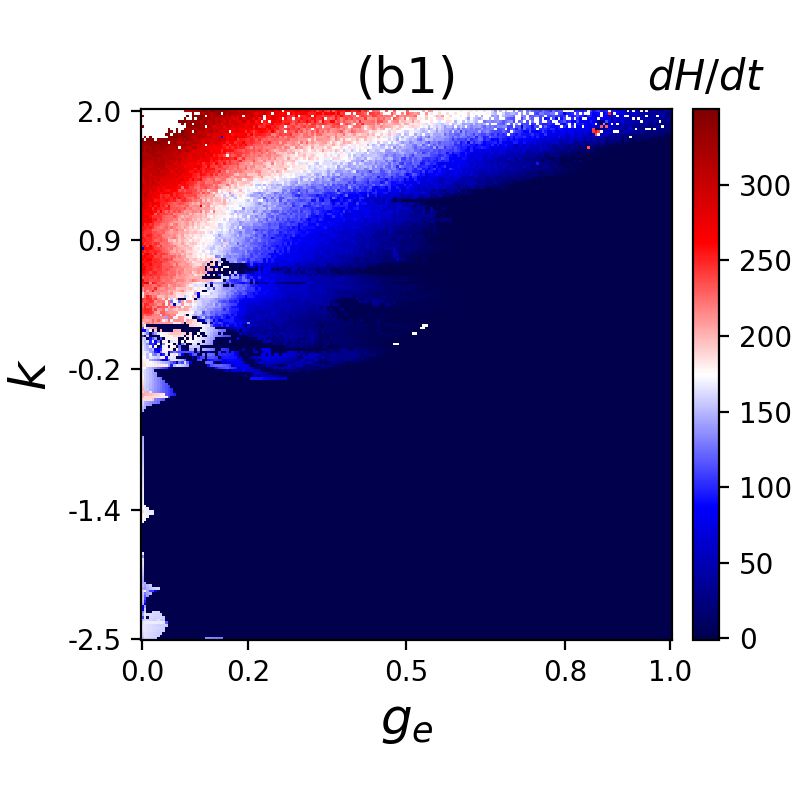}
  \hspace{0.01\textwidth}%
  \includegraphics[width=5cm,height=4cm]{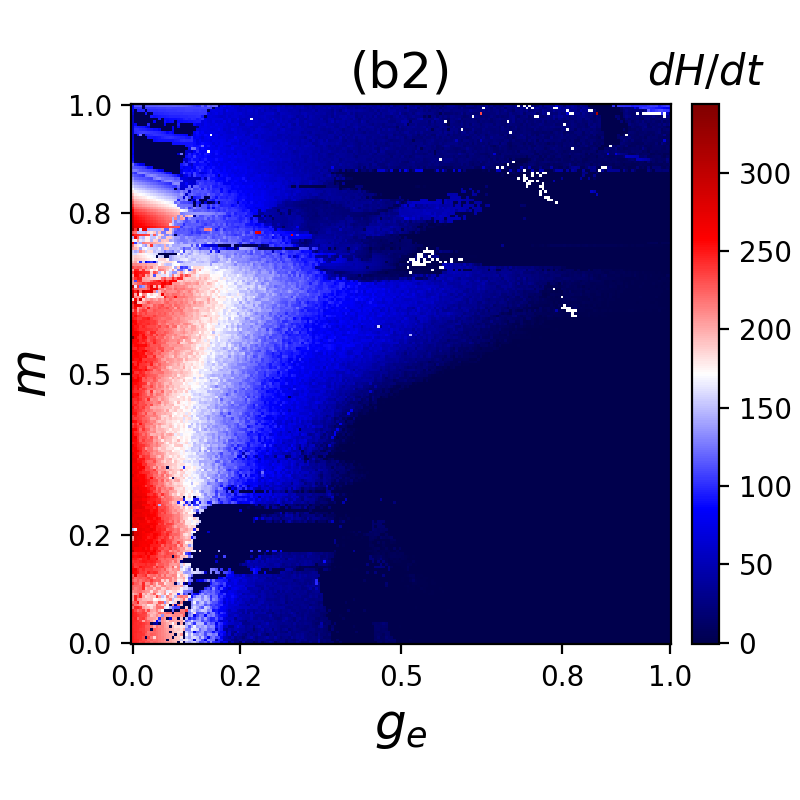}
  \hspace{0.01\textwidth}%
  \includegraphics[width=5cm,height=4cm]{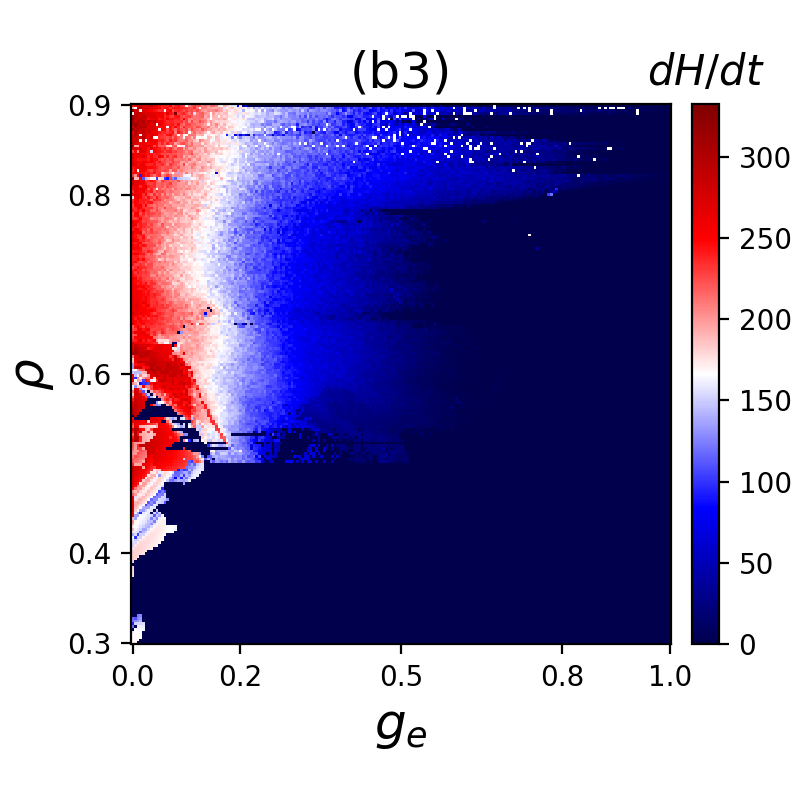}

  \caption{Heatmaps of the time derivatives of the Lyapunov function (a1)–(a3) and the corresponding synchronization Hamiltonian function (b1)–(b3) versus the coupling \(g_e\) and the system parameters \(k\), \(m\), and \(\rho\), respectively. Across all parameter pairs, \(\dot{V}\) and \(\dot{H}\) display consistent qualitative trends, demonstrating a robust correspondence between Lyapunov-based and Hamiltonian-based stability indicators.}
  \label{fig:VHdot2D}
\end{figure}

Figure~\ref{fig:VHdot2D} presents heatmaps of $\dot V$ (first row) and and the corresponding $\dot H$ (second row) as functions of the coupling strength $g_e$ and the intrinsic parameters $k$, $m$, and $\rho$. The color gradients represent the mean post-transient energy-dissipation rate. In all panels, dark blue regions correspond to parameter combinations yielding rapid and stable synchronization. The close agreement between the $\dot V$ and $\dot H$ maps confirms that both measures capture the same dissipative structure of the system. Efficient synchronization occurs for intermediate $m$ and sufficiently large $g_e$, whereas strong flux coupling ($k,\rho$ large) can destabilize the manifold by injecting excess energy. These results emphasize how electrical, magnetic, and memristive couplings jointly determine both the rate and energetic cost of neuronal synchrony.

In summary, the numerical results  across the tested parameter ranges support the analytical pictures from Sections \ref{sec_3.1} and \ref{sec:synchr_dynamics_Hamilton}: (1) the linearized transverse dynamics are stable under the derived sufficient conditions, and (2) the Hamiltonian function decays to zero as $t\to\infty$ and the Lyapunov and Hamiltonian formulations yielding consistent measures of stability and energetics.
These findings provide a basis for the pH-PINN framework introduced in Section~\ref{sec:pHNN_PINN}, which exploits the equivalence between Lyapunov and Hamiltonian formalisms to learn the underlying \marius{Hamiltonian} structure directly from data, bridging analytical theory with machine-learning discovery of \marius{Hamiltonian}-conserving and dissipative dynamical systems.

\section{A port-Hamiltonian physics-informed
neural network (pH-PINN) approach}
\label{sec:pHNN_PINN}
To study \marius{synchronization Hamiltonian} of the coupled HR neurons in Eq. \eqref{coupled_eq}  from a data-driven perspective, we employ a pH–PINN. This framework is well-suited to our system, whose synchronization behavior admits a Hamiltonian structure consistent with the pH formulation of conservative and dissipative flows. Analytical Hamiltonian approaches can establish stability but often become infeasible for high-dimensional or partially known systems. In such settings, the governing physical laws, such as energy conservation and dissipation, remain valid even when the explicit equations are unknown. The pH–PINN incorporates these physical constraints in the learning process, allowing the \marius{synchronization Hamiltonian} and its dynamics to be inferred directly from data. Here, the known analytical \marius{Hamiltonian} serves as a benchmark to quantitatively validate the learned representation. In the sequel, we briefly outline the principles of PINNs, HNNs, and pHNNs, and describe how these are integrated to construct the proposed pH-PINN framework.

\subsection{Physics-informed neural networks (PINNs)}  
PINN~\cite{Raissi2019} embeds the governing differential equations directly into the training objective.  
For a system  
\begin{equation}
\dot{\mathbf{x}} = \mathbf{f}(\mathbf{x}), 
\quad
\mathbf{x}\in\mathbb{R}^n,\;
\mathbf{f}:\mathbb{R}^n \to \mathbb{R}^n,
\end{equation}
let \(\dot{\mathbf{x}}_\theta(\mathbf{x}) \equiv \mathbf{f}_\theta(\mathbf{x})\) denote a neural approximation of the vector field.  
Given data or collocation samples \(\{(\mathbf{x}_i,\dot{\mathbf{x}}_i)\}_{i=1}^B\), a PINN minimizes the residual of the governing equation:
\begin{equation}
\mathcal{L}_{\mathrm{\tiny PDE}}
=
\frac{1}{B}\sum_{i=1}^B
\big\|
\dot{\mathbf{x}}_i - \mathbf{f}_\theta(\mathbf{x}_i)
\big\|^2,
\end{equation}
where the equations act as {soft physical constraints} that bias learning toward dynamically consistent solutions and improve generalization beyond purely data-driven fitting.  
This residual formulation also includes structure-generated vector fields derived from learned scalar potentials, leading naturally to HNNs.

\subsection{Hamiltonian neural networks (HNNs)}  
HNNs~\cite{Greydanus2019HNN} employ a scalar-valued network \(H_\theta:\mathbb{R}^n\!\to\!\mathbb{R}\) that maps the state vector \(\mathbf{x}\) to a \marius{Hamiltonian} \(H_\theta(\mathbf{x})\).  
Automatic differentiation provides the gradient \(\nabla H_\theta(\mathbf{x})\), representing partial derivatives of \(H_\theta\) with respect to the state components.  
Rather than predicting the vector field directly, the dynamics are recovered through a (possibly state-dependent) skew-symmetric matrix \(\mathbf{J}(\mathbf{x}) \in \mathbb{R}^{n\times n}\):
\begin{equation}
\dot{\mathbf{x}}_\theta(\mathbf{x})
= 
\mathbf{J}(\mathbf{x})\,\nabla H_\theta(\mathbf{x}).
\label{eq:hnn_basic}
\end{equation}
Training minimizes the mismatch between the induced field \(\dot{\mathbf{x}}_\theta\) and reference derivatives \(\dot{\mathbf{x}}\) (from data or automatic differentiation) as in the PINN framework.  
Classical HNNs use the canonical symplectic matrix \(\mathbf{J}_c\), yielding the standard Hamiltonian form  
\(\dot{\mathbf{x}} = (\partial H/\partial \mathbf{p},\, -\,\partial H/\partial \mathbf{q})^\top\).  
For complex or noncanonical systems, however, \(\mathbf{J}(\mathbf{x})\) may be learned as a separate neural network to capture generalized symplectic structures~\cite{Chen2021}. 

\subsection{Port-Hamiltonian neural networks (pHNN)}
Hamiltonian neural networks can be extended to the pH setting by augmenting the conservative Hamiltonian flow with an additional symmetric matrix \(\mathbf{R}\). While HNNs describe ideally conservative systems, many dynamical systems exhibit extra non-conservative effects, which the pH formalism represents together with the conservative part~\cite{vanderSchaft2014}. See also~\cite{Desai2021PortHNN} for a neural implementation.
\begin{equation}
\begin{aligned}
\dot{\mathbf{x}} &= \big[\mathbf{J}(\mathbf{x}) - \mathbf{R}(\mathbf{x})\big]\,\nabla H_\theta(\mathbf{x}).
\end{aligned}
\label{eq:ph_standard}
\end{equation}
Here, \(\mathbf{J}\) encodes power-conserving interconnections, and \(\mathbf{R}\in\mathbb{R}^{n\times n}\) represents a non-conservative part of the dynamics. The \marius{Hamiltonian} rate satisfies
\begin{equation}
\dot{H}_\theta 
=
\nabla H_\theta^{\!\top}(\mathbf{J}-\mathbf{R})\nabla H_\theta
=
-\,\nabla H_\theta^{\!\top}\mathbf{R}\,\nabla H_\theta,
\label{eq:ph_energy_rate}
\end{equation}
since \(\nabla H_\theta^{\!\top}\mathbf{J}\nabla H_\theta = 0\) for any skew-symmetric \(\mathbf{J}\).  
Thus, the pH formulation makes the energetic structure explicit: the conservative flow preserves \(H_\theta\) and the non-conservative part determines its time evolution.

\subsection{Port--Hamiltonian physics-informed neural network (pH-PINN)}
\label{sec:ph-pinn}
We extend the conventional pHNN (Eq.~\eqref{eq:ph_standard}) to a physics-informed framework by embedding the Helmholtz decomposition of the error dynamics as explicit training constraints. Specifically, we incorporate the analytical conservative and non-conservative fields of the linearized error system in Eq.~\eqref{cons_diss} (derived from Eq.~\eqref{linearized_error_Sys}) into the pH factorization following the generalized Hamiltonian approach~\cite{chun2016calculation,ma2017calculation}.

Let \(H_\theta\), \(\mathbf{J}_\phi(\mathbf{e})\), and \(\mathbf{R}_\psi(\mathbf{e})\) define the pHNN.  
The resulting pH-PINN minimizes the composite loss
\begin{equation}
\mathcal{L}_{\mathrm{\tiny Total}}
=
\lambda_{\mathrm{\tiny DynRes}}\mathcal{L}_{\mathrm{\tiny DynRes}}
+\lambda_{\tiny J}\mathcal{L}_{\tiny J}
+\lambda_{\tiny R}\mathcal{L}_{\tiny R}
+\lambda_{\mathrm{\tiny Cons}}\mathcal{L}_{\mathrm{\tiny Cons}}
+\lambda_{\tiny{\dot H}}\mathcal{L}_{\tiny{\dot H}},
\label{eq:total_loss}
\end{equation}
where, over a batch \(\{(\mathbf{e}_i,\dot{\mathbf{e}}^{\mathrm{true}}_i)\}_{i=1}^B\),
\begin{equation}
\left\{
\begin{aligned}
\mathcal{L}_{\mathrm{\tiny DynRes}}
&=\frac{1}{B}\sum_{i=1}^B
\!\big\|
\dot{\mathbf{e}}^{\mathrm{true}}_i
-\big[\mathbf{J}_{\!\phi}(\mathbf{e}_i)-\mathbf{R}_{\!\psi}(\mathbf{e}_i)\big]\nabla H_\theta(\mathbf{e}_i)
\big\|^2,\\[1mm]
\mathcal{L}_{\tiny J}
&=\frac{1}{B}\sum_{i=1}^B
\!\big\|
\mathbf{F}_c(\mathbf{e}_i)
- \mathbf{J}_{\!\phi}(\mathbf{e}_i)\nabla H_\theta(\mathbf{e}_i)
\big\|^2,\\[1mm]
\mathcal{L}_{\tiny R}
&=\frac{1}{B}\sum_{i=1}^B
\!\big\|
\mathbf{F}_d(\mathbf{e}_i)
+\mathbf{R}_{\!\psi}(\mathbf{e}_i)\nabla H_\theta(\mathbf{e}_i)
\big\|^2,\\[1mm]
\mathcal{L}_{\mathrm{\tiny Cons}}
&=\frac{1}{B}\sum_{i=1}^B
\!\big(\nabla H_\theta(\mathbf{e}_i)^{\!\top}\mathbf{F}_c(\mathbf{e}_i)\big)^{\!2},\\[1mm]
\mathcal{L}_{\tiny{\dot H}}
&=\frac{1}{B}\sum_{i=1}^B
\!\Big(
\nabla H_\theta(\mathbf{e}_i)^{\!\top}\dot{\mathbf{e}}^{\mathrm{true}}_i
-\!\nabla H_\theta(\mathbf{e}_i)^{\!\top}\mathbf{R}_{\!\psi}(\mathbf{e}_i)\nabla H_\theta(\mathbf{e}_i)
\Big)^{\!2}.
\end{aligned}
\right.
\end{equation}

Each term enforces a distinct physical law:  
\(\mathcal{L}_{\mathrm{\tiny DynRes}}\) ensures dynamic consistency;  
\(\mathcal{L}_{\tiny J}\) and \(\mathcal{L}_{\tiny R}\) align learned and analytical conservative/{non-conservative} flows;  
\(\mathcal{L}_{\mathrm{\tiny Cons}}\) enforces \marius{Hamiltonian} invariance along conservative trajectories;  
and \(\mathcal{L}_{\tiny{\dot H}}\) enforces the \marius{Hamiltonian}--rate identity (Eq. \ref{eq:ph_energy_rate}).

The \marius{synchronization Hamiltonian} is parameterized by a multilayer perceptron (MLP) \(f_{\theta}:\mathbb{R}^n\!\to\!\mathbb{R}\) and anchored at the synchronization manifold to remove gauge freedom and ensure a strict local minimum:
\begin{equation}
H_{\theta}(\mathbf{e})
=
f_{\theta}(\mathbf{e})
-\big(f_{\theta}(\mathbf{0})
+\nabla f_{\theta}(\mathbf{0})^{\!\top}\mathbf{e}\big)
+\varepsilon\|\mathbf{e}\|_2^2,
\end{equation}
so that \(H_{\theta}(\mathbf{0})=0\) and \(\nabla H_{\theta}(\mathbf{0})=\mathbf{0}\).

Following Eq.~\eqref{eq:ph_standard} and the analytical decomposition in Eq.~\eqref{cons_diss}, we impose sparse interconnection and diagonal {non-conservative} structures for \(\mathbf{J}_{\!\phi}\) and \(\mathbf{R}_{\!\psi}\), respectively, while learning both via separate neural networks:
\[
\mathbf{J}_{\!\phi}=
\begin{pmatrix}
0 & \ast & \color{gray}{?} & \ast & \ast\\
\ast & 0 & 0 & 0 & 0\\
\color{gray}{?} & 0 & 0 & 0 & 0\\
\ast & 0 & 0 & 0 & 0\\
\ast & 0 & 0 & 0 & 0
\end{pmatrix},
\quad
\mathbf{R}_{\!\psi}=
\begin{pmatrix}
\ast & 0 & 0 & 0 & 0\\
0 & \ast & 0 & 0 & 0\\
0 & 0 & \ast & 0 & 0\\
0 & 0 & 0 & \ast & 0\\
0 & 0 & 0 & 0 & \ast
\end{pmatrix},
\]

This structure, inferred from \(\mathbf{F}_c\) and \(\mathbf{F}_d\) (Eq.~\eqref{cons_diss}), reveals a sparse coupling pattern.  
We thus mask \(\mathbf{J}_{\!\phi}\) and constrain \(\mathbf{R}_{\!\psi}\) to be diagonal, consistent with standard pH learning~\cite{Desai2021PortHNN}.  
Because the sparsity pattern depends on the Hamiltonian gradient \(\nabla H\), we adopt a {separable-gradient} assumption, noting that many physical systems possess approximately separable \marius{Hamiltonian} functions~\cite{Greydanus2019HNN,Zhong2019SymODEN,Chen2021}.  
Each component \(\partial H / \partial e_i\) thus depends primarily on its own coordinate (and possibly a small subset), consistent with the weak cross-partials observed in \(\mathbf{F}_c\) and \(\mathbf{F}_d\).  
Under this assumption, rows \(2\)--\(5\) of \(\mathbf{F}_c\) depend only on \(e_x\), while \(F_{c,x}\) depends on \(e_y, e_u,\) and \(e_\phi\) (but not \(e_z\)), motivating a masked, skew-symmetric \(\mathbf{J}_{\!\phi}\) (\(\mathbf{J}_{\!\phi}^{\!\top}=-\mathbf{J}_{\!\phi}\)).  
Inspection of \(\mathbf{F}_d\) further reveals primarily self-coupling (coordinate-wise damping) terms, motivating a diagonal parameterization of \(\mathbf{R}_{\!\psi}\).

\subsection{Training and Results}
\label{sec:training_results}

We train a pH\textendash PINN by minimizing the total loss in Eq.~\eqref{eq:total_loss}. 
\marius{The ground-truth derivatives needed for the supervised learning are computed \emph{analytically} by evaluating the model (Eq. \eqref{linearized_error_Sys}) right-hand sides at the states saved during ODE integration (no numerical/finite-difference differentiation is used on these states). Concretely, after integrating the coupled HR system in Eq.~\eqref{coupled_eq} together with the linearized error dynamics in Eq.~\eqref{linearized_error_Sys} and storing the states on the solver time grid $\{t_k\}_{k=1}^{N}$, we compute $\dot{\mathbf e}^{\mathrm{true}}(t_k)$ by direct substitution of the saved states into the right-hand side of Eq.~\eqref{linearized_error_Sys} (and similarly for the decomposition $\mathbf F=\mathbf F_c+\mathbf F_d$ in Eq.~\eqref{cons_diss}).}

\marius{Each sample comprises the error state \(\mathbf e\in\mathbb{R}^5\), its ground\textendash truth derivative \(\dot{\mathbf e}^{\mathrm{true}}\), the full neuron states of both units (used only to build a short {context} vector \(\mathbf c=(x_1,u_1,u_2,\phi_1)^\top\in\mathbb{R}^4\)), and the analytical Hamiltonian and \marius{Hamiltonian} rate \((H,\dot H)\) from Eqs.~\eqref{eq:hamiltonian} and \eqref{eq:Hdot} for monitoring.
The context vector \(\mathbf c\) is required because the analytical conservative/dissipative fields in Eq.~\eqref{cons_diss} are \emph{state-dependent} through coefficients such as \(N(x_1,u_1,\phi_1)\) and the memristive switching terms \(\alpha(u_1,u_2)\), \(\beta(u_1,u_2)\); accordingly, the learned pH--PINN must be conditioned on \((x_1,u_1,u_2,\phi_1)\) to represent the same non-autonomous dynamics.
Importantly, \(\mathbf c\) contains only observed state variables needed to parameterize the right-hand side and does not include any supervision targets \textit{e.g.,} \(\dot{\mathbf e}^{\mathrm{true}}\), \(H\), or \(\dot H\)).}

\begin{figure}
\includegraphics[width=\textwidth]{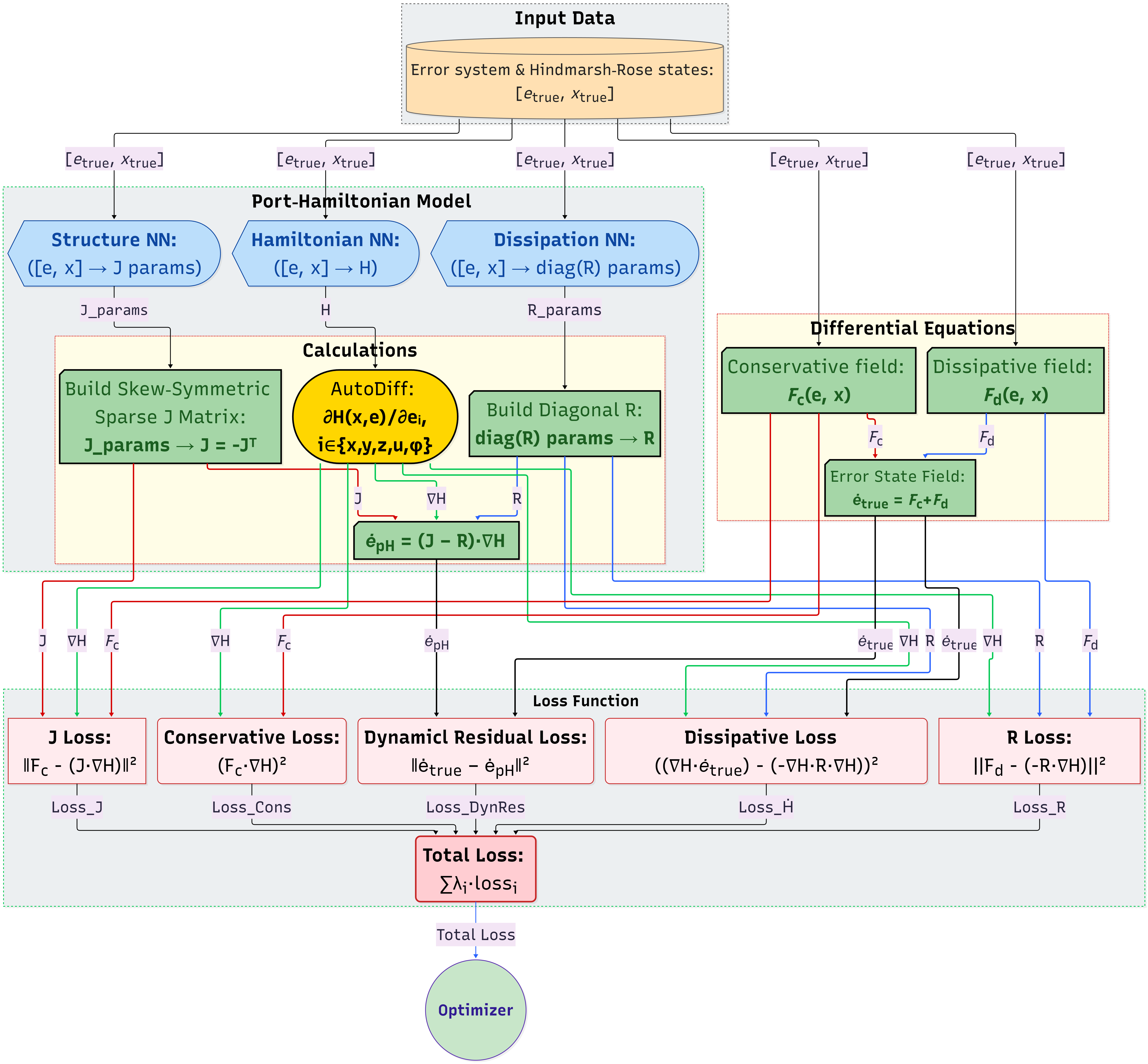}
    \caption{Schematic diagram of the proposed pH-PINN architecture.}
    \label{fig:pinn_diagram}
\end{figure}

\marius{All quantities are kept in physical units and no normalization is used: applying a state normalization would require a consistent rescaling of \(\mathbf e\), \(\dot{\mathbf e}\), and the learned pH factors \((\mathbf J,\mathbf R,H)\) to preserve the scale of the induced prediction \(\dot{\mathbf e}_{\mathrm{pred}}=(\mathbf J-\mathbf R)\nabla H\) in the residual loss \(\|\dot{\mathbf e}^{\mathrm{true}}-\dot{\mathbf e}_{\mathrm{pred}}\|^2\).
Given that the simulated trajectories are bounded over the training window, we found training stable without additional scaling.
All networks consume the concatenated \([\mathbf e,\mathbf c]\) as input. The overall architecture is shown in Fig.~\ref{fig:pinn_diagram}, and the implementation details are summarized in Table~\ref{tab:ph_pinn_impl}.}

\begin{table}[H]
\centering
\footnotesize
\setlength{\tabcolsep}{3pt}
\renewcommand{\arraystretch}{1.2}
\caption{Implementation details of the pH-PINN blocks.}
\label{tab:ph_pinn_impl}
\begin{tabular}{p{1cm} p{2.7cm} p{6.5cm} p{4.2cm}}
\hline
\textbf{Block} & \textbf{Inputs (dim)} & \textbf{Architecture} & \textbf{Outputs} \\
\hline
\(H_\theta\)
& \([\mathbf e,\mathbf c]\) (9)
& \begin{tabular}[t]{@{}l@{}}
  width: 2048,\\
  depth: 1,\\
  activation: \texttt{tanh},\\
  gauge anchoring at \(\mathbf e=\mathbf 0\),\\
  quadratic well \(\varepsilon=0.01\)
  \end{tabular}
& scalar: \(H\) \\
\hline
\(J_\phi\)
& \([\mathbf e,\mathbf c]\) (9)
& \begin{tabular}[t]{@{}l@{}}
  width: 32,\\
  depth: 4,\\
  activation: \texttt{tanh},\\
  skew\textendash symmetric \(J\), sparse pattern
  \end{tabular}
& 4 params: \([j_{12},j_{13},j_{14},j_{15}]\) \\
\hline
\(R_\psi\)
& \([\mathbf e,\mathbf c]\) (9)
& \begin{tabular}[t]{@{}l@{}}
  width: 2,\\
  depth: 2,\\
  activation: \texttt{tanh},\\
  diagonal \(R\)
  \end{tabular}
& 5 params: \(\operatorname{diag}(R_{11},\ldots,R_{55})\) \\
\hline
\end{tabular}
\end{table}

\marius{
The dataset consists of \(N=200{,}000\) time samples \(\{(\mathbf e(t_k),\mathbf c(t_k),\dot{\mathbf e}^{\mathrm{true}}(t_k))\}_{k=1}^{N}\).
We randomly permute the samples once and perform a fixed \(80/20\) split into a training set and a held-out validation set (used only for evaluation and model selection).
An \emph{epoch} denotes one full pass through the training set. We use a batch size of \(160{,}000\), which in our setup matches the training-set size after the split, so each epoch corresponds to a single full-batch AdamW update (more generally, the number of batches per epoch is \(\lceil N_{\mathrm{train}}/\text{batch\_size}\rceil\)).
At the end of every epoch, we evaluate the total loss (and its components) on the \emph{entire} validation set.} 

\begin{figure}
    \centering
    \includegraphics[width=\textwidth]{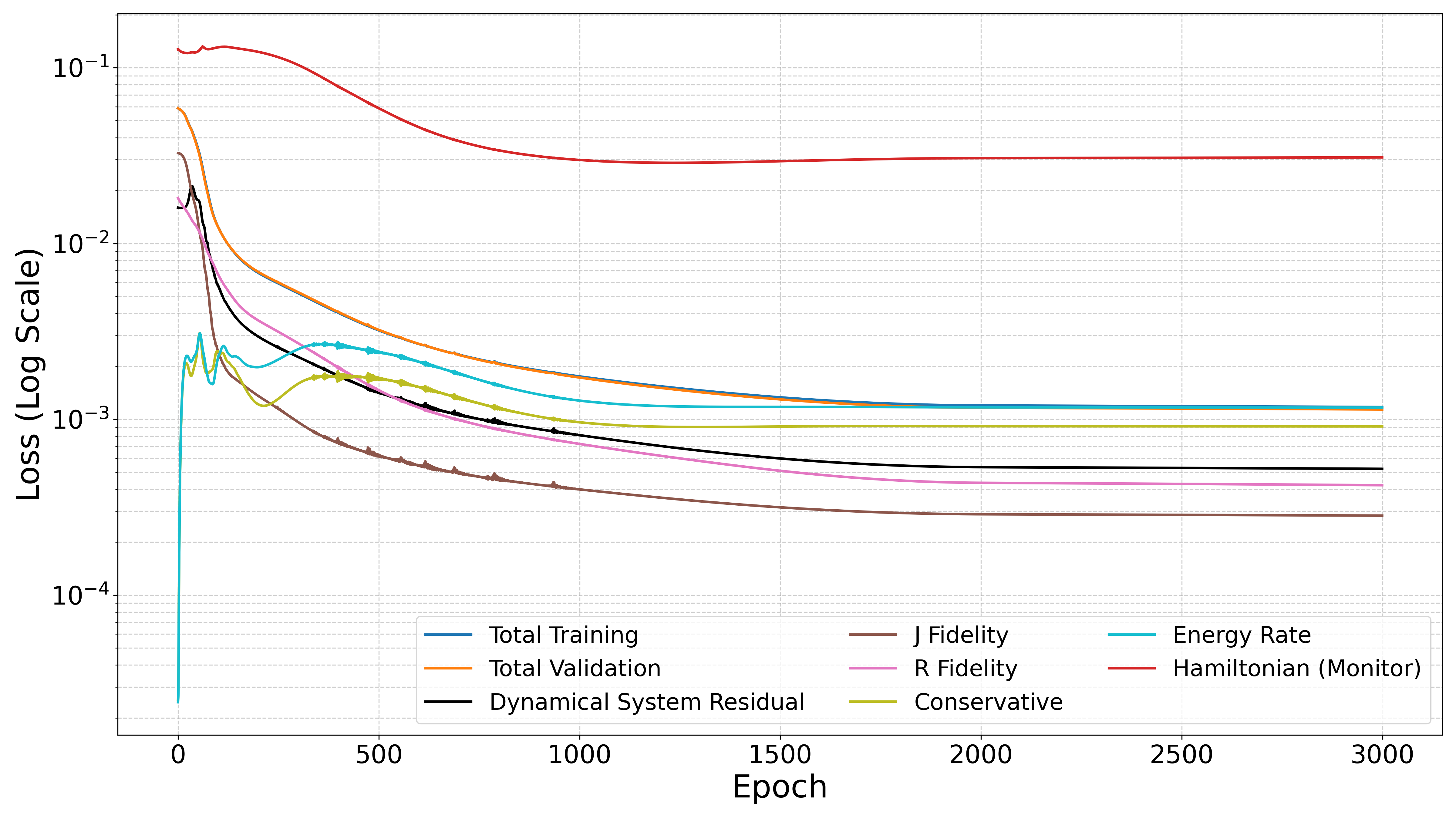}
    \caption{Training and validation loss curves over epochs. The convergence of the total loss, alongside the individual physics\textendash based loss components, demonstrates successful model training and enforcement of physical constraints.}
    \label{fig:pinn_losses}
\end{figure}

We train using the \texttt{AdamW} optimizer~\cite{Loshchilov2019AdamW} with a linear learning-rate decay from \(1\times10^{-3}\) to \(1\times10^{-5}\) during epochs 1--2000, and keep \(1\times10^{-5}\) thereafter \marius{until} epoch 3000. All experiments are performed in \texttt{JAX}~\cite{Bradbury2018JAX} (64\textendash bit precision) using \texttt{Equinox}~\cite{Kidger2021Equinox} modules and \texttt{Optax} optimizers; hyperparameters are tuned with \texttt{Optuna}~\cite{Akiba2019Optuna}. Figure~\ref{fig:pinn_losses} reports the training and validation trajectories for the total loss together with the {raw} component losses from Eq.~\eqref{eq:total_loss} (plotted before multiplying by their weights \(\lambda_{\mathrm{\tiny{DynRes}}}{=}0.5\), \(\lambda_{\tiny{J}}{=}1\), \(\lambda_{\tiny{R}}{=}1\), \(\lambda_{\tiny{Cons}}{=}0.1\), \(\lambda_{\tiny{\dot H}}{=}0.1\)). For monitoring, we also track a Hamiltonian loss (against the analytical \(H\)) that does not contribute to optimization.

Two components, the conservative\textendash invariance term \(\mathcal{L}_{\mathrm{\tiny{Cons}}}\) and the \marius{Hamiltonian}\textendash rate consistency term \(\mathcal{L}_{\tiny{\dot H}}\), start near zero, rise, and then descend alongside the others. This \marius{behavior} is expected as at initialization, the network can satisfy constraints involving \(\nabla H_\theta\) with a nearly flat Hamiltonian (so \(\nabla H_\theta\!\approx\!0\)), yielding trivially small \(\mathcal{L}_{\mathrm{\tiny{Cons}}}\) and \(\mathcal{L}_{\tiny{\dot H}}\). As the dynamics and {non-conservative} terms (\(\mathcal{L}_{\mathrm{\tiny{DynRes}}}\), \(\mathcal{L}_{\tiny R}\)) pull \(H_\theta\) away from this degenerate solution, \(\nabla H_\theta\) becomes nonzero; these losses temporarily increase and then decrease as all constraints are reconciled. Overall, the total loss decreases smoothly with a small train–validation gap. Overall, the total loss decreases smoothly with a small train--validation gap. We retain the checkpoint achieving the lowest validation loss (evaluated at the end of each epoch on the full validation set) and report all results from this checkpoint. Training is run for a fixed budget of 3000 epochs (no early-stopping criterion is used).

\begin{figure}[H]
\includegraphics[width=0.5\linewidth]{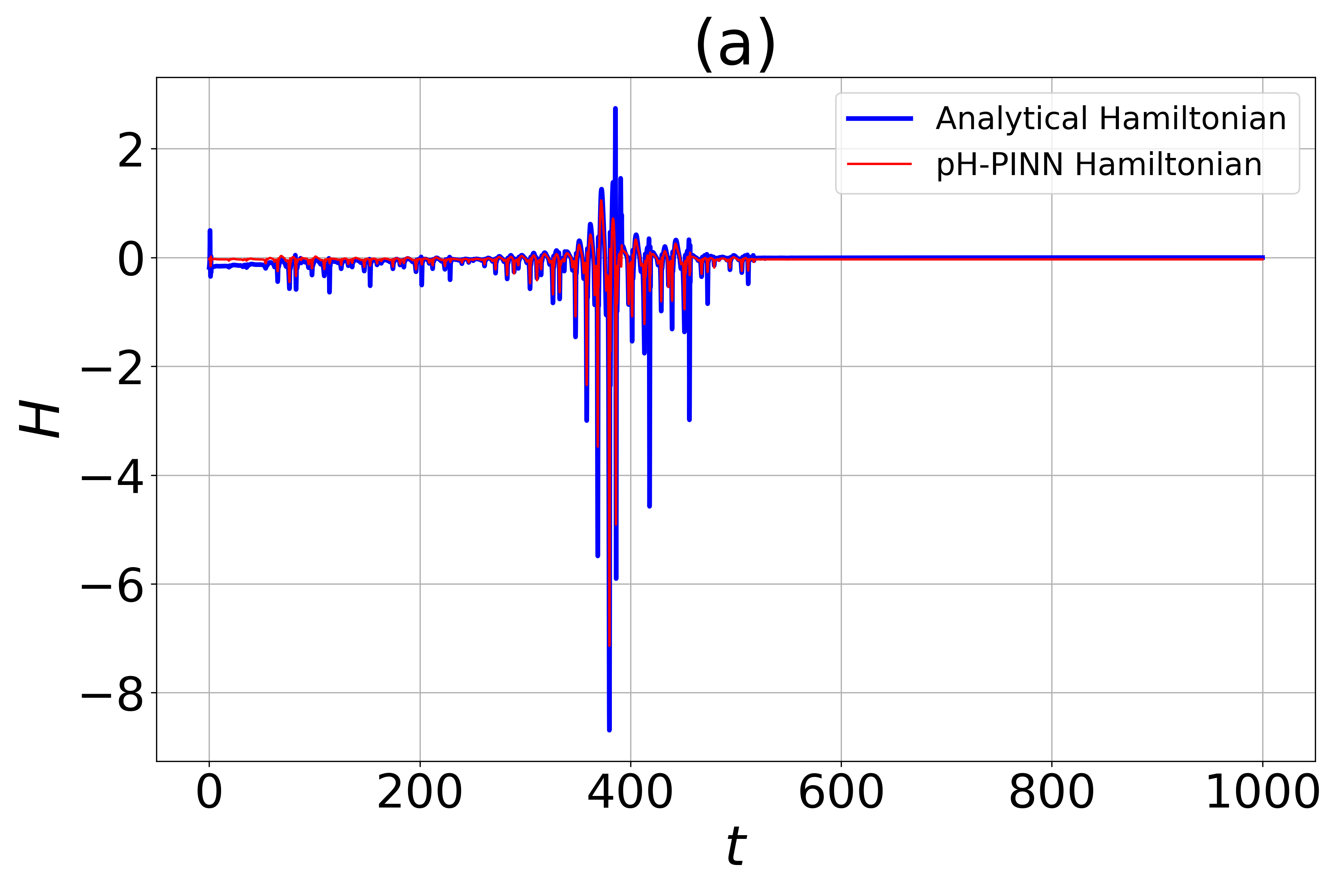}\hspace{0.01\textwidth}\includegraphics[width=0.5\linewidth]{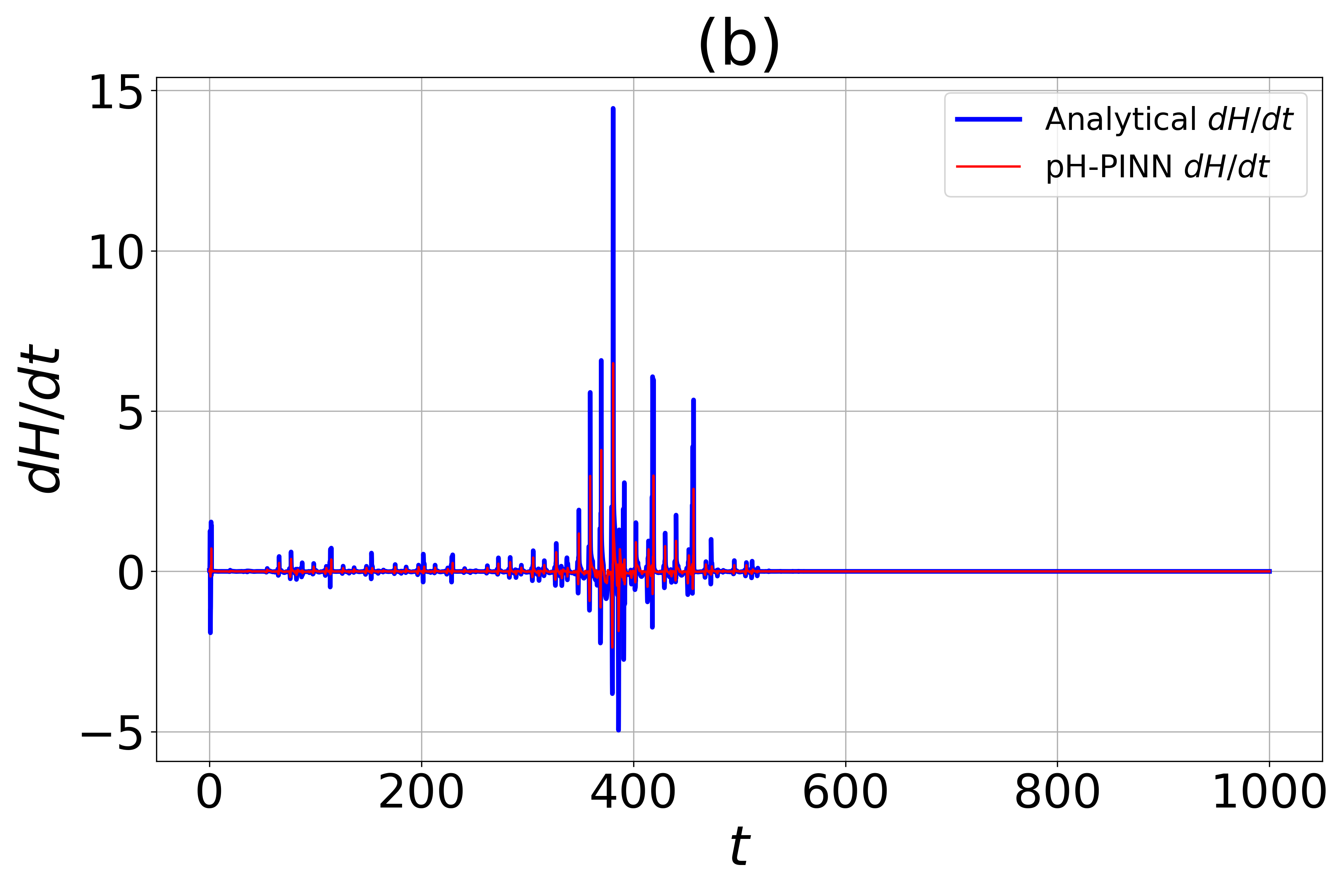}
\caption{Comparison of analytically derived quantities and those discovered by the pH\textendash PINN. Panel \textbf{(a)} shows Hamiltonian trajectories; panel \textbf{(b)} shows the time derivative of the Hamiltonian. The close agreement indicates the model has learned the underlying \marius{Hamiltonian} structure from data.}
\label{fig:pinn_hamiltonian}
\end{figure}

Figure~\ref{fig:pinn_hamiltonian} compares the learned and analytical energies.  
Panel (a) plots the pH\textendash PINN Hamiltonian \(H_\theta(\mathbf e)\) against the analytical \(H\) from Eq.~\eqref{eq:hamiltonian}, while panel (b) compares the pH\textendash PINN \marius{Hamiltonian} rate from Eq.~\eqref{eq:ph_energy_rate}, with the analytical \(\dot H\) in Eq.~\eqref{eq:Hdot}. The close alignment across trajectories shows that the model has recovered both the \marius{Hamiltonian} landscape and the \(\mathbf{J}/\mathbf{R}\) structure directly from data, without using the analytical expressions of the \marius{Hamiltonian} during training.

\begin{figure}[H]
\centering
\includegraphics[width=0.35\linewidth]{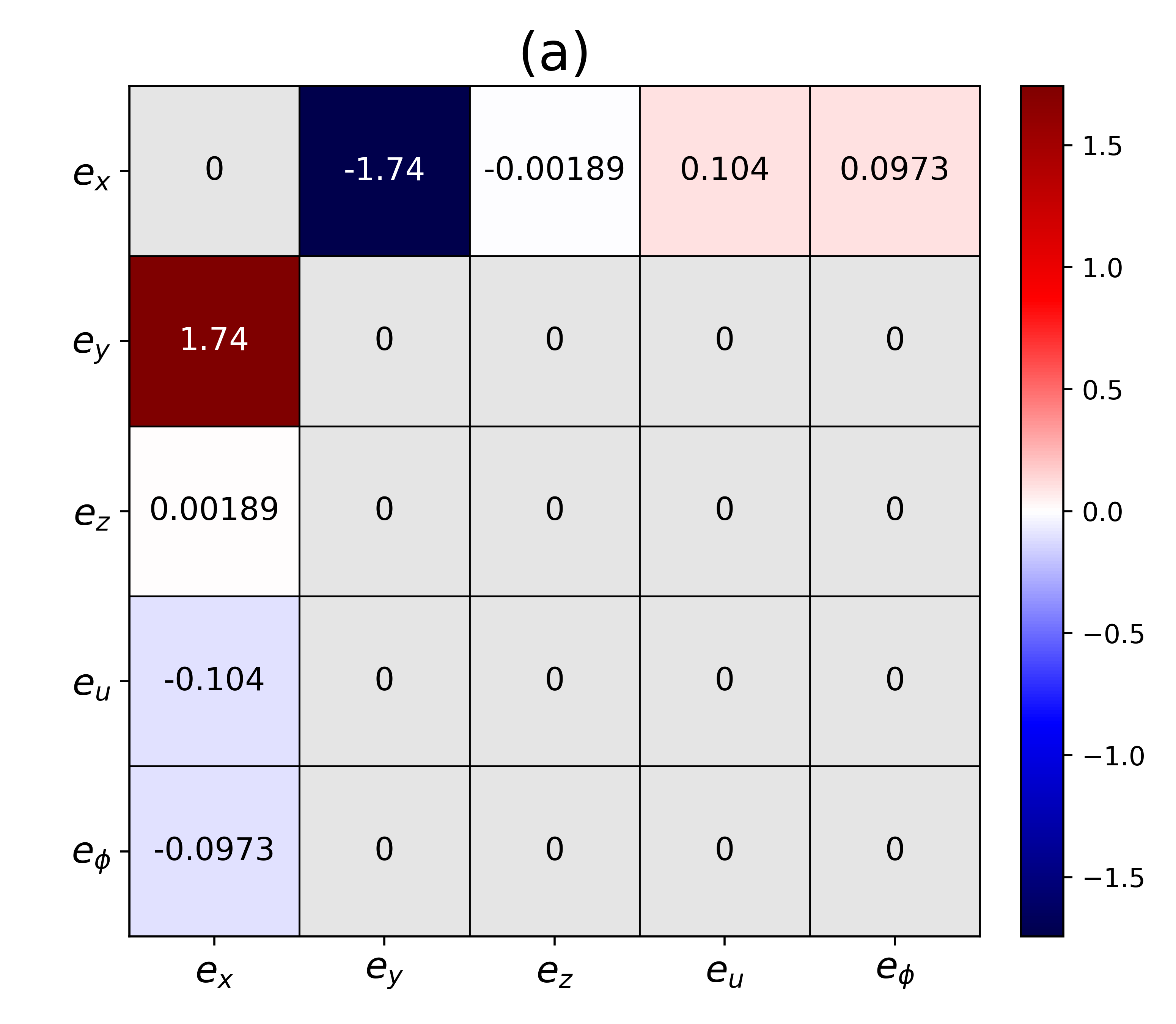}\hspace{0.05\textwidth}\includegraphics[width=0.35\linewidth]{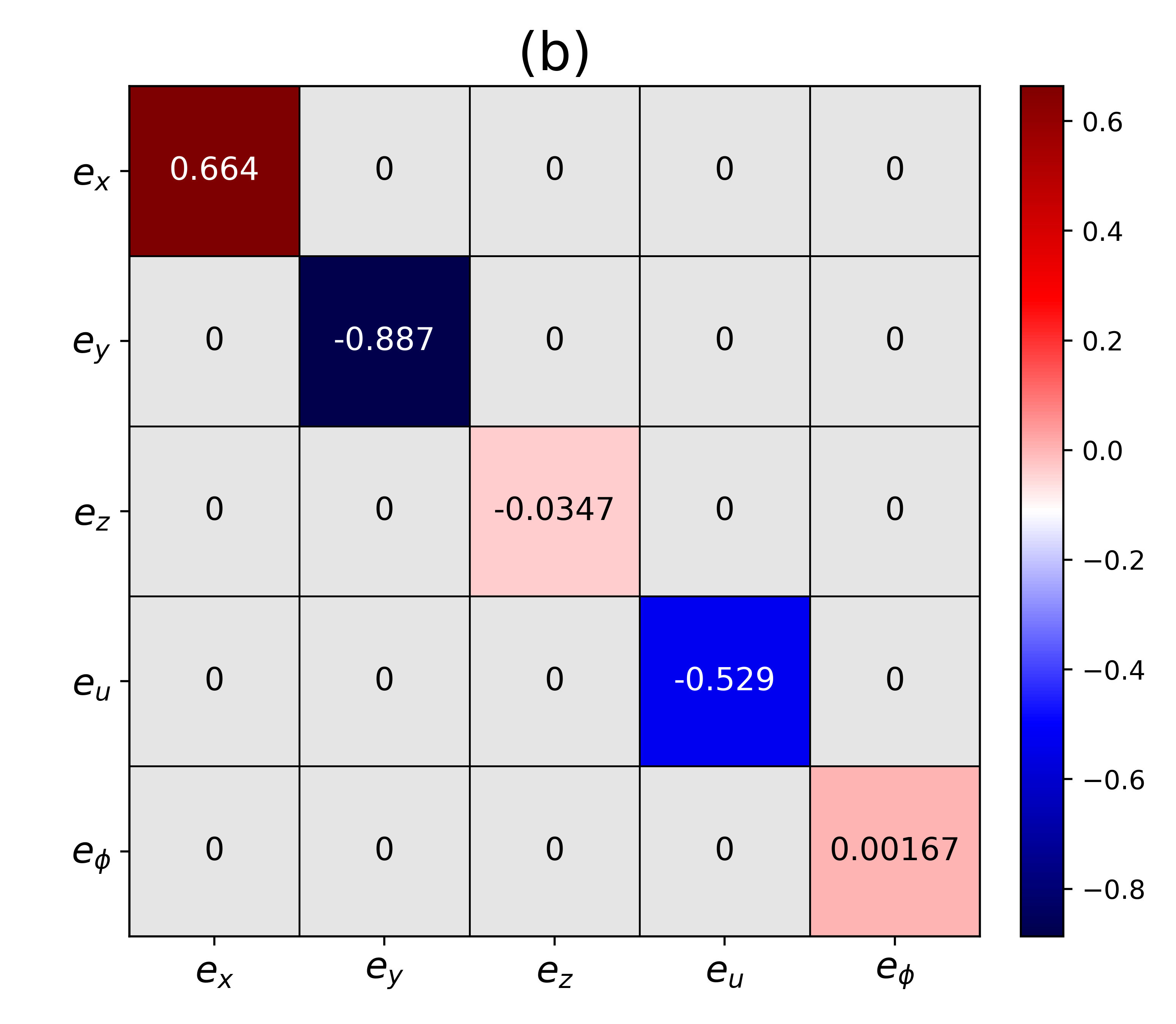}\hspace{0.02\textwidth}
\caption{Time\textendash averaged learned pH matrices. Panel \textbf{(a)}: interconnection matrix \(\langle \mathbf{J}_{\text{pred}} \rangle\); panel \textbf{(b)}: symmetric matrix \(\langle \mathbf{R}_{\text{pred}} \rangle\). The colormaps show the mean value of each entry over the entire validation dataset.}
\label{fig:pinn_matrices}
\end{figure}

Figure~\ref{fig:pinn_matrices} summarizes the learned pH factors.  
Panel~\textbf{(a)} shows the time\textendash averaged interconnection matrix \(\langle \JJ_{\text{pred}}\rangle\), which is skew\textendash symmetric and sparse. Entries related to \(e_z\) are relatively small compared to the others, consistent with the conservative field in Eq.~\eqref{cons_diss}, where no \(e_z\) dependence appears (even though, {a priori}, \(\partial H/\partial e_z\) need not vanish).  
Eq.~\eqref{cons_diss} further shows that the \(e_u\) and \(e_\phi\) channels couple conservatively only through \(e_x\); accordingly, the network learns similar magnitudes for the \(e_x\!\leftrightarrow\!e_u\) and \(e_x\!\leftrightarrow\!e_\phi\) couplings.  
Panel~\textbf{(b)} presents the symmetric matrix \(\langle \RR_{\text{pred}}\rangle\), which is diagonal as implied by {separable-gradient} assumption and Eq.~\eqref{cons_diss}.

\section{Summary and conclusions}
\label{sec:summary_conclusions}
In this work, we established a rigorous and data-driven framework for chaotic synchronization in a five-dimensional Hindmarsh--Rose neuron model with electromagnetic induction and switchable memristive autapse. Our contributions are threefold.

\smallskip
\noindent
\textit{(i) \marius{Local} synchronization via Lyapunov analysis.}
\marius{For two diffusively coupled neurons, we derived the linearized transverse error system and analyzed the stability of the synchronization manifold \(\mathcal{M}\) associated with this linearized dynamics using a quadratic Lyapunov function. By bounding the time-varying coefficients along bounded reference trajectories and enforcing positivity of an explicit diagonal Lyapunov matrix, we obtained verifiable sufficient conditions for transverse stability. In dissipative memristive switching regimes, these conditions yield \(\dot V\le -\lambda_{\min}(\mathbf M)\|\mathbf e\|^2\) and, via a Rayleigh bound and Barbalat’s lemma, asymptotic convergence \(\|\mathbf e(t)\|\to 0\) (complete synchronization for the linearized error system). When the switching visits non-dissipative regimes, the same framework leads to a practical stability estimate of the form \(\dot V\le -2\alpha V + C\), implying uniform boundedness and an explicit ultimate bound on the synchronization error (practical synchronization). The analysis makes explicit how electrical coupling and intrinsic parameters jointly suppress transverse perturbations. A fully global nonlinear synchronization proof would require explicit bounds on the neglected nonlinear remainder terms, which we leave for future work.}

\smallskip
\noindent\textit{(ii) \marius{Synchronization Hamiltonian} via Helmholtz decomposition.}
We decomposed the error vector field into conservative and dissipative components and derived a closed-form \marius{synchronization Hamiltonian} $H$ and its rate identity $\dot H$ for the linearized error flow. This yields a quantitative measure of the energetic cost of synchrony and clarifies the respective roles of interconnection, damping, and the memristive offset in shaping energy exchange.

\smallskip
\noindent\textit{(iii) Learning the \marius{synchronization Hamiltonian} and its time rate with a port--Hamiltonian PINN.}
We proposed a port--Hamiltonian physics-informed neural network (pH-PINN) that embeds the conservative/{non-conservative} structure into training through physically motivated losses. The model learns $(H_\theta,\JJ_\phi,\RR_\psi)$ from data while enforcing skew-symmetry of $\JJ_\phi$, diagonality of $\RR_\psi$, and the \marius{Hamiltonian} balance $\dot H_\theta=-\nabla H_\theta^\top \RR_\psi \nabla H_\theta$. Numerically, the learned Hamiltonian and its rate closely match their analytical counterparts.

Direct simulations confirm complete synchronization and \marius{show decay of the synchronization Hamiltonian to zero as $t\to\infty$}. Parameter sweeps across $(k,\rho,m,g_e)$ reveal consistent trends between Lyapunov and Hamiltonian diagnostics: stronger electrical coupling accelerates synchronization, whereas excessive flux coupling can inject energy and delay convergence. Heatmaps of $\dot V$ and $\dot H$ agree across the parameter planes, supporting the equivalence of the two perspectives.

Our proof and learned structure are derived for the linearized error dynamics in a two-neuron setting with a separable-gradient prior. Extending the analysis to (a) nonidentical neurons and parameter mismatch, (b) larger networks with arbitrary topology, (c) stochastic forcing and measurement noise, and (d) partial observability and latent-state identification are natural next steps. Methodologically, lifting the separability prior via adaptive masking, handling the piecewise memristive nonlinearity beyond linearization, and incorporating uncertainty quantification into the pH-PINN would broaden applicability. \marius{Morever,} deploying the framework on experimental data and leveraging it for structure-preserving control design (\textit{e.g.,} energy shaping and damping injection) are promising directions.

In summary, the paper provides (I) verifiable conditions for synchronization in a biophysically enriched Hindmarsh–Rose model, (II) a closed-form \marius{synchronization Hamiltonian} function together with an explicit rate law for its evolution, and (III) a principled learning architecture that recovers an equivalent Hamiltonian-based structure from data (up to scaling and additive constants). Collectively, these results connect rigorous dynamical systems analysis with physics-informed machine learning for nonlinear neuronal synchronization and provide a template for a port-Hamiltonian-aware modeling in broader classes of complex systems.

\section*{Acknowledgements} This work was funded by the Department of Data Science (DDS), Friedrich-Alexander-Universit\"at Erlangen-Nürnberg, Germany, and the Deutsche Forschungsgemeinschaft (DFG, German Research Foundation) via the grant YA 764/1-1 to M.E.Y—Project No. 456989199. 


\section*{Data availability}
The code developed and used to generate the data and train the models supporting the findings of this study is publicly available \cite{code}.


\begin{thebibliography}{10}
\expandafter\ifx\csname url\endcsname\relax
  \def\url#1{\texttt{#1}}\fi
\expandafter\ifx\csname urlprefix\endcsname\relax\def\urlprefix{URL }\fi
\expandafter\ifx\csname href\endcsname\relax
  \def\href#1#2{#2} \def\path#1{#1}\fi

\bibitem{neustadter2016eeg}
E.~Neustadter, K.~Mathiak, B.~I. Turetsky, {EEG} and {MEG} probes of schizophrenia pathophysiology, in: T.~Abel, T.~Nickl-Jockschat (Eds.), The Neurobiology of Schizophrenia, Academic Press, 2016, Ch.~13, pp. 213--236.
\newblock \href {https://doi.org/10.1016/B978-0-12-801829-3.00021-5} {\path{doi:10.1016/B978-0-12-801829-3.00021-5}}.

\bibitem{lehnertz2009synchronization}
K.~Lehnertz, S.~Bialonski, M.-T. Horstmann, D.~Krug, A.~Rothkegel, M.~Staniek, T.~Wagner, Synchronization phenomena in human epileptic brain networks, Journal of Neuroscience Methods 183~(1) (2009) 42--48.
\newblock \href {https://doi.org/10.1016/j.jneumeth.2009.05.015} {\path{doi:10.1016/j.jneumeth.2009.05.015}}.

\bibitem{borges2023intermittency}
F.~S. Borges, E.~C. Gabrick, P.~R. Protachevicz, G.~S.~V. Higa, E.~L. Lameu, P.~X.~R. Rodriguez, M.~S.~A. Ferraz, J.~Szezech, J.~D., A.~M. Batista, A.~H. Kihara, Intermittency properties in a temporal lobe epilepsy model, Epilepsy \& Behavior 139 (2023) 109072.
\newblock \href {https://doi.org/10.1016/j.yebeh.2022.109072} {\path{doi:10.1016/j.yebeh.2022.109072}}.

\bibitem{protachevicz2019bistable}
P.~R. Protachevicz, F.~S. Borges, E.~L. Lameu, P.~Ji, K.~C. Iarosz, A.~H. Kihara, I.~L. Caldas, J.~Szezech, Jose~D., M.~S. Baptista, E.~E.~N. Macau, C.~G. Antonopoulos, A.~M. Batista, J.~Kurths, Bistable firing pattern in a neural network model, Frontiers in Computational Neuroscience 13 (2019) 19.
\newblock \href {https://doi.org/10.3389/fncom.2019.00019} {\path{doi:10.3389/fncom.2019.00019}}.

\bibitem{boccaletti2006complex}
S.~Boccaletti, V.~Latora, Y.~Moreno, M.~Chavez, D.-U. Hwang, Complex networks: Structure and dynamics, Physics Reports 424~(4-5) (2006) 175--308.
\newblock \href {https://doi.org/10.1016/j.physrep.2005.10.009} {\path{doi:10.1016/j.physrep.2005.10.009}}.

\bibitem{osipov2007synchronization}
G.~V. Osipov, J.~Kurths, C.~Zhou, Synchronization in Oscillatory Networks, Springer Series in Synergetics, Springer Berlin Heidelberg, Berlin, Heidelberg, 2007.
\newblock \href {https://doi.org/10.1007/978-3-540-71269-5} {\path{doi:10.1007/978-3-540-71269-5}}.

\bibitem{rosenblum1996phase}
M.~G. Rosenblum, A.~S. Pikovsky, J.~Kurths, Phase synchronization of chaotic oscillators, Physical Review Letters 76~(11) (1996) 1804--1807.
\newblock \href {https://doi.org/10.1103/PhysRevLett.76.1804} {\path{doi:10.1103/PhysRevLett.76.1804}}.

\bibitem{pikovsky1996synchronization}
A.~S. Pikovsky, M.~G. Rosenblum, J.~Kurths, Synchronization in a population of globally coupled chaotic oscillators, Europhysics Letters 34~(3) (1996) 165--170.
\newblock \href {https://doi.org/10.1209/epl/i1996-00433-3} {\path{doi:10.1209/epl/i1996-00433-3}}.

\bibitem{parlitz1996experimental}
U.~Parlitz, L.~Junge, W.~Lauterborn, L.~Kocarev, Experimental observation of phase synchronization, Physical Review E 54~(2) (1996) 2115--2117.
\newblock \href {https://doi.org/10.1103/PhysRevE.54.2115} {\path{doi:10.1103/PhysRevE.54.2115}}.

\bibitem{pietras2019network}
B.~Pietras, A.~Daffertshofer, Network dynamics of coupled oscillators and phase reduction techniques, Physics Reports 819 (2019) 1--105.
\newblock \href {https://doi.org/10.1016/j.physrep.2019.06.001} {\path{doi:10.1016/j.physrep.2019.06.001}}.

\bibitem{fell2011role}
J.~Fell, N.~Axmacher, The role of phase synchronization in memory processes, Nature Reviews Neuroscience 12~(2) (2011) 105--118.
\newblock \href {https://doi.org/10.1038/nrn2979} {\path{doi:10.1038/nrn2979}}.

\bibitem{dahms2012cluster}
T.~Dahms, J.~Lehnert, E.~Sch{\"o}ll, Cluster and group synchronization in delay-coupled networks, Physical Review E 86~(1) (2012) 016202.
\newblock \href {https://doi.org/10.1103/PhysRevE.86.016202} {\path{doi:10.1103/PhysRevE.86.016202}}.

\bibitem{jalan2003self}
S.~Jalan, R.~Amritkar, Self-organized and driven phase synchronization in coupled maps, Physical Review Letters 90~(1) (2003) 014101.
\newblock \href {https://doi.org/10.1103/PhysRevLett.90.014101} {\path{doi:10.1103/PhysRevLett.90.014101}}.

\bibitem{amritkar2003self}
R.~Amritkar, S.~Jalan, Self-organized and driven phase synchronization in coupled map networks, Physica A: Statistical Mechanics and its Applications 321~(1-2) (2003) 220--225.
\newblock \href {https://doi.org/10.1016/S0378-4371(02)01750-8} {\path{doi:10.1016/S0378-4371(02)01750-8}}.

\bibitem{abarbanel1996generalized}
H.~D. Abarbanel, N.~F. Rulkov, M.~M. Sushchik, Generalized synchronization of chaos: The auxiliary system approach, Physical Review E 53~(5) (1996) 4528.
\newblock \href {https://doi.org/10.1103/PhysRevE.53.4528} {\path{doi:10.1103/PhysRevE.53.4528}}.

\bibitem{zheng2000generalized}
Z.~Zheng, G.~Hu, Generalized synchronization versus phase synchronization, Physical Review E 62~(6) (2000) 7882.
\newblock \href {https://doi.org/10.1103/PhysRevE.62.7882} {\path{doi:10.1103/PhysRevE.62.7882}}.

\bibitem{femat2002synchronization}
R.~Femat, G.~Sol{'i}s-Perales, Synchronization of chaotic systems with different order, Physical Review E 65~(3) (2002) 036226.
\newblock \href {https://doi.org/10.1103/PhysRevE.65.036226} {\path{doi:10.1103/PhysRevE.65.036226}}.

\bibitem{bowong2006adaptive}
S.~Bowong, P.~V.~E. McClintock, Adaptive synchronization between chaotic dynamical systems of different order, Physics Letters A 358~(2) (2006) 134--141.
\newblock \href {https://doi.org/10.1016/j.physleta.2006.05.006} {\path{doi:10.1016/j.physleta.2006.05.006}}.

\bibitem{bowong2004stability}
S.~Bowong, Stability analysis for the synchronization of chaotic systems with different order: application to secure communications, Physics Letters A 326~(1-2) (2004) 102--113.
\newblock \href {https://doi.org/10.1016/j.physleta.2004.04.004} {\path{doi:10.1016/j.physleta.2004.04.004}}.

\bibitem{kobiolka2025reduced}
J.~Kobiolka, J.~Habermann, M.~E. Yamakou, Reduced-order adaptive synchronization in a chaotic neural network with parameter mismatch: a dynamical system versus machine learning approach, Nonlinear Dynamics 113~(10) (2025) 10989--11008.
\newblock \href {https://doi.org/10.1007/s11071-024-10821-6} {\path{doi:10.1007/s11071-024-10821-6}}.

\bibitem{qing2009increasing}
Q.-Y. Miao, J.-A. Fang, Y.~Tang, A.-H. Dong, Increasing-order projective synchronization of chaotic systems with time delay, Chinese Physics Letters 26~(5) (2009) 050501.
\newblock \href {https://doi.org/10.1088/0256-307X/26/5/050501} {\path{doi:10.1088/0256-307X/26/5/050501}}.

\bibitem{al2011adaptive}
M.~M. Al-sawalha, M.~S.~M. Noorani, Adaptive increasing-order synchronization and anti-synchronization of chaotic systems with uncertain parameters, Chinese Physics Letters 28~(11) (2011) 110507.
\newblock \href {https://doi.org/10.1088/0256-307X/28/11/110507} {\path{doi:10.1088/0256-307X/28/11/110507}}.

\bibitem{pecora2015synchronization}
L.~M. Pecora, T.~L. Carroll, Synchronization of chaotic systems, Chaos: An Interdisciplinary Journal of Nonlinear Science 25~(9) (2015).
\newblock \href {https://doi.org/10.1063/1.4917383} {\path{doi:10.1063/1.4917383}}.

\bibitem{yamakou2023synchronization}
M.~E. Yamakou, M.~Desroches, S.~Rodrigues, Synchronization in stdp-driven memristive neural networks with time-varying topology, Journal of Biological Physics 49~(4) (2023) 483--507.
\newblock \href {https://doi.org/10.1007/s10867-023-09642-2} {\path{doi:10.1007/s10867-023-09642-2}}.

\bibitem{fujisaka1983stability}
H.~Fujisaka, T.~Yamada, Stability theory of synchronized motion in coupled-oscillator systems, Progress of Theoretical Physics 69~(1) (1983) 32--47.
\newblock \href {https://doi.org/10.1143/PTP.69.32} {\path{doi:10.1143/PTP.69.32}}.

\bibitem{yamakou2016ratcheting}
E.~M. Yamakou, E.~M. Inack, F.~M. Moukam~Kakmeni, Ratcheting and energetic aspects of synchronization in coupled bursting neurons, Nonlinear Dynamics 83~(1-2) (2016) 541--554.
\newblock \href {https://doi.org/10.1007/s11071-015-2346-0} {\path{doi:10.1007/s11071-015-2346-0}}.

\bibitem{tang2014synchronization}
Y.~Tang, F.~Qian, H.~Gao, J.~Kurths, Synchronization in complex networks and its application---a survey of recent advances and challenges, Annual Reviews in Control 38~(2) (2014) 184--198.
\newblock \href {https://doi.org/10.1016/j.arcontrol.2014.09.003} {\path{doi:10.1016/j.arcontrol.2014.09.003}}.

\bibitem{boccaletti2002synchronization}
S.~Boccaletti, J.~Kurths, G.~Osipov, D.~L. Valladares, C.~Zhou, The synchronization of chaotic systems, Physics Reports 366~(1-2) (2002) 1--101.
\newblock \href {https://doi.org/10.1016/S0370-1573(02)00137-0} {\path{doi:10.1016/S0370-1573(02)00137-0}}.

\bibitem{arenas2008synchronization}
A.~Arenas, A.~D{'\i}az-Guilera, J.~Kurths, Y.~Moreno, C.~Zhou, Synchronization in complex networks, Physics Reports 469~(3-4) (2008) 93--153.
\newblock \href {https://doi.org/10.1016/j.physrep.2008.09.002} {\path{doi:10.1016/j.physrep.2008.09.002}}.

\bibitem{protachevicz2021emergence}
P.~R. Protachevicz, M.~Hansen, K.~C. Iarosz, I.~L. Caldas, A.~M. Batista, J.~Kurths, Emergence of neuronal synchronisation in coupled areas, Frontiers in Computational Neuroscience 15 (2021) 663408.
\newblock \href {https://doi.org/10.3389/fncom.2021.663408} {\path{doi:10.3389/fncom.2021.663408}}.

\bibitem{borges2017synchronised}
F.~S. Borges, P.~R. Protachevicz, E.~L. Lameu, R.~C. Bonetti, K.~C. Iarosz, I.~L. Caldas, M.~S. Baptista, A.~M. Batista, Synchronised firing patterns in a random network of adaptive exponential integrate-and-fire neuron model, Neural Networks 90 (2017) 1--7.
\newblock \href {https://doi.org/10.1016/j.neunet.2017.03.005} {\path{doi:10.1016/j.neunet.2017.03.005}}.

\bibitem{Raissi2019}
M.~Raissi, P.~Perdikaris, G.~E. Karniadakis, Physics-informed neural networks: A deep learning framework for solving forward and inverse problems involving nonlinear partial differential equations, Journal of Computational Physics 378 (2019) 686--707.
\newblock \href {https://doi.org/10.1016/j.jcp.2018.10.045} {\path{doi:10.1016/j.jcp.2018.10.045}}.

\bibitem{Wang2022CausalityPINN}
S.~Wang, S.~Sankaran, P.~Perdikaris, Respecting causality is all you need for training physics-informed neural networks, arXiv preprint arXiv:2203.07404 (2022).
\newblock \href {https://doi.org/10.48550/arXiv.2203.07404} {\path{doi:10.48550/arXiv.2203.07404}}.

\bibitem{Cuomo2022PINNReview}
S.~Cuomo, V.~Schiano Di~Cola, F.~Giampaolo, G.~Rozza, M.~Raissi, F.~Piccialli, Scientific machine learning through physics-informed neural networks: Where we are and what’s next, Journal of Scientific Computing 92~(3) (2022) 88.
\newblock \href {https://doi.org/10.1007/s10915-022-01939-z} {\path{doi:10.1007/s10915-022-01939-z}}.

\bibitem{savaliya2026self}
D.~Savaliya, M.~E. Yamakou, Self-induced stochastic resonance: A physics-informed machine learning approach, Chaos, Solitons \& Fractals 207 (2026) 117998.
\newblock \href {https://doi.org/10.1016/j.chaos.2026.117998} {\path{doi:10.1016/j.chaos.2026.117998}}.

\bibitem{Greydanus2019HNN}
S.~Greydanus, M.~Dzamba, J.~Yosinski, Hamiltonian neural networks, in: H.~Wallach, H.~Larochelle, A.~Beygelzimer, F.~d'Alch{'e} Buc, E.~Fox, R.~Garnett (Eds.), Advances in Neural Information Processing Systems, Vol.~32, Curran Associates, Inc., Vancouver, Canada, 2019, pp. 15353--15363.
\newblock \href {http://arxiv.org/abs/1906.01563} {\path{arXiv:1906.01563}}.

\bibitem{finzi2020simplifying}
M.~Finzi, K.~A. Wang, A.~G. Wilson, Simplifying hamiltonian and lagrangian neural networks via explicit constraints, in: H.~Larochelle, M.~Ranzato, R.~Hadsell, M.~F. Balcan, H.~Lin (Eds.), Advances in Neural Information Processing Systems, Vol.~33, Curran Associates, Inc., 2020, pp. 13880--13889.
\newblock \href {http://arxiv.org/abs/2010.13581} {\path{arXiv:2010.13581}}.

\bibitem{Chen2021}
Y.~Chen, T.~Matsubara, T.~Yaguchi, Neural symplectic form: Learning hamiltonian equations on general coordinate systems, in: M.~Ranzato, A.~Beygelzimer, Y.~Dauphin, P.~S. Liang, J.~Wortman~Vaughan (Eds.), Advances in Neural Information Processing Systems, Vol.~34, Curran Associates, Inc., 2021, pp. 16659--16670, \url{https://proceedings.neurips.cc/paper/2021/hash/8b519f198dd26772e3e82874826b04aa-Abstract.html}.

\bibitem{vanderSchaft2014}
A.~van~der Schaft, D.~Jeltsema, Port-Hamiltonian Systems Theory: An Introductory Overview, Vol.~1 of Foundations and Trends in Systems and Control, Now Publishers Inc., Boston--Delft, 2014.
\newblock \href {https://doi.org/10.1561/2600000002} {\path{doi:10.1561/2600000002}}.

\bibitem{Desai2021PortHNN}
S.~Desai, M.~Mattheakis, D.~Sondak, P.~Protopapas, S.~Roberts, Port-hamiltonian neural networks for learning explicit time-dependent dynamical systems, arXiv preprint arXiv:2107.08024, journal ref: Phys. Rev. E 104, 034312 (2021) (2021).
\newblock \href {https://doi.org/10.48550/arXiv.2107.08024} {\path{doi:10.48550/arXiv.2107.08024}}.

\bibitem{neary2023phnn}
C.~Neary, U.~Topcu, Compositional learning of dynamical system models using port-hamiltonian neural networks, in: N.~Matni, M.~Morari, G.~J. Pappas (Eds.), Proceedings of The 5th Annual Learning for Dynamics and Control Conference (L4DC), Vol. 211 of Proceedings of Machine Learning Research, PMLR, Philadelphia, PA, USA, 2023, pp. 679--691.
\newblock \href {http://arxiv.org/abs/2212.00893} {\path{arXiv:2212.00893}}.

\bibitem{Moradi2025OePHNN}
S.~Moradi, G.~I. Beintema, N.~Jaensson, R.~T{\'o}th, M.~Schoukens, Port-hamiltonian neural networks with output error noise models. Preprint submitted to \emph{Automatica} (2025).
\newblock \href {https://doi.org/10.48550/arXiv.2502.14432} {\path{doi:10.48550/arXiv.2502.14432}}.

\bibitem{di2025port}
L.~Di~Persio, M.~Ehrhardt, Y.~Outaleb, S.~Rizzotto, Port-hamiltonian neural networks: From theory to simulation of interconnected stochastic systems, arXiv preprint arXiv:2509.06674 (2025).
\newblock \href {https://doi.org/10.48550/arXiv.2509.06674} {\path{doi:10.48550/arXiv.2509.06674}}.

\bibitem{ortega2024learnability}
J.-P. Ortega, D.~Yin, Learnability of linear port-hamiltonian systems, Journal of Machine Learning Research 25~(68) (2024) 1--56.
\newblock \href {https://doi.org/10.48550/arXiv.2303.15779} {\path{doi:10.48550/arXiv.2303.15779}}.

\bibitem{hindmarsh1984model}
J.~L. Hindmarsh, R.~M. Rose, A model of neuronal bursting using three coupled first order differential equations, Proceedings of the Royal Society of London. Series B, Biological Sciences 221~(1222) (1984) 87--102.
\newblock \href {https://doi.org/10.1098/rspb.1984.0024} {\path{doi:10.1098/rspb.1984.0024}}.

\bibitem{selverston2000reliable}
A.~I. Selverston, M.~I. Rabinovich, H.~D.~I. Abarbanel, R.~Elson, A.~Sz{"u}cs, R.~D. Pinto, R.~Huerta, P.~Varona, Reliable circuits from irregular neurons: a dynamical approach to understanding central pattern generators, Journal of Physiology-Paris 94~(5-6) (2000) 357--374.
\newblock \href {https://doi.org/10.1016/S0928-4257(00)01101-3} {\path{doi:10.1016/S0928-4257(00)01101-3}}.

\bibitem{lv2016model}
M.~Lv, C.~Wang, G.~Ren, J.~Ma, X.~Song, Model of electrical activity in a neuron under magnetic flow effect, Nonlinear Dynamics 85~(3) (2016) 1479--1490.
\newblock \href {https://doi.org/10.1007/s11071-016-2773-6} {\path{doi:10.1007/s11071-016-2773-6}}.

\bibitem{yamakou2020chaotic}
M.~E. Yamakou, Chaotic synchronization of memristive neurons: Lyapunov function versus hamilton function, Nonlinear Dynamics 101~(1) (2020) 487--500.
\newblock \href {https://doi.org/10.1007/s11071-020-05715-2} {\path{doi:10.1007/s11071-020-05715-2}}.

\bibitem{zhang2024switchable}
J.~Zhang, Z.~Li, Switchable memristor-based hindmarsh--rose neuron under electromagnetic radiation, Nonlinear Dynamics 112~(8) (2024) 6647--6662.
\newblock \href {https://doi.org/10.1007/s11071-024-09399-w} {\path{doi:10.1007/s11071-024-09399-w}}.

\bibitem{ma2017mode}
J.~Ma, Y.~Wang, C.~Wang, Y.~Xu, G.~Ren, Mode selection in electrical activities of myocardial cell exposed to electromagnetic radiation, Chaos, Solitons \& Fractals 99 (2017) 219--225.
\newblock \href {https://doi.org/10.1016/j.chaos.2017.04.016} {\path{doi:10.1016/j.chaos.2017.04.016}}.

\bibitem{Bradbury2018JAX}
J.~Bradbury, R.~Frostig, P.~Hawkins, M.~J. Johnson, C.~Leary, D.~Maclaurin, S.~Wanderman{-}Milne, {JAX}: Composable transformations of python+numpy programs, \url{https://github.com/google/jax} (2018).

\bibitem{diffrax}
P.~Kidger, Diffrax, \url{https://github.com/patrick-kidger/diffrax}, gitHub repository.

\bibitem{Tsitouras2011}
C.~Tsitouras, Runge--kutta pairs of order 5(4) satisfying only the first column simplifying assumption, Computers \& Mathematics with Applications 62~(2) (2011) 770--775.
\newblock \href {https://doi.org/10.1016/j.camwa.2011.06.002} {\path{doi:10.1016/j.camwa.2011.06.002}}.

\bibitem{krasovskii1963stability}
N.~N. Krasovskii, Stability of Motion: Applications of {Lyapunov}'s Second Method to Differential Systems and Equations with Delay, Stanford University Press, Stanford, CA, 1963, english translation of the 1959 Russian edition.

\bibitem{kobe1986helmholtz}
D.~H. Kobe, Helmholtz's theorem revisited, American Journal of Physics 54~(6) (1986) 552--554.
\newblock \href {https://doi.org/10.1119/1.14562} {\path{doi:10.1119/1.14562}}.

\bibitem{chun2016calculation}
C.-N. Wang, Y.~Wang, J.~Ma, Calculation of hamilton energy function of dynamical system by using helmholtz theorem, Acta Physica Sinica 65~(24) (2016) 240501.
\newblock \href {https://doi.org/10.7498/aps.65.240501} {\path{doi:10.7498/aps.65.240501}}.

\bibitem{ma2017calculation}
J.~Ma, F.~Wu, W.~Jin, P.~Zhou, T.~Hayat, Calculation of hamilton energy and control of dynamical systems with different types of attractors, Chaos: An Interdisciplinary Journal of Nonlinear Science 27~(5) (2017).
\newblock \href {https://doi.org/10.1063/1.4983469} {\path{doi:10.1063/1.4983469}}.

\bibitem{Zhong2019SymODEN}
Y.~D. Zhong, B.~Dey, A.~Chakraborty, Symplectic {ODE}-net: Learning hamiltonian dynamics with control, arXiv (sep 2019).
\newblock \href {https://doi.org/10.48550/arXiv.1909.12077} {\path{doi:10.48550/arXiv.1909.12077}}.

\bibitem{Loshchilov2019AdamW}
I.~Loshchilov, F.~Hutter, Decoupled weight decay regularization, in: International Conference on Learning Representations, New Orleans, LA, USA, 2019.
\newblock \href {https://doi.org/10.48550/arXiv.1711.05101} {\path{doi:10.48550/arXiv.1711.05101}}.

\bibitem{Kidger2021Equinox}
P.~Kidger, Equinox: Neural networks in {JAX} via functional programming, \url{https://github.com/patrick-kidger/equinox} (2021).

\bibitem{Akiba2019Optuna}
T.~Akiba, S.~Sano, T.~Yanase, T.~Ohta, M.~Koyama, Optuna: A next-generation hyperparameter optimization framework, in: Proceedings of the 25th ACM SIGKDD International Conference on Knowledge Discovery \& Data Mining, KDD '19, Association for Computing Machinery, Anchorage, AK, USA, 2019, pp. 2623--2631.
\newblock \href {https://doi.org/10.1145/3292500.3330701} {\path{doi:10.1145/3292500.3330701}}.

\bibitem{code}
\url{https://github.com/Behnam2553/port-Hamiltonian-PINN-and-HR-neuron.git} (2025).

\end{thebibliography}

\end{document}